\documentclass[twocolumn, twocolappendix]{aastex701}
\usepackage{xspace}
\usepackage{amsmath,amssymb,amstext}
\usepackage{xcolor}
\usepackage{gensymb}
\usepackage{multirow}
\usepackage{rotating}
\usepackage{tabularx}
\usepackage{comment}
\usepackage[version=4]{mhchem} 
\usepackage{enumitem}
\setlist[itemize]{noitemsep, topsep=0pt, parsep=0pt, partopsep=0pt}

\newcommand{\Gaia}{\textit{Gaia}\xspace}

\newcommand{\bjdtdb}{\ensuremath{\mathrm{BJD}_\mathrm{TDB}}\xspace}
\newcommand{\Rstar}{\ensuremath{R_{\star}}\xspace} 
\newcommand{\Mstar}{\ensuremath{M_{\star}}\xspace}
\newcommand{\Rjup}{\ensuremath{R_\mathrm{J}}\xspace} 
\newcommand{\Mjup}{\ensuremath{M_\mathrm{J}}\xspace}

\newcommand{\Rsun}{\ensuremath{R_\odot}\xspace} 
\newcommand{\Msun}{\ensuremath{M_\odot}\xspace}
\newcommand{\Rp}{\ensuremath{R_p}\xspace}
\newcommand{\Mp}{\ensuremath{M_p}\xspace}

\newcommand{\Teff}{\ensuremath{T_\mathrm{eff}}\xspace}
\newcommand{\logg}{\ensuremath{\log{g}}\xspace}
\newcommand{\mh}{\ensuremath{\mathrm{[M/H]}}\xspace}
\newcommand{\feh}{\ensuremath{\mathrm{[Fe/H]}}\xspace}
\newcommand{\afe}{\ensuremath{\mathrm{[}\alpha\mathrm{/Fe]}}\xspace}

\newcommand{\ms}{\ensuremath{\mathrm{m}\,\mathrm{s}^{-1}}\xspace}

\newcommand{\kms}{\ensuremath{\mathrm{km}\,\mathrm{s}^{-1}}\xspace}

\newcommand{\ums}{\texttt{UberMS}}

\begin{document}

\title{An Ancient Brown Dwarf Transiting a Metal-Poor Thick Disk Star}

\author[0000-0002-4489-3168]{Jéa~Adams Redai}
\affiliation{Center for Astrophysics $\vert$ Harvard \& Smithsonian, 60 Garden Street, Cambridge, MA 02138, USA}
\email[show]{jea.adams@cfa.harvard.edu}  

\author[0000-0002-0572-8012]{Vedant Chandra} 
\affiliation{Center for Astrophysics $\vert$ Harvard \& Smithsonian, 60 Garden Street, Cambridge, MA 02138, USA}
\email{vedant.chandra@cfa.harvard.edu}

\author[0000-0001-7961-3907]{Samuel W. Yee}
\altaffiliation{51 Pegasi b Fellow}
\affiliation{Center for Astrophysics $\vert$ Harvard \& Smithsonian, 60 Garden Street, Cambridge, MA 02138, USA}
\affiliation{Department of Physics \& Astronomy, University of California Los Angeles, Los Angeles, CA 90095, USA}
\email{syee@astro.ucla.edu}

\author[0000-0003-0741-7661]{Victoria DiTomasso} 
\affiliation{Center for Astrophysics $\vert$ Harvard \& Smithsonian, 60 Garden Street, Cambridge, MA 02138, USA}
\email{victoria.ditomasso@cfa.harvard.edu}

\author[0000-0003-2253-2270]{Sean Andrews} 
\affiliation{Center for Astrophysics $\vert$ Harvard \& Smithsonian, 60 Garden Street, Cambridge, MA 02138, USA}
\email{sandrews@cfa.harvard.edu}

\author[0000-0001-8798-1347]{Karin Öberg} 
\affiliation{Center for Astrophysics $\vert$ Harvard \& Smithsonian, 60 Garden Street, Cambridge, MA 02138, USA}
\email{koberg@cfa.harvard.edu}

\author[0000-0002-9003-484X]{David Charbonneau}
\affiliation{Center for Astrophysics $\vert$ Harvard \& Smithsonian, 60 Garden Street, Cambridge, MA 02138, USA}
\email{dcharbonneau@cfa.harvard.edu}

\author[0000-0002-0721-6715]{Rebecca Woody} 
\affiliation{Center for Astrophysics $\vert$ Harvard \& Smithsonian, 60 Garden Street, Cambridge, MA 02138, USA}
\email{rebecca.woody@cfa.harvard.edu}

\author[0000-0001-9911-7388]{David W. Latham}
\affiliation{Center for Astrophysics $\vert$ Harvard \& Smithsonian, 60 Garden Street, Cambridge, MA 02138, USA}
\email{dlatham@cfa.harvard.edu}

\author[0000-0001-6637-5401]{Allyson Bieryla}
\affiliation{Center for Astrophysics $\vert$ Harvard \& Smithsonian, 60 Garden Street, Cambridge, MA 02138, USA}
\email{abieryla@cfa.harvard.edu}

\author[0000-0002-8964-8377]{Samuel~N.~Quinn}
\affiliation{Center for Astrophysics $\vert$ Harvard \& Smithsonian, 60 Garden Street, Cambridge, MA 02138, USA}
\email{squinn@cfa.harvard.edu}

\author[0000-0001-6416-1274]{Theron W. Carmichael}
\affiliation{Institute for Astronomy, University of Hawai'i, 2680 Woodlawn Drive, Honolulu, HI 96822, USA}
\email{tcarmich@hawaii.edu}

\author[0000-0002-5370-7494]{Chih-Chun Hsu}
\affiliation{Center for Interdisciplinary Exploration and Research in Astrophysics (CIERA), Northwestern University, 1800 Sherman Ave, Evanston, IL, 60201, USA}
\email{chsu@northwestern.edu}

\author[0000-0002-0701-4005]{Noah Vowell}
\affiliation{Center for Data Intensive and Time Domain Astronomy, Department of Physics and Astronomy, Michigan State University, East
Lansing, MI 48824, USA}
\email{vowellno@msu.edu}

\author[0000-0003-0774-6502]{Jason J. Wang}
\affiliation{Center for Interdisciplinary Exploration and Research in Astrophysics (CIERA) and Department of Physics and Astronomy, Northwestern University, Evanston, IL 60208, USA}
\email{jason.wang@northwestern.edu}

\author[0000-0003-0562-6750]{Sebastian Zieba}
\altaffiliation{NASA Sagan Fellow}
\affiliation{Center for Astrophysics $\vert$ Harvard \& Smithsonian, 60 Garden Street, Cambridge, MA 02138, USA}
\email{sebastian.zieba@cfa.harvard.edu}

%% SG1 Authors
\author[0000-0001-6981-8722]{Paul~Benni}
\affiliation{Acton Sky Portal (Private Observatory), Acton, MA, USA}
\email{pbenni@verizon.net}

\author[0000-0001-6588-9574]{Karen~A.~Collins}
\affiliation{Center for Astrophysics $\vert$ Harvard \& Smithsonian, 60 Garden Street, Cambridge, MA 02138, USA}
\email{karen.collins@cfa.harvard.edu}

%% SG3 Authors
\author[0000-0002-5741-3047]{David~R.~Ciardi}
\affiliation{Caltech/IPAC-NASA Exoplanet Science Institute, 770 S. Wilson Avenue, Pasadena, CA 91106, USA}
\email{ciardi@ipac.caltech.edu}

\author[0000-0003-2192-5371]{Julian~van~Eyken}
\affiliation{Caltech/IPAC-NASA Exoplanet Science Institute, 770 S. Wilson Avenue, Pasadena, CA 91106, USA}
\email{ciardi@ipac.caltech.edu}

\author[0000-0003-0241-2757]{William~Fong}
\affiliation{Department of Physics and Kavli Institute for Astrophysics and Space Research, Massachusetts Institute of Technology, Cambridge, MA 02139, USA}
\email{willfong@mit.edu}

\author[0000-0003-2527-1598]{Michael~B.~Lund}
\affiliation{Caltech/IPAC-NASA Exoplanet Science Institute, 770 S. Wilson Avenue, Pasadena, CA 91106, USA}
\email{ciardi@ipac.caltech.edu}

\author[ 0000-0002-4398-6258]{Andrei M. Tatarnikov}
\affiliation{Sternberg Astronomical Institute, M.V. Lomonosov Moscow State University, 13, Universitetskij pr., 119234, Moscow, Russia}
\email{andrey.tatarnikov@gmail.com}

% ADD PHILL CARGILE? 

% \author{TFOP authors: Paul Benni, Karen Collins -- attempted a transit observation but weathered out?, Boris Safonov, David Ciardi and friends}
% \email{fakeemail2@google.com}


\begin{abstract}
\noindent
We report the discovery of TOI-7019b, the first transiting brown dwarf (BD) known to orbit a star that is part of the Milky Way's ancient thick disk, as defined chemically ($\feh = -0.79 \pm 0.05$~dex, $\afe = +0.26 \pm 0.05$~dex, $\mh = -0.59 \pm 0.06$~dex) and kinematically ($v_\mathrm{\perp} \approx 150 \pm 1$~\kms{}). 
We estimate a system age $\tau = 12 \pm 2$~Gyr by fitting the host star's spectrum and spectral energy distribution to alpha-enhanced isochrones, and independently using the age-metallicity relation of the thick disk. 
This makes TOI-7019 by far the most metal-poor and ancient BD host known to date. 
We measure a BD mass of $61.3 \pm 2.1$~\Mjup and radius of $0.82 \pm 0.02$~\Rjup from a joint analysis of transit photometry and radial velocity measurements, along with an orbital period of  $48.2592 \pm 0.0001$ days and an orbital eccentricity of $0.403 \pm 0.002$. The measured radius appears $12.3\% \pm 2.8\%$ larger than predicted relative to standard evolutionary models for old, metal-poor brown dwarfs, hinting at missing physics like the magnetic inhibition of convection.  
TOI-7019b lowers the probed metallicity regime for transiting BDs by over a factor of two, making it a benchmark system to test evolutionary models in the low-metallicity regime.
Future measurements of TOI-7019b's atmosphere will test whether a brown dwarf’s atmospheric composition tracks its host star's abundances, as expected for binary-like co-formation.

\end{abstract}

\section{Introduction}

Brown dwarfs, with masses between those of giant planets and hydrogen-burning stars, are central to understanding the transition between star formation and planet formation processes. Occupying a mass range of roughly 13–80 M$_{\mathrm{J}}$, these substellar objects are generally not massive enough to sustain stable hydrogen fusion in their cores, and thus they cool and fade over time \citep{Burrows2001_BDTheory,Chabrier2001}. As they are unable to sustain stable fusion, their evolution is dominated by continuous cooling and contraction. Consequently, their physical properties, particularly their radii and temperatures, are strongly dependent not only on their mass, but also on their age \citep{Baraffe2003_MassRad,Marley2021_Sonora}. 

Transiting brown dwarfs are exceptionally valuable because they allow for the direct measurement of both mass and radius \citep{Carmichael2023}. These measurements provide the most stringent tests of substellar evolutionary models, which predict the rate at which a brown dwarf should cool, and its radius should contract based on its fundamental properties \citep[e.g.,][]{Marley2021_Sonora, Morley24_Diamondback}. However, the population of well-characterized transiting brown dwarfs remains small, at just over 50 confirmed objects with well-measured masses and radii \citep{Vowell2025_BD}. This is largely because brown dwarfs are intrinsically rarer than either giant planets or stellar companions at the short orbital periods where transits are most readily detected \citep{Grether_BDDesert, Sahlmann2011}.

A significant challenge in testing substellar evolution models has been obtaining precise ages for the host stars. While ages for young stars in clusters or associations can be determined with some confidence, estimating the age of older, isolated field stars is notoriously uncertain. However, stellar kinematics provide a powerful complementary approach: stars with thick-disk kinematics, identifiable through their Galactic space velocities, are known to be significantly older (typically $>$8 Gyr) and more metal-poor than the thin-disk population \citep{Bensby2003,Haywood2013,Bensby2014}. Furthermore, thick disk stars also show a stronger correlation between age and metallicity than their thin disk counterparts \citep{Hayden2015_agemetals, Xiang2022, Imig2023_agemetals}. A transiting brown dwarf around a thick disk star would provide a critical age anchor, but no such objects are currently known.

% TOI-7019b, as the first transiting, dynamically measured brown dwarf in the thick disk, now provides a crucial empirical link between those isolated field objects and the metallicity–age sequence of the Milky Way’s old populations

Beyond providing an age anchor, the characteristically low metallicity of these ancient thick-disk stars offers an opportunity to probe another crucial but poorly constrained factor: the role of chemical composition in a brown dwarf's formation and evolution. Theoretical models predict that metal-rich brown dwarfs should have larger radii at a given age and mass due to enhanced opacity in their atmospheres \citep{Burrows2011, Marley2021_Sonora}, so metal-poor brown dwarfs in the thick disk might be expected to be amongst the smallest such objects.
Formation mechanisms are likely also intimately connected to host star metallicity. An object formed via core accretion, as gas giant planets are thought to do, requires slow buildup of a solid core followed by rapid gas accretion, making the viability of this pathway highly sensitive to available solid material \citep{RiceArmitage_Formation, Alibert2004_formation, Mordasini_Formation}. On the other hand, objects formed through gravitation instability are a direct result of the rapid collapse of massive, unstable disk regions, so this pathway is less dependent on metallicity \citep{B0ss1997_formation, Boss2006_formation}. Observational evidence suggests that lower-mass brown dwarfs ($\lesssim$ 10--40\Mjup, with the exact boundary uncertain) are more commonly found around host stars with enhanced metallicities, like their giant planet counterparts, reinforcing the theory that these objects form via similar mechanisms \citep{Ma_Ge,Schlaufman,Vowell2025_BD}. On the other hand, higher-mass brown dwarfs, appear to be found around host stars with a wider range of metallicities (e.g., \citealt{Vowell2025_BD}). Free-floating brown dwarfs with low metallicities have been identified in the field, including the extremely metal-poor objects uncovered by Backyard Worlds and related surveys \citep[e.g.,][]{Schneider2020_ExtremeSD, Meisner2020_BW, Kirkpatrick2021_theaccidenht, Lodieu2022, Zhang2025}. These discoveries, extending to \feh $\leq -1.6$, demonstrate that substellar objects can indeed form in the chemically primitive environments of the early Galaxy. However, because we cannot directly measure the physical properties of these isolated brown dwarfs, their evolutionary states remain model-dependent. Transiting objects in this metal-poor regime would be helpful for constraining those models, but nearly all well-characterized transiting brown dwarfs orbit solar or near-solar metallicity hosts ($-0.4 \lesssim \feh \lesssim 0.5$~dex), leaving the highly metal-poor regime essentially unexplored. This observational trend may be due to inherent formation physics, strong observational bias, or some combination of the two.

In this paper, we present the discovery and characterization of TOI-7019b, a transiting brown dwarf orbiting an ancient, metal-poor star with a metallicity of $\feh=-0.8$~dex. 
% TOI-7019 is confidently associated with the thick disk of the Milky Way based on its kinematics and $\alpha$-abundance measurement, with an estimated stellar age $\tau \gtrsim 12$~Gyr. 
% TOI-7019 is by far the most ancient metal-poor brown dwarf host known, making it a valuable benchmark for evolutionary models. 
We describe the host star TOI-7019 in $\S$\ref{sec:toistar}, from its unusual kinematics and metallicity in \textit{Gaia}~DR3 data to stellar abundance analysis with high-resolution spectroscopy, confidently associating it with the thick disk of the Milky Way.
We then measure the properties of the transiting companion TOI-7019b in $\S$\ref{sec:toibd}, using a combination of TESS and ground-based photometry, and radial velocities, finding it to be a massive brown dwarf. We contextualize our discovery in $\S$\ref{sec:discuss}, and list our conclusions in $\S$\ref{sec:conclusions}. 
% TOI-7019 is the first transiting BD host confidently associated with the Milky Way's thick disk, and is by far the most metal-poor and ancient BD host known. 
% This system provides an unprecedented opportunity to anchor evolutionary models at the extreme of low metallicity and old age, addressing a critical gap in our understanding of how chemical composition affects substellar structure and evolution.

\section{TOI-7019: An Ancient Thick-Disk Star}\label{sec:toistar}
\subsection{\Gaia Metallicity and Velocities}\label{sec:gaiadata}\label{sec:thickdisk}

\begin{figure*}[!t]
    \centering
    \includegraphics[height=0.45\linewidth]{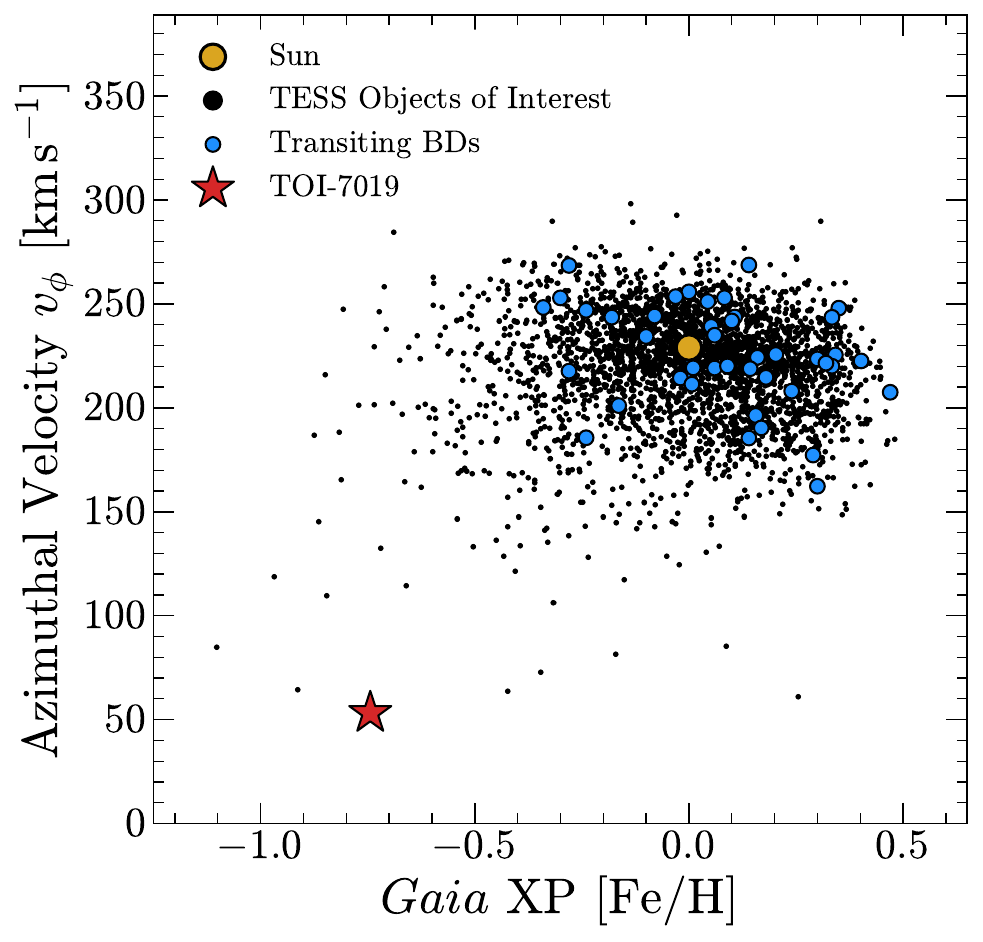}
\includegraphics[height=0.45\linewidth]{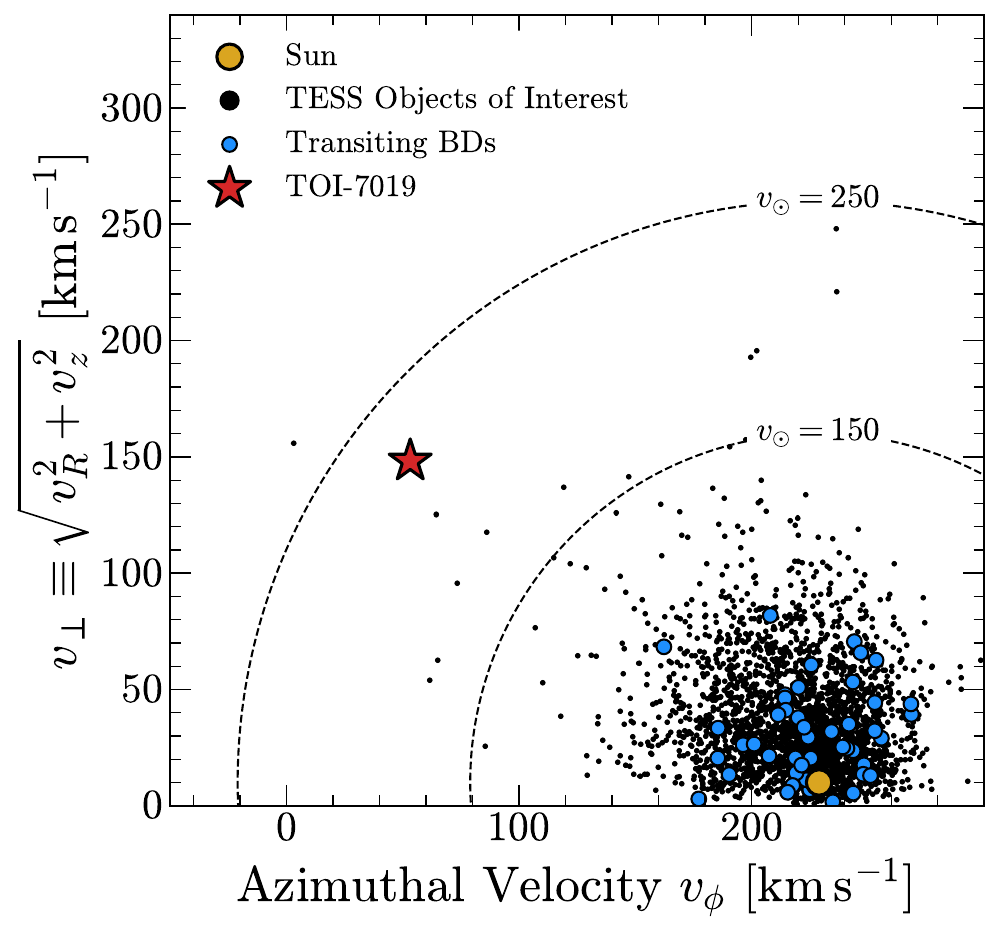}
    \caption{Left: \textit{Gaia} XP metallicities from \cite{Andrae2023} for TESS objects of interest, versus the Galactocentric azimuthal velocity. 
    The Sun is shown for reference. 
    Whereas most TOIs are thin disk stars in the Solar neighborhood, TOI-7019 stands out as an extreme outlier in this space.
    Right: The Toomre diagram, showing the Galactocentric azimuthal velocity versus the perpendicular (radial and vertical) velocity.  Dashed lines indicate constant total Galactocentric velocity.
    TOI-7019 lies in a sparsely populated regime of the Toomre diagram typically associated with the kinematic thick disk.}
    \label{fig:discovery}
\end{figure*}

The third data release of the \textit{Gaia} space observatory has delivered parallax-based distances, proper motions, radial velocities, and spectro-photometric metallicities for millions of stars \citep{GaiaEDR3_Brown2021, GaiaDR3_Vallenari2022, Andrae2023}. 
These data provide the 6D kinematics and chemical abundance information required to distinguish stars belonging to the various sub-components of our galaxy, like the ancient stellar halo, old `thick disk', and younger `thin disk' \citep[e.g.,][]{Hayden2015_agemetals, Imig2023_agemetals, Chandra2024}. 
We cross-matched the list of TESS objects of interest \citep[TOIs;][]{TESS_TOIs_Guerrero2021} with \textit{Gaia} parameters to the spectro-photometric metallicity catalog of \cite{Andrae2023}. 
This catalog utilizes low-resolution ($R\sim20$--70 \textit{Gaia} `XP' spectrophotometry to predict stellar parameters and metallicities for over a hundred millions stars. 
We also incorporated the prior-informed distance catalog of \cite{Bailer-Jones2020}, which provides more reliable distances for stars with low parallax SNR, compared to simple parallax inversion \citep[e.g.,][]{Astraatmadja2016}.

\begin{figure*}
 \centering
    \includegraphics[width=0.99\textwidth]{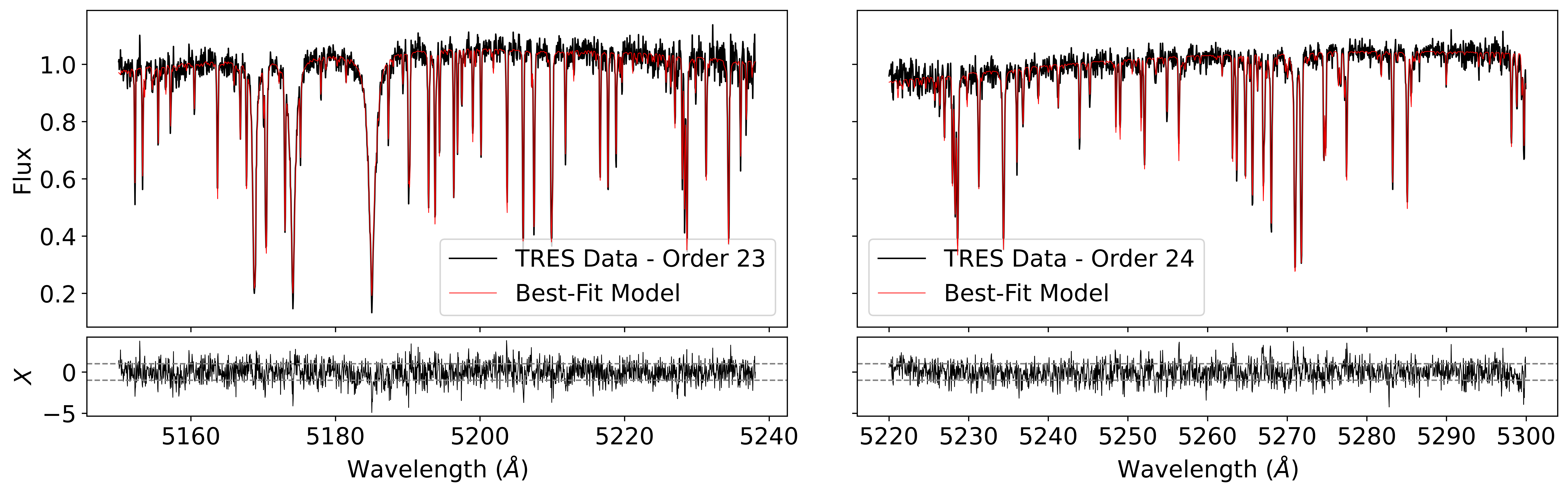} 
    \caption{TRES spectrum, orders 23 and 24, as used with the stellar characterization code \texttt{UberMS}. Data shown in black and best-fit model shown in red. Residuals divided by the spectral uncertainties are plotted in the lower panel, with horizontal dashed lines marking $\chi=1$ and $\chi=-1$.}
    \label{fig:uberms_tres}
\end{figure*}

With 6D phase-space information from \textit{Gaia} in-hand, we calculated kinematic parameters for each star, including Galactocentric velocities, angular momenta, and orbital properties. 
We used \texttt{astropy} \citep{AstropyCollaboration2013, AstropyCollaboration2018, AstropyCollaboration2022} and \texttt{gala} \citep{gala,adrian_price_whelan_2020_4159870} for these calculations. 
A right-handed Galactocentric frame is assumed with a solar position $\mathbf{x}_\odot = (-8.12, 0.00, 0.02)$~kpc, and solar velocity $\mathbf{v}_\odot = (12.9, 245.6, 7.8)$~$\mathrm{km\,s^{-1}}$ \citep{Reid2004,Drimmel2018,GravityCollaboration2019}. 
We adopted the \texttt{MilkyWayPotential2022} from \texttt{gala}, which has a circular velocity at the solar position $v_\mathrm{c}(R_\odot) = 229~\textrm{km}~\textrm{s}^{-1}$ informed by the measurements of \citep{Eilers2019}.

% \subsection{Galactic Kinematics \& Thick Disk Membership}\label{sec:thickdisk}\

The left panel of Figure~\ref{fig:discovery} shows \textit{Gaia} metallicities from \cite{Andrae2023} for TESS object of interest (TOI) candidates, versus the Galactocentric azimuthal velocity $v_\phi$. 
We also show known transiting brown dwarfs compiled by \cite{Vowell2025_BD}. 
Most TOIs are thin-disk stars, with metallicities and velocities similar to the Sun, and consistent with being part of the thin-disk. However, a few TOIs lie outside this distribution. TOI-7019 stands out as one of the most extreme outliers, with $v_\phi \approx 50$~\kms{} and \textit{Gaia} XP-inferred metallicity [Fe/H] $\approx -0.75$. 
This flagged TOI-7019 as a potential thick disk or halo star, and motivated us to conduct follow-up observations to investigate the TESS transits. 

% \begin{figure}
%  \centering\includegraphics[height=0.45\textwidth]{orbit.pdf} 
%     \caption{
%     Galactocentric $R$--$z$ plane showing orbits integrated for $1$~Gyr, for transiting BD hosts (black) and TOI-7019 (red). 
%     TOI-7019 stands starkly apart from any known transiting BD, in the kinematically-hot, metal-poor thick disk population.}
%     \label{fig:bdkin}
% \end{figure}

% The kinematics of TOI-7019 place it firmly outside the thin disk population.
% Figure~\ref{fig:discovery} compares the kinematics of TOI-7019 to all TESS objects of interest. 
% A background sample of randomly-selected stars from \textit{Gaia} DR3 is also shown, for which we use metallicities from \cite{Andrae2023}. 
The right panel of Figure~\ref{fig:discovery} shows the Toomre diagram (azimuthal velocity versus tangential velocity), a useful diagnostic that can distinguish between thin disk, thick disk, and halo stars \citep[e.g.,][]{Bensby2014, Haywood2018}. 
TOI-7019 is an outlier in this space too with its anomalous $v_\phi$ and high $v_\perp$, firmly in the regime typically associated with thick disk stars. 

Furthermore, we integrated the star's Galactic orbit using the \texttt{gala} package \citep{pricewhelan2017_gala} with a standard Milky Way potential model. The resulting orbit is highly eccentric ($e = 0.81$) and reaches a maximum vertical distance from the Galactic plane $\mathrm{Z_{max}} = 0.7\ \mathrm{kpc}$, much larger than the $\sim 0.25$~kpc scale height of the thin disk \citep{Imig2025}. 
Such an orbit, which deviates from a circular path and carries the star far from the mid-plane, is characteristic of the dynamically hot and ancient thick-disk population. 

% This robust kinematic classification provides a strong age prior of $>$8 Gyr for the TOI-7019 system, confirming that TOI-7019 is an ancient star. 

\subsection{TRES Spectroscopy}\label{ssec:tres_spectra}

TOI-7019 was initially observed using the Tillinghast Reflector Echelle Spectrograph \citep[TRES;][]{TRES_Furesz2008} on the 1.5m Tillinghast Reflector at Fred Lawrence Whipple Observatory (FLWO) as part of the TESS Follow-up Observing Program (TFOP) Working Group Sub Group 2 (SG2). These observations were made with the goal of determining spectroscopic stellar properties (\S\ref{ssec:stellar_char}) as well as measuring radial velocities (RVs) to trace the stellar reflex motion caused by the transiting object.
The initial TRES high-resolution spectra confirmed the high systemic RV and low stellar metallicity, motivating our continued observation of the system.
We obtained 15 observations between 2025 Apr 15 and 2025 Jul 23. Each observation had a total integration time of between 40--60 minutes split over multiple exposures, achieving a signal-to-noise ratio of $\sim 30$ per resolution element. 

The TRES data were reduced using the standard pipeline described in \citet{TRES_Buchhave2010,TRES_Quinn2012}.
We measured relative radial velocity shifts between each observation by first constructing a median template from all the TRES observations, and cross-correlating this template against each of the observations on an order-by-order basis.
Instrumental RV zero-point offsets were corrected for using nightly standard star observations.
The internal RV precision achieved for these observations was 30--50~\ms. We provide the relative TRES RVs in Table \ref{tab:tres_rvs}.
We additionally calculated the offset required to shift these relative RVs onto an absolute scale, as determined by TRES observations of minor planets, which have precisely and accurately known barycentric velocities \citep{Bieryla2022}.
We added this offset ($-150021.8 \pm 20.2$~\ms) to the center-of-mass velocity on the scale of the relative TRES RVs, $\mu_\mathrm{TRES} = 5656 \pm 17$~\ms as determined by our RV fit (\S\ref{sec:toibd}) to determine an absolute systemic velocity of $\gamma = -144.366 \pm 0.023$~\kms, consistent with the value of $-142.9\pm2.9$~\kms reported by \Gaia DR3.

\subsection{Stellar Characterization \label{ssec:stellar_char}}

We characterized TOI-7019 using the stellar inference code \texttt{UberMS}\footnote{\url{https://github.com/pacargile/uberMS/}}. Built upon the frameworks of \texttt{The Payne} \citep{Ting2019_Payne} and \texttt{MINESweeper} \citep{Cargile2020_MINESweeper}, \texttt{UberMS} employs neural networks to simultaneously fit Kurucz stellar atmospheric models \citep{Kurucz1992, Kurucz1993} to both high-resolution spectra and broadband photometry, in conjunction with MIST stellar isochrones \citep{MISTI_Choi2016}. In particular, \texttt{UberMS} uses an updated set of MIST isochrones that account for non-solar $\alpha-$element abundances ($\afe$; \citealt{Park2024}.

We applied \texttt{UberMS} to the TRES spectra and Gaia G, BP, RP \citep{GaiaEDR3_Riello2021}, 2MASS J, H, K \citep{TMASS_Cutri2003} and WISE W1, W2 photometry \citep{WISE_Cutri2012}, following the methodology of \citet{Pass2025_UberMS}, briefly summarized here.
We considered the subsection of the TRES spectra from 5150-5300$\mathring{\text{A}}$ (orders 23 and 24), which crucially includes the Mg b triplet region for constraining $\afe$. In addition to the calibrations and corrections applied during the standard spectral data reduction, we applied an additional blaze correction to each observation by fitting a 10th order Chebyshev polynomial to standard star observations taken nearly nightly of Sirius/Vega and removing that trend. We then cross-correlated each spectrum with a co-added template to shift each observation to the rest frame. After these corrections, we coadded the available spectra via an error-weighted mean.

% ====== Side-by-side version of Tables 1 and 2 ======
\begin{table*}[t]
\centering
\caption{TOI-7019 Stellar and System Parameters}
\label{tab:stellar_system_side_by_side}

% ---------- LEFT: former Table 1 ----------
\begin{minipage}[t]{0.48\textwidth}
\centering
\textbf{Table 1a.}\label{tab:results_star} TOI-7019 stellar parameters inferred by \texttt{UberMS}.\\[-2pt]
\begin{tabular}{ll}
\hline
\hline
\multicolumn{1}{c}{Stellar Properties} & \multicolumn{1}{c}{Value} \\
\hline
Gaia DR3 ID & 2123995644386597888 \\
RA (J2000, epoch 2015.5)\textsuperscript{a} & 18:15:09.34 \\
Dec (J2000, epoch 2015.5)\textsuperscript{a} & +50:51:45.86 \\
$G$ (mag)\textsuperscript{b} & 13.13 \\
$\Teff$ (K)\textsuperscript{c} & $5800 \pm 100$ \\
$[$Fe/H$]$ (dex)\textsuperscript{c} & $-0.79 \pm 0.05$ \\
$[\alpha/\text{Fe}]$ (dex)\textsuperscript{c} & $0.26 \pm 0.05$ \\
$[$M/H$]$ (dex)\textsuperscript{c} & $-0.59 \pm 0.06$ \\
$\log g$ (cgs)\textsuperscript{c} & $4.4 \pm 0.1$ \\
$\Mstar$ (M$_\odot$)\textsuperscript{c} & $0.78 \pm 0.01$ \\
$\Rstar$ (R$_\sun$)\textsuperscript{c} & $0.92 \pm 0.01$ \\
Heliocentric Distance (pc)\textsuperscript{b} & $435 \pm 2$ \\
Age (Gyr, spectro-photometric)\textsuperscript{c} & $13.5 \pm 1$ \\
Age (Gyr, age-metallicity relation)\textsuperscript{c} & $12 \pm 1.5$ \\
\hline
\end{tabular}

\vspace{6pt}
\footnotesize
\textit{Notes.} \textsuperscript{a} \cite{Stassun2019} \;
\textsuperscript{b} GAIA. \;
\textsuperscript{c} This work. \\
\footnotesize Adopted error bars for $\Teff$, $\feh$, $\afe$, and $\logg$ follow \citealt{Pass2025_UberMS}; other uncertainties are purely statistical from \texttt{UberMS} and do not include model systematics.\\[4pt]
\end{minipage}
\hfill
% ---------- RIGHT: former Table 2 ----------
\begin{minipage}[t]{0.48\textwidth}
\centering
\textbf{Table 1b.}\label{tab:parameters} TOI-7019 system parameters from joint fit.\\[4pt]
\begin{tabular}{lr}
\hline
\hline
\multicolumn{2}{c}{\textit{Fit Parameters}} \\
\hline
$P$ (days) & $48.2592 \pm 0.0001$ \\
$t_0$ (\bjdtdb)\textsuperscript{a} & $2942.3067 \pm 0.0012$ \\
$R_b/R_*$ & $0.0896 \pm 0.0017$ \\
$b$ & $0.24 \pm 0.14$ \\
$e$ & $0.4025 \pm 0.0015$ \\
$\omega$ (deg) & $-171.50 \pm 0.54$ \\
$K$ (\ms) & $4417 \pm 17$ \\
$\gamma_\mathrm{sys}$ (\kms) & $-144.366 \pm 0.023$\\
\hline
\multicolumn{2}{c}{\textit{Derived Companion Parameters}} \\
\hline
$M_b$ (\Mjup) & $61.3 \pm 2.1$ \\
$R_b$ (\Rjup) & $0.821 \pm 0.015$ \\
$\rho_b$ ($\mathrm{g/cm^{3}}$) & $141.7 \pm 9.6$ \\
$a$ (AU) & $0.2535 \pm 0.009$ \\
$i$ (deg) & $89.77 \pm 0.16$ \\
$T_{\mathrm{eq}}$ (K) & $479 \pm 10$ \\
\hline
\end{tabular}

\vspace{6pt}
\footnotesize
\textit{Notes.} \textsuperscript{a} BJD offset of 2457000.\\
\footnotesize Parameters derived from joint analysis of TESS photometry, ground-based photometry, and TRES radial velocities using \texttt{juliet}. The stellar mass and radius used for the derived parameters are $M_* = 0.779$~\Msun\ and $R_* = 0.922$~\Rsun, as derived in Section~\ref{ssec:stellar_char}. Uncertainties in the stellar mass and radius assume the error floor in \cite{Tayar2022}.
\end{minipage}

\end{table*}

With these spectra and the photometry listed above, we used \texttt{UberMS} to fit for stellar parameters. Again, we followed the methodology of \citet{Pass2025_UberMS}, with the following modifications: (1) broadened the prior on extinction ($A_V$) to account for TOI-7019's greater distance when compared to the \citet{Pass2025_UberMS} sample, with the upper bound set by the \citep{Schlegel1998} extinction maps and a Gaussian prior informed by the Bayestar19 3D dust maps \citep{Green2019_Bayestar}, adopting a truncated normal prior centered on $0.06 \pm 0.02$ with bounds of [0.000, 0.206]; and (2) we fit a second-order polynomial rather than a linear function to the continuum of the TRES spectrum to account for remaining imperfections in the blaze correction.
The resulting stellar parameters are summarized in Table \ref{tab:results_star}, adopting the suggested error bars from \citet{Pass2025_UberMS} for $\Teff$, $\feh$, $\afe$, and $\logg$ based on their comparison with various calibration samples, and the computed statistical errors from \texttt{UberMS} for the other parameters. The TRES spectra and their best-fit models are shown in Figure \ref{fig:uberms_tres}.

\begin{figure}
 \centering
    \includegraphics[width=0.99\columnwidth]{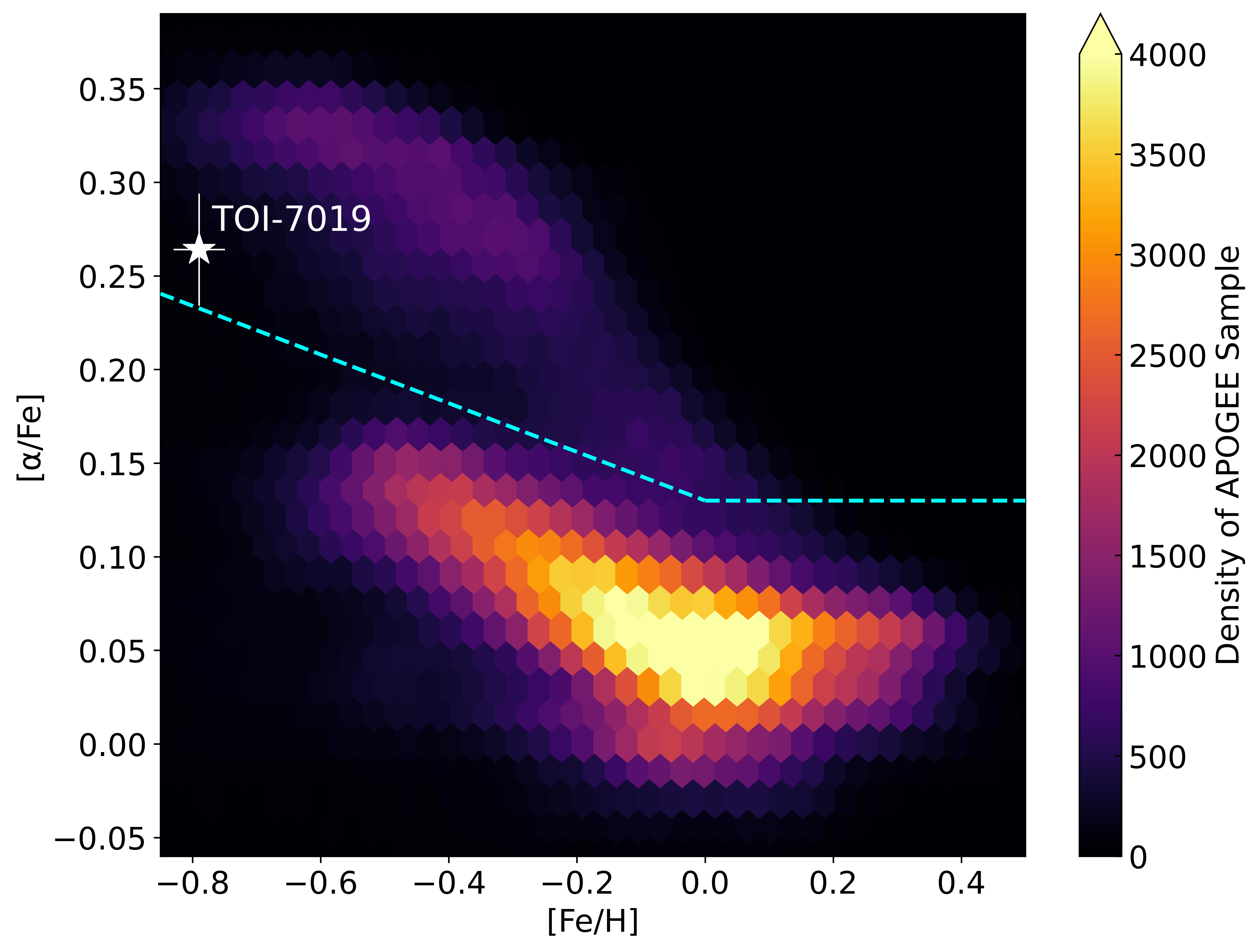} 
    \caption{$\feh$ and $\afe$ of TOI-7019 in comparison to the allStarLite-dr17 APOGEE sample. We have shifted the APOGEE sample by +0.04 in $\afe$ to account for a systematic offset between the $\afe$ from APOGEE and that measured by \texttt{UberMS} \citep{Pass2025_UberMS}.
    The dashed blue line indicates a typical selection that separates the high-$\alpha$ (thick) and low-$\alpha$ (thin) disk.}
    \label{fig:comp_to_apogee}
\end{figure}

The \ums{} fit to the TRES spectrum confirms the metal-poor nature of TOI-7019. 
The best-fitting iron metallicity and alpha-abundance is [Fe/H]~$= -0.79 \pm 0.05$ and [$\alpha$/Fe]~$= 0.26 \pm 0.05$.  Accounting for the non-solar $\afe$, the \emph{total} metals to hydrogen metallicity is [M/H]~$=-0.59 \pm 0.06$.  The TRES metallicity is consistent with the [Fe/H]~$= -0.75 \pm 0.10$ measurement from \cite{Andrae2023} based on low-resolution \textit{Gaia} XP data. 
Figure~\ref{fig:comp_to_apogee} compares the TRES measurements to a broad sample of disk stars from the APOGEE survey \citep{Majewski2017}. 
In this abundance space, there is a clear separation between the thick disk (predominantly high-$\alpha$)
and the thin disk (predominantly low-$\alpha$; e.g., \citealt{Hayden2015}). 
The metallicity of TOI-7019 is well below the typical locus of thin disk stars. 
Furthermore, its alpha-abundance is significantly higher than that of the low-$\alpha$ disk stars in APOGEE. 
Therefore, these abundances, combined with the kinematics discussed in \S\ref{sec:gaiadata}, confidently identify TOI-7019 as a member of the Milky Way's kinematically thick, high-$\alpha$ disk. 

\subsection{System Age}

Although it is challenging to assign absolute ages to individual main sequence stars given the slow changes they experience over this phase of stellar evolution \citep[e.g.,][]{Tayar2022, Woody2025}, several lines of evidence point towards TOI-7019 being an ancient system that formed over ten billion years ago. 
First, the spectro-photometric \ums{} fit produces an estimate of the stellar age from the MIST isochrones. 
Formally, the best-fitting age is $\tau_\mathrm{iso} = 13.5 \pm 1$~Gyr. 
Note that solutions older than the age of the Universe are not excluded by the fitting routine, since the absolute age scale of the models is systematically uncertain. 

\begin{figure}
    \centering
    \includegraphics[width=\columnwidth]{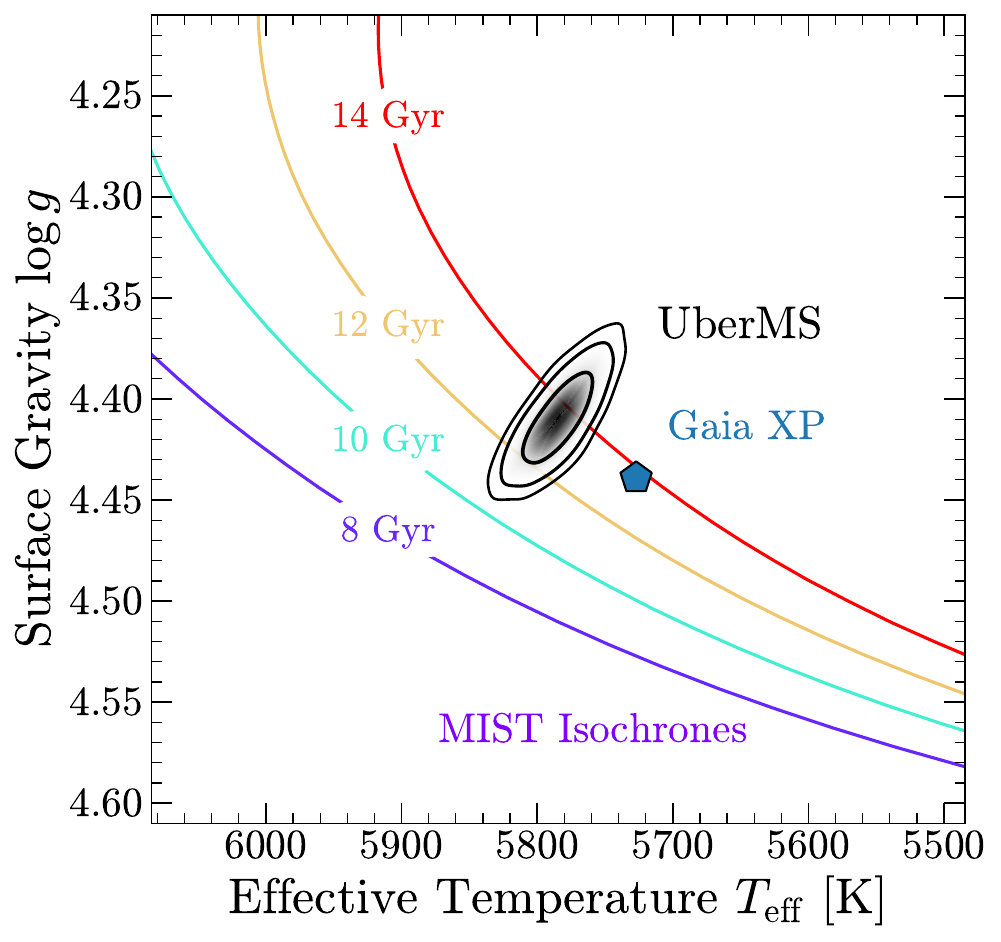}
    \caption{The observational HR diagram --- effective temperature versus surface gravity --- showing the \texttt{UberMS} fit to TOI-7019's TRES spectrum and broadband photometry. 
    Black contours delineate the $1\sigma-3\sigma$ statistical error regimes. 
    The \cite{Andrae2023} measurement using low-resolution \textit{Gaia} XP data is also shown. \textit{Gaia} XP metallicites have a reported uncertainty of 0.1 dex \citep{Andrae2023}.
    MIST isochrones are overlaid for a range of stellar ages, with other parameters fixed to the adopted values for TOI-7019.}
    \label{fig:kiel}
\end{figure}

To visualize the spectro-photometric age constraint, we show the observational H-R diagram in Figure~\ref{fig:kiel}. 
Black contours show the \ums{} fit, and the \cite{Andrae2023} measurement using \textit{Gaia} XP data is also shown. 
MIST isochrones are overlaid for a range of ages, with the metallicity and alpha-abundance fixed to the adopted parameters for TOI-7019. 
The stellar parameters of TOI-7019 are most consistent with ages in the range of $\tau \approx 12-14$~Gyr. 
Ages younger than $10$~Gyr are excluded by the $3\sigma$ contours, although we emphasize that only statistical errors are shown here, and the isochrones themselves have substantial systematic uncertainties \citep[e.g.,][]{Tayar2022}. 

Another way to estimate TOI-7019's age is using the age-metallicity relation of the high-$\alpha$ disk. 
The high-$\alpha$ disk is thought to have formed most of its stars before $\approx 8$~Gyr in the past \citep{Haywood2013, Bonaca2020, Xiang2022}. 
TOI-7019's origin in the high-$\alpha$ thick disk therefore already suggests a relatively old age for the system. 
Fortuitously, the high-$\alpha$ disk has a well-defined age-metallicity relation (AMR; unlike the low-$\alpha$ disk, whose AMR is complicated by radial migration). 

\begin{figure}
    \centering
    \includegraphics[width=\columnwidth]{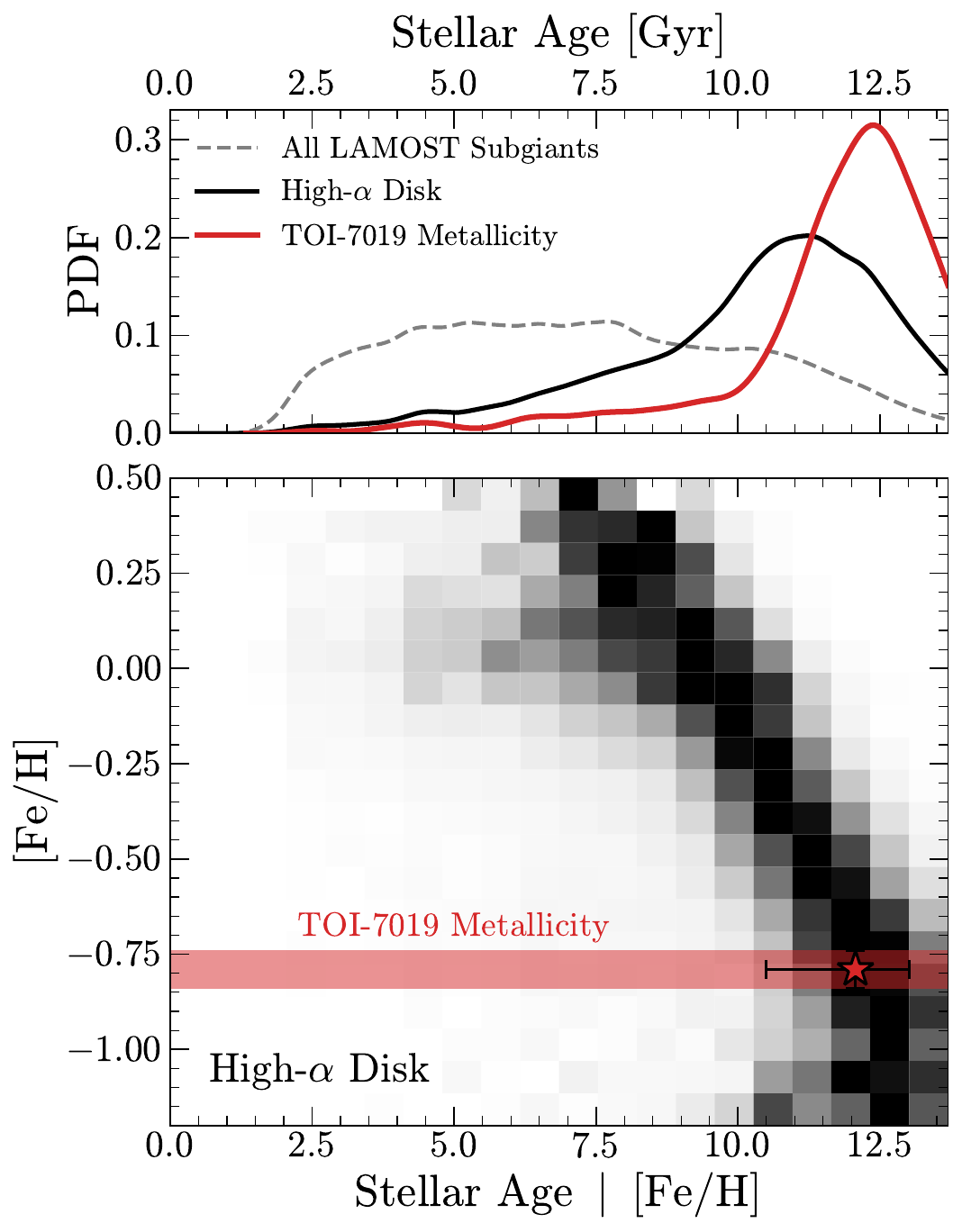}
    \caption{The age-metallicity relation (AMR) of the Milky Way's high-$\alpha$ disk, using subgiant stars from \cite{Xiang2022}. 
    The row-normalized black histogram shows the marginal distribution of stellar ages at each metallicity. 
    The shaded red band indicates the metallicity of TOI-7019. 
    Marginal age distributions are shown in the top panel for the entire subgiant sample, the high-$\alpha$ disk selection, and high-$\alpha$ disk stars with metallicities similar to TOI-7019.}
    \label{fig:amr}
\end{figure}

Figure~\ref{fig:amr} shows the AMR of the high-$\alpha$ disk from \cite{Xiang2022}. 
Specifically, that work compiled subgiant stars from the LAMOST survey, for which accurate isochrone ages can be determined. 
We follow the high-$\alpha$ disk selection from \cite{Xiang2022}, selecting alpha-enhanced stars with modest angular momentum around the galactic center. 
The black histogram in Figure~\ref{fig:amr} shows, at each metallicity, the marginal distribution of stellar ages. 
The tail of stars with ages younger than $10$~Gyr likely represents contamination from the low-$\alpha$ disk. 
By selecting subgiants with similar metallicities as TOI-7019, we can derive the typical age of high-$\alpha$ disk stars at this metallicity: $\tau_\mathrm{AMR} \approx 12.5_{-1.5}^{+1.0}$~Gyr (red histogram in the top panel of Figure~\ref{fig:amr}). 

Combining the isochrone and AMR constraints, we conservatively assign a stellar age $\tau = 12 \pm 2$~Gyr to the TOI-7019 system. 
It is challenging to ascribe absolute ages at these early epochs of the MW's history \citep[e.g.,][]{Tayar2022, Woody2025}. 
Regardless, several independent lines of evidence confirm that TOI-7019 is an ancient system formed in the first few billion years after the Big Bang, and thus likely older than most stars known to host substellar companions, which are part of the MW's younger low-$\alpha$ disk. 

\section{TOI-7019 B: A Transiting Brown Dwarf}\label{sec:toibd}

In this section, we describe TESS measurements that detected the presence of a transiting companion around TOI-7019. 
Using follow-up photometry and radial velocity measurements, we determined the physical properties of this companion, which turns out to be a dense brown dwarf.

\subsection{TESS Photometry}\label{tess_photom}
TOI-7019 was first detected as a transit candidate by NASA's TESS mission \citep{TESS_Ricker15}.
TESS observed TOI-7019 in 19 different sectors (Table \ref{tab:tess}), primarily in the Full Frame Images (FFIs). The FFI data were recorded at cadences of 1800s, 600s, and 200s during the TESS prime mission, first extended mission, and second extended mission respectively.
Light curves were produced from the FFIs using the MIT Quick-Look Pipeline \citep[QLP;][]{TESS_QLP_Huang2020a,TESS_QLP_Huang2020b,TESS_QLP_Kunimoto2021,TESS_QLP_Kunimoto2022b}.
The QLP light curves were searched for transit events as part of the ``faint-star search'' \citep{TESS_Faint_Kunimoto2022a}, which led to the identification of TOI-7019.01 as a $P = 48.259$~day transit candidate.
Following its release as a TESS Object of Interest, TOI-7019 was added to the list of targets pre-selected for high-cadence (120s) observations in Sector 85, with light curves produced by the TESS Science Processing Operations Center \citep[SPOC;][]{TESS_SPOC_Jenkins2016}.
The TESS observations of TOI-7019 are summarized in Table \ref{tab:tess} in the appendix.

We used the Simple Aperture Photometry (SAP) light curves for the sectors where only QLP data were available, and the Presearch Data Conditioning \citep[PDC;][]{TESS_PDC_Smith2012,TESS_PDC_Stumpe2012,TESS_PDC_Stumpe2014} light curves produced by the SPOC for the Sector 85 observations.
We detrended the light curves for each sector by first masking out the transit events and then fitting a basis spline to the out-of-transit data using the \texttt{Keplerspline} code \citep{Keplerspline_Vanderburg2014,Keplerspline_Shallue2018}.
For our subsequent analysis, we only retained the data spanning one transit duration before and after each transit event.
We also noted that the transits during Sectors 74 and 79 occurred during intervals where the TESS data were flagged for poor data quality, so we excluded those, leaving us with eight separate transit events detected by TESS.

\subsection{Ground-Based Photometry}

Following the identification of TOI-7019.01 as a promising transit candidate orbiting a star with unusual galactic kinematics, we performed a ground-based follow-up campaign to confirm it as a substellar companion and measure its properties.
This included the TRES spectroscopy as described in Section \ref{ssec:tres_spectra}.
We also obtained time-series photometry of TOI-7019 from ground-based facilities. The higher angular resolution of these observations compared with TESS allowed us to confirm the identity of the star on which the transit events were occurring. We also observed in filters different from that of TESS to check for chromatic depth variations, the presence of which would indicate that the transiting companion may be a self-luminous stellar object. 

We observed transits of TOI-7019 with the 0.36m telescope at the Acton Sky Portal private observatory on 2024 Sep 19 using an $r^\prime$ filter, as well as with KeplerCam on the 1.2m telescope at FLWO on 2025 May 15, in alternating $B$ and $i^\prime$ filters. The data were calibrated and reduced using \texttt{AstroImageJ} \citep{AstroImageJ_Collins17}. We performed aperture photometry to extract light curves for the target star.
We recorded relevant parameters used to jointly detrend the light curves with the transit fit --- these were the observation airmass, width of the target point spread function, as well as the target pixel location. We note that because of the relatively long and infrequent transits of TOI-7019, we were only able to observe the transit ingress on these two occassions; however these were still able to confirm that the transits were on-target and achromatic.

\subsection{High Angular Resolution Imaging}\label{imaging}
To check for the presence of even closer companions that are unresolved by both TESS and seeing-limited observations, we obtained high angular-resolution imaging of TOI-7019.
TOI-7019 was observed on January 31, 2025 with the speckle polarimeter on the 2.5-m telescope at the Caucasian Observatory of Sternberg Astronomical Institute (SAI) of Lomonosov Moscow State University. A low--noise CMOS detector Hamamatsu ORCA--quest (\citep{SAI_Safonov2017, SAI_Strakhov2023}) was used as a detector. The atmospheric dispersion compensator was active, which allowed us to use the wide $I_\mathrm{c}$ band, the respective angular resolution is $0.083^{\prime\prime}$ arcseconds. Long exposure atmospheric seeing was $0.68^{\prime\prime}$ at the moment of observation. No companion was detected. The detection limits at distances $0.25$ and $1.0^{\prime\prime}$ from the star are $\Delta=3.8^m$ and $5.1^m$

% We obtained $I$-band speckle imaging using the Speckle Polarimeter \citep{SAI_Safonov2017,SAI_Strakhov2023} instrument on the 2.5m telescope at the Caucasian Mountain Observatory (CMO) of the Sternberg Astronomical Institute (SAI) in Russia.
We also observed TOI-7019 with the Palomar High Angular Resolution Observer \citep[PHARO;][]{PHARO_Hayward2001}, an adaptive optics imaging instrument on the 200-in Hale telescope at Palomar Observatory. The PHARO observations were made in the $K\mathrm{cont}$ filter and reduced using the pipeline described in \citet{PHARO_Furlan2017}.
Both observations revealed no close companions down to the sensitivity limits, as shown in Figure \ref{fig:ao_imaging}. The PHARO observations, which were more sensitive, rule out companions with contrast ratios less than $\Delta = 6.5$~mag at separations of $>0\farcs5$ from the primary.

\begin{figure}
 \centering
    \includegraphics[width=0.99\columnwidth]{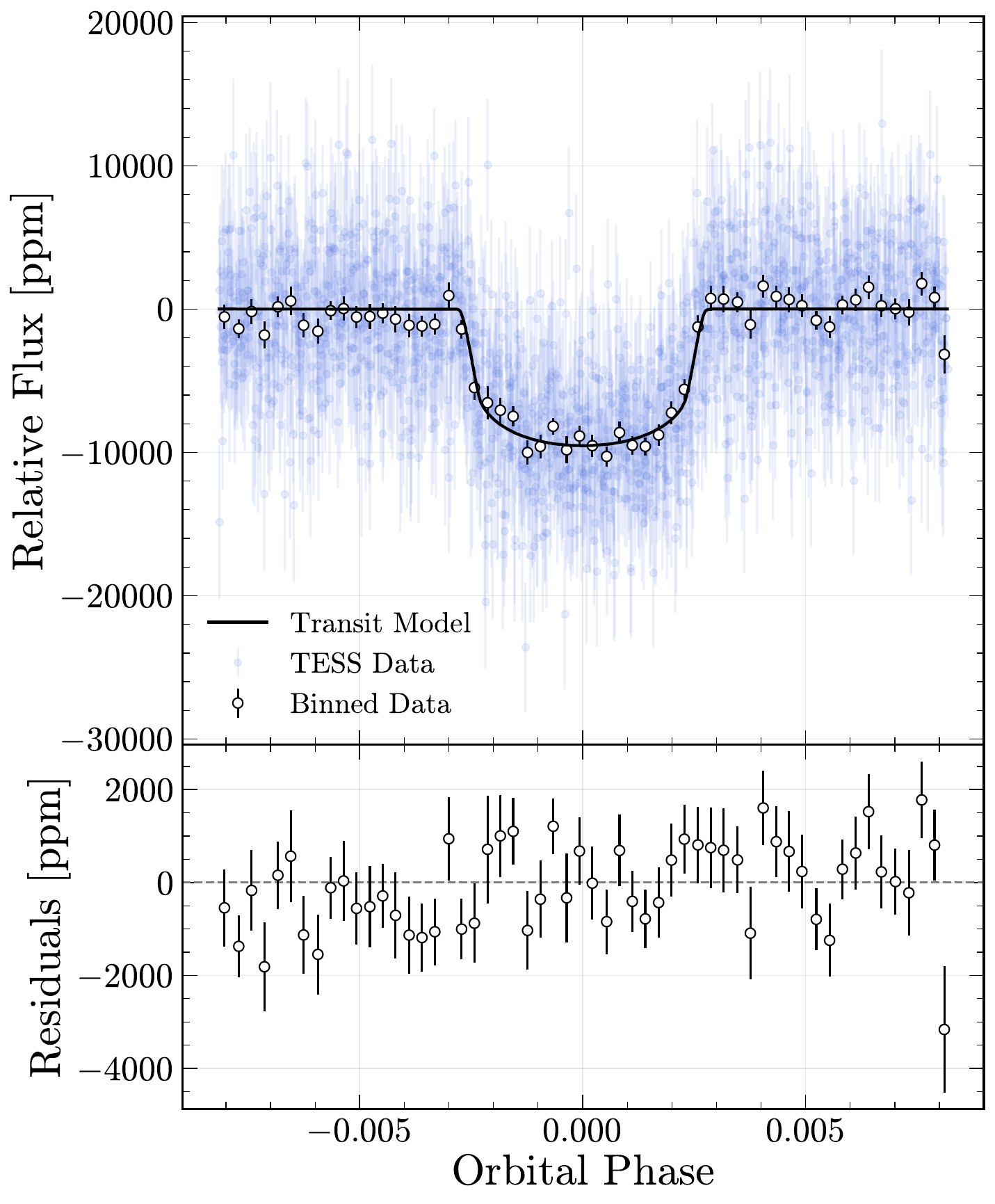} 
    \caption{Phase-folded TESS light curve of TOI-7019b from eight transit events observed in Sectors 14, 26, 40, 53, 55, 58, 81, and 85. The data are binned for clarity (white points), with the best-fit transit model from our joint analysis shown in black.  Lower panel shows the residuals after subtracting the best-fit model.}
    \label{fig:TESS_lc}
\end{figure}

\begin{figure*}
 \centering
    \includegraphics[width = \textwidth]{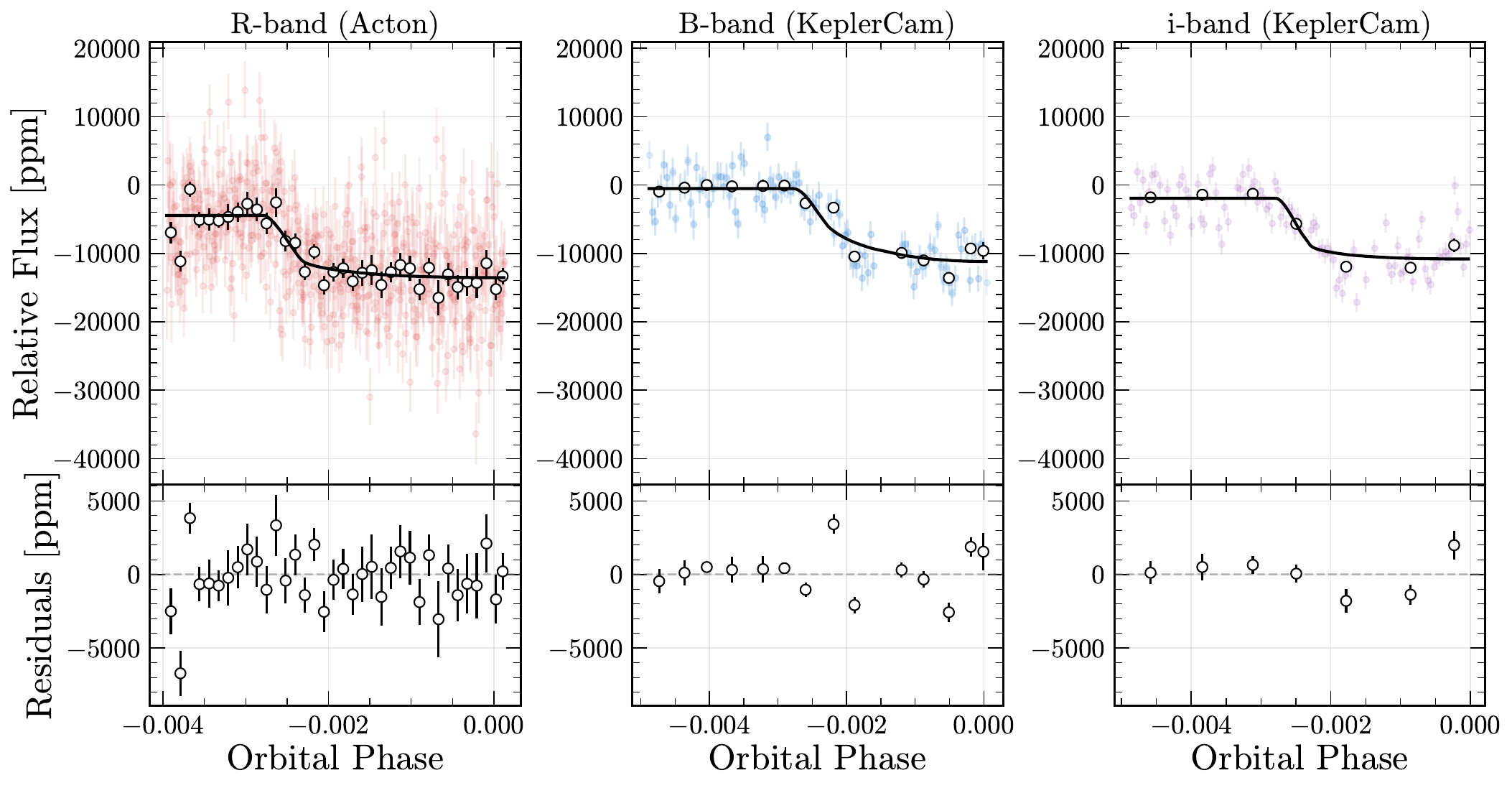} 
    \caption{Ground-based transit light curves of TOI-7019b obtained with KeplerCam on the FLWO 1.2m telescope in $B$ and $i'$ filters, and with the Acton Sky Portal 0.36m telescope in $r'$ band. The individual light curves are shown in color, with the best-fit transit model from our joint analysis shown in black. All observations capture the transit ingress and confirm that the events occur on the target star and are achromatic. Residuals after subtracting the best-fit model are shown below each panel.}
    \label{fig:groundbased_lc}
\end{figure*}

\subsection{Joint Transit \& Radial Velocity Analysis}

To derive a comprehensive set of physical parameters for the TOI-7019 system, we performed a global analysis of the available photometry and radial velocity data using the open-source \texttt{juliet} package \citep{Espinoza2019_juliet}. This package facilitates a joint analysis, simultaneously modeling the TESS photometry, the three ground-based transit light curves (in B, Sloan $i^\prime$, and Sloan $r^\prime$ filters), and the TRES radial velocity measurements using nested sampling \citep{Speagle2020_dynesty}.

Our model consists of a single transiting brown dwarf, modeled using the \texttt{batman} package \citep{Batman_Kreidberg15}, parameterized by the orbital period (P), time of transit center ($t_0$), eccentricity (e), and argument of periastron ($\omega$). We placed a uniform prior on the semi-major axis ($a$), and utilized the parameterization of \cite{Espinoza2018_sampling} to fit for the companion-to-star radius ratio ($p= \frac{R_p}{R_*}$) and impact parameter (b). We adopted a quadratic limb-darkening law for each instrument, fitting for the coefficients (q$_1$ and q$_2$) using the parameterization of \cite{Kipping2013_limbdarkening}.

\begin{figure}
 \centering
    \includegraphics[width=0.99\columnwidth]{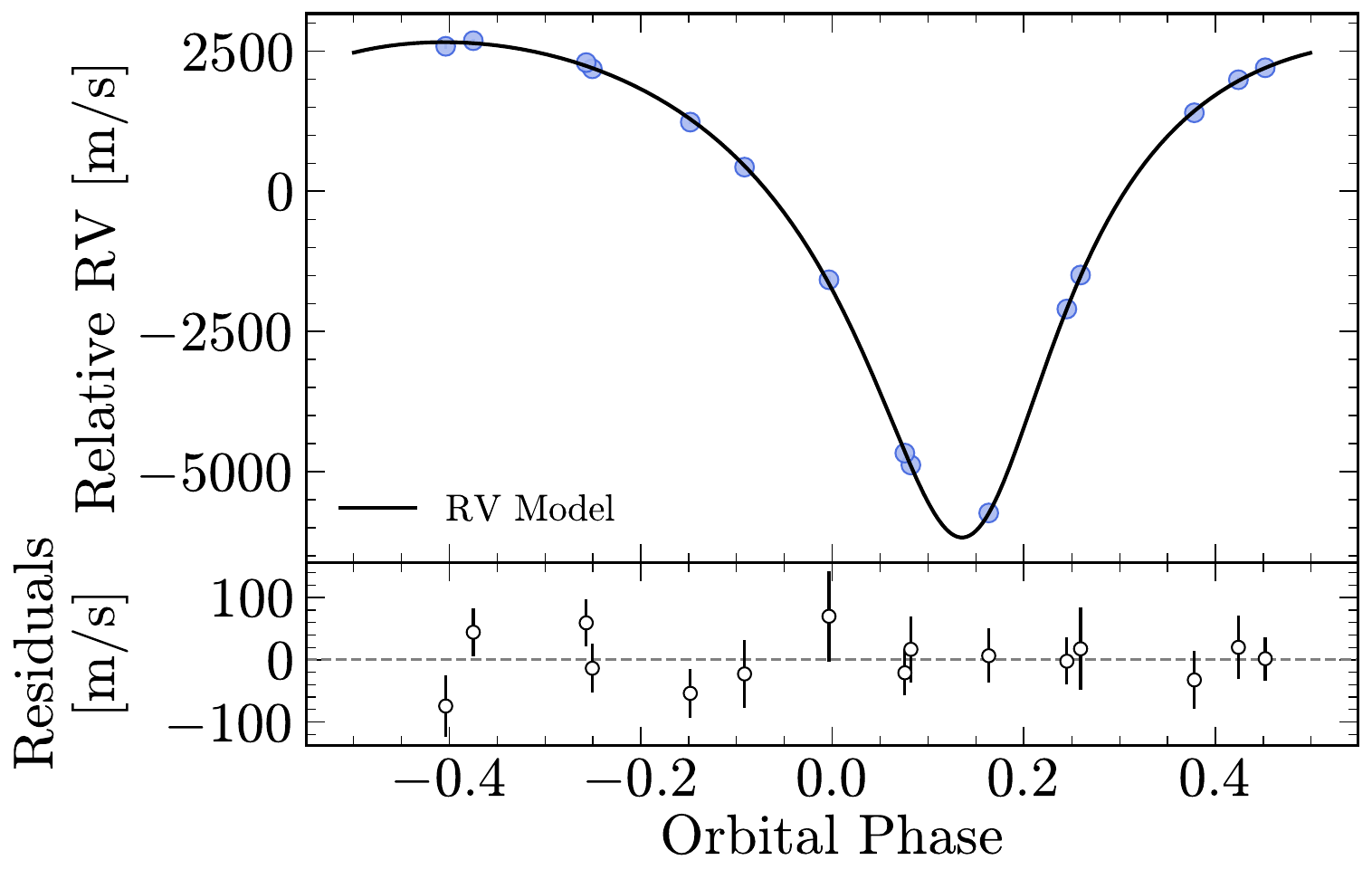}
    \includegraphics[width=0.99\columnwidth]{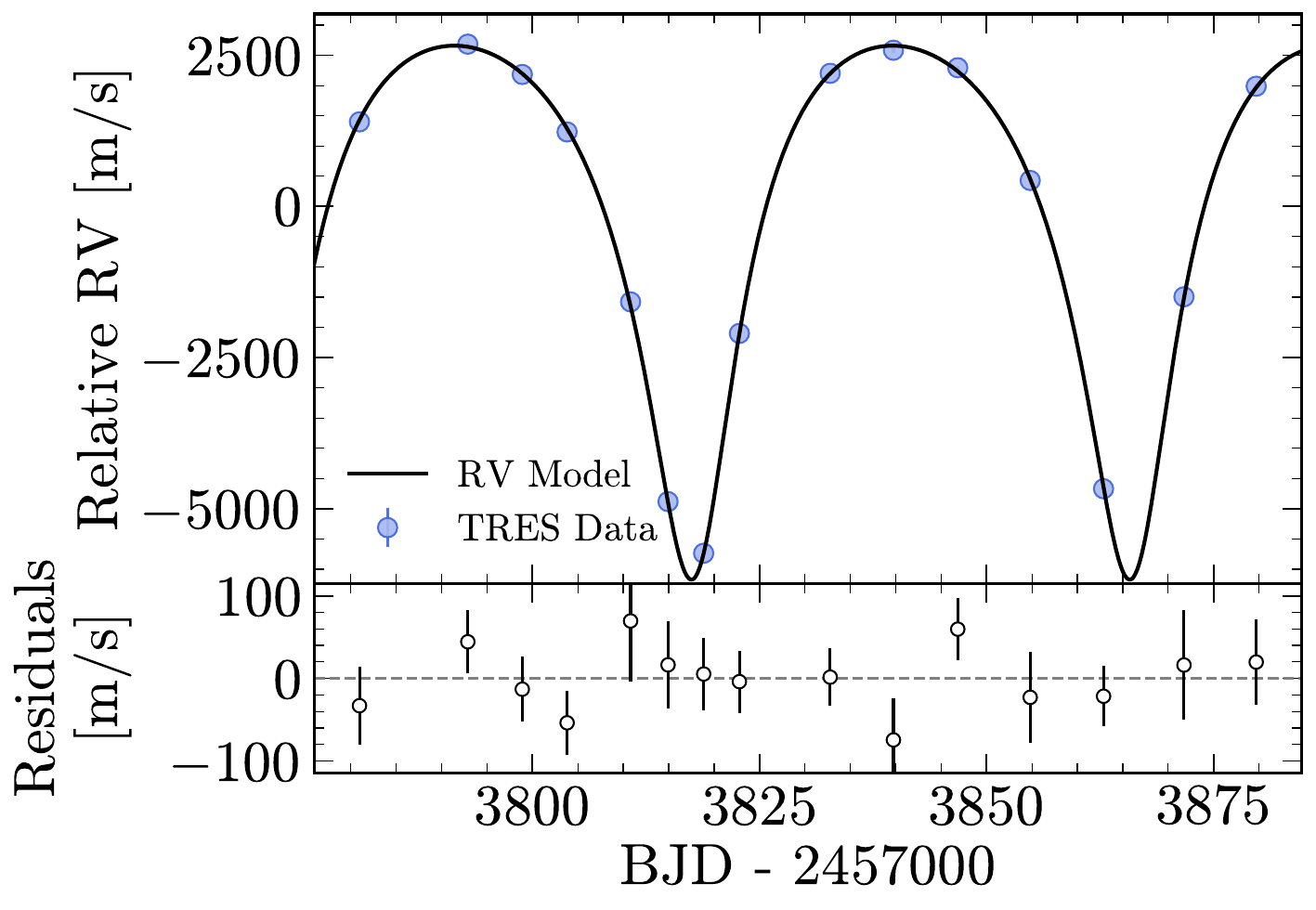} 
    \caption{TRES radial velocity measurements of TOI-7019 showing the motion induced by the transiting brown dwarf. The best-fit model (black line) yields a semi-amplitude of $K = 4417 \pm 17$ \ms, corresponding to a companion mass of $61.3 \pm 2.1$ \Mjup. The top panel shows the RV time-series phase-folded to the orbital period of the brown dwarf, while the lower panel shows the full RV time-series. The center-of-mass velocity $\mu_\mathrm{TRES}$ has been subtracted from the TRES RVs in this figure.}
    \label{fig:TRES_rv}
\end{figure}

The RV model fits for the semi-amplitude (K) of the companion's signal and a center-of-mass velocity offset for the TRES instrument ($\mathrm{\mu_{TRES}}$). To account for potential instrumental or stellar noise not captured by the formal errors, we included a jitter term ($\sigma_w$) for each photometric and RV instrument, which was added in quadrature to the uncertainties. 

For the ground-based photometry, we incorporated a linear detrending model into the fit to account for atmospheric conditions and instrumentals. This model simultaneously solved for the transit shape and systematic noise correlated with external parameters. For each of the three ground-based light curves, we used the measured airmass and the star's X-pixel position on the detector as linear regressors.

We defined uniform or log-uniform priors for most fitted parameters, which are detailed in Table \ref{tab:priors}. We explored the full parameter space and derived the posterior probability distributions using the \texttt{dynesty} nested sampling algorithm \citep{Speagle2020_dynesty}, as implemented in \texttt{juliet}. The median and 68\% confidence intervals of the marginalized posterior distributions were adopted as the final best-fit parameters and their uncertainties.

From our joint analysis of the TESS transits, ground-based transits, and TRES radial velocity measurements, we fit a companion-to-star radius ratio of $p=\frac{R_p}{R_\star}$ of 0.0896 $\pm$ 0.0017, an eccentricity of $0.4025 \pm 0.0015$, an RV semi-amplitude of $4417 \pm 17$ m/s, and a period of $48.25919 \pm 0.00009$ days.
To derive the companion mass and radii, we used the stellar properties derived in \S\ref{ssec:stellar_char} (Table \ref{tab:results_star}). 
We imposed error floors of 4\% in the stellar radius and 5\% in stellar mass to account for systematic uncertainties in the stellar models, as suggested by \citet{Tayar2022}.
We derived a brown dwarf mass of $61.3 \pm 2.1\,\Mjup$ and a radius of $R_b=0.82 \pm 0.02\,\Rjup$.

\begin{figure*}
 \centering
    \includegraphics[width=0.9\textwidth]{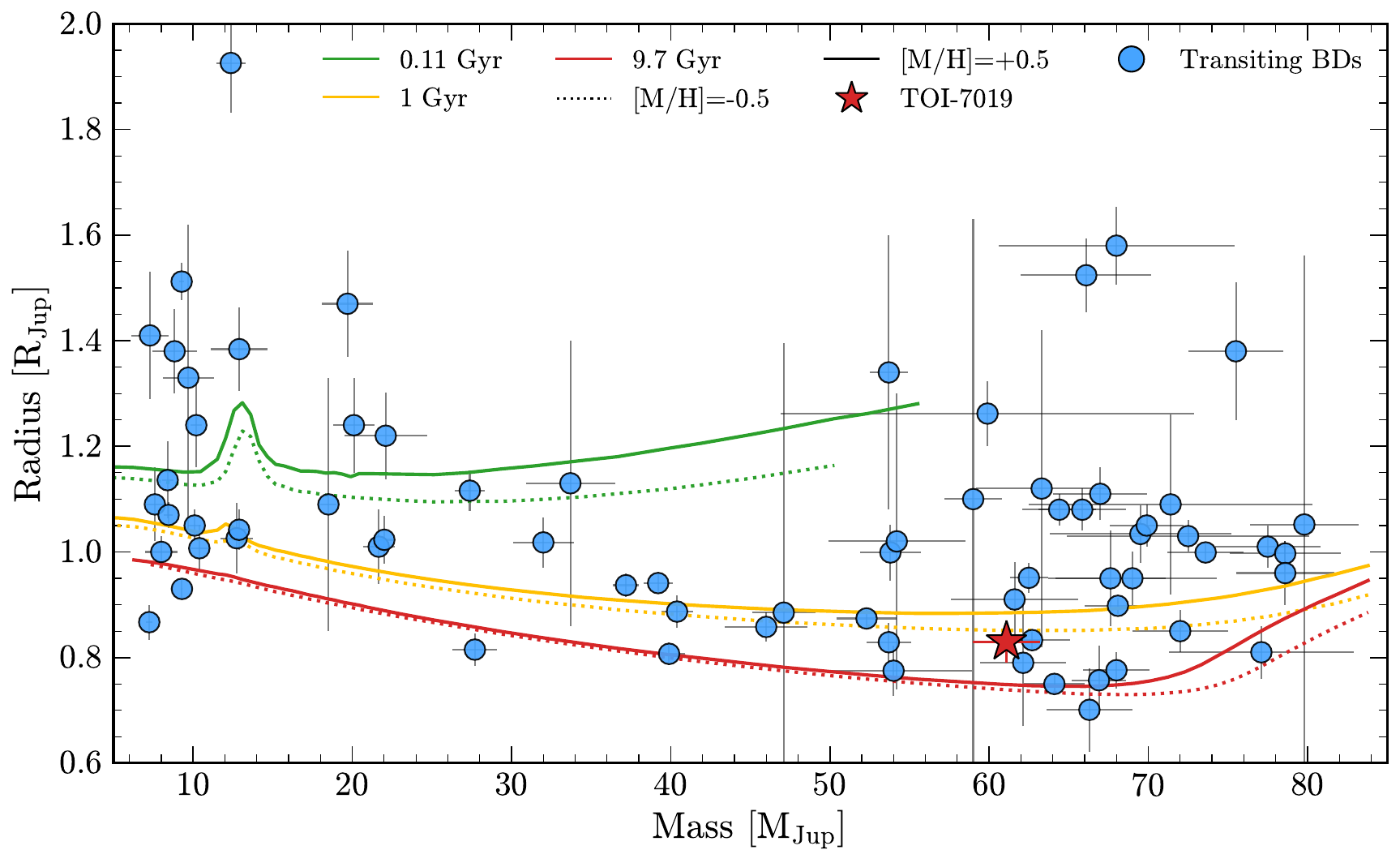} 
    \caption{Mass-Radius diagram for transiting brown dwarfs ($<80$\Mjup). TOI-7019b (red star) occupies a unique position as a transiting brown dwarf orbiting the most metal-poor known host. Theoretical isochrones from the Sonora models \citep{Morley24_Diamondback} are shown for solar metallicity at ages of 0.1, 1, 4, and 10 Gyr.}
    \label{fig:mr}
\end{figure*}

\begin{figure*}
 \centering
    \includegraphics[width=0.9\textwidth]{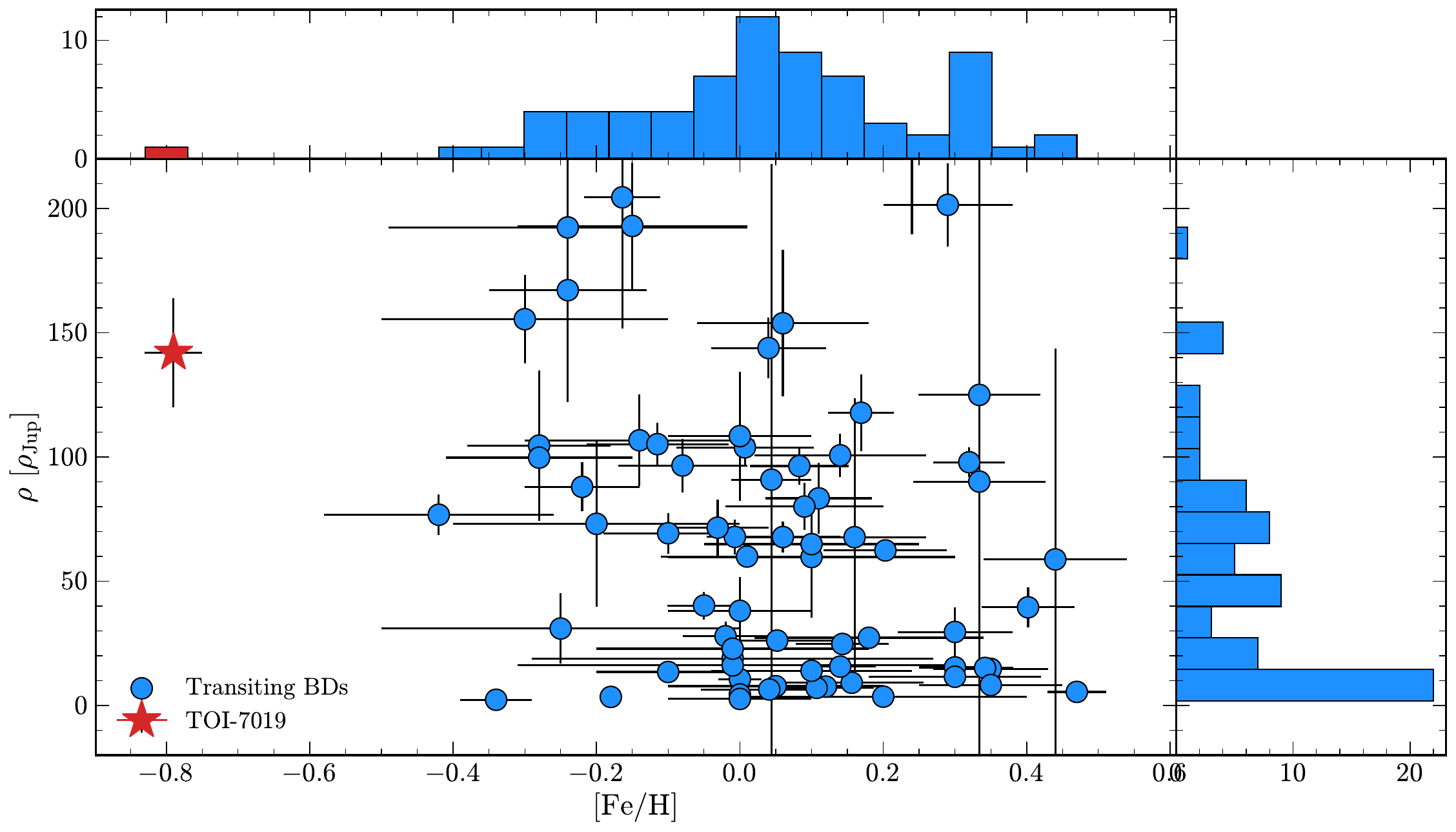} 
    \caption{Host star metallicity as a function of mean companion density for transiting brown dwarfs. TOI-7019b (red) at [Fe/H] = $-0.79$ dramatically extends the metallicity range probed by transiting brown dwarfs, and is among the densest brown dwarfs discovered. Histograms show the metallicity and mass distributions for companions.}
    \label{fig:metals}
\end{figure*}

\section{Discussion}\label{sec:discuss}

\subsection{TOI-7019: A Unique Laboratory for Substellar Evolution}

The discovery of TOI-7019b represents a convergence of extremes that makes it uniquely valuable for understanding brown dwarf evolution. With a metallicity of [Fe/H]~$ = -0.8$~dex, TOI-7019 is by far the most metal-poor host of a transiting brown dwarf discovered to date. It extends the metallicity baseline for  transiting brown dwarfs by 0.4 dex (below the previous record holder TOI-5610 at $-0.42$ dex; \citealt{Larsen_prevrec}).
This is a regime where theoretical models predict the strongest metallicity-dependent effects on substellar structure \citep{Burrows2011, Marley2021_Sonora, Morley24_Diamondback}. Simultaneously, its host star's robust thick-disk membership provides an age constraint of $\tau = 12 \pm 2$~Gyr, making this likely the oldest well-characterized transiting brown dwarfs known.\\

\subsection{Constraints on Mass, Radius, and Cooling Models}

% Our measured radius of \Rp = 0.83 $\pm$ 0.02 \Rjup for the 61.09 $\pm$ 2.1 \Mjup TOI-7019b provides a critical test of brown dwarf evolution models. 
As shown in Figure \ref{fig:mr}, TOI-7019b occupies a unique position in the mass-radius diagram, with a notably smaller radius than the bulk of the transiting brown dwarf population \citep{Carmichael2023, Vowell2025_BD}. While current evolutionary models like Sonora and ATMO \citep{Marley2021_Sonora, Phillips2020_ATMO, Morley24_Diamondback} only extend to $\mh = -0.5$, the theoretical framework suggests that atmospheric opacity decreases with metallicity, resulting in more efficient cooling and smaller radii for metal-poor brown dwarfs. Before interpreting the comparison between TOI-7019b and existing evolutionary model grids, it is important to clarify our assumptions regarding the relationship between the metallicity of the brown dwarf and that of its host star. If TOI-7019b formed through gravitational instability or cloud fragmentation, we might expect its bulk composition to closely mirror that of the host star, since both objects condensed from the same chemically well-mixed reservoir (similar to binary formation; \citealt{Kratter2010_gi, Hawkins_widebinary}). In contrast, if it formed via core accretion, the companion’s envelope may become enriched in heavy elements relative to the host due to the accretion of solids and disk chemistry effects \citep{Thorngren2016_enrichment, Atreya2016_saturn}. Since no direct measurement of TOI-7019b's atmospheric abundances yet exists, we assume as a working hypothesis that its bulk metallicity reflects that of the host star to first order, while noting that future spectroscopic measurements will be critical to test this assumption. With this assumption, at $\feh = -0.79$ or $\mh = -0.59$, this companion pushes beyond the current model grids, making it a critical test case for extending theoretical predictions into the extremely metal-poor regime. 

Although TOI-7019b receives minimal stellar irradiation ($T_\mathrm{Teq} \approx 480 K$ at 0.25 AU), its radius of 0.82 \Rjup is $12.3 \pm 5\%$ larger than predicted for a 9 Gyr, $\mh \approx -0.5$ brown dwarf compared to the \cite{Morley24_Diamondback} models. This is somewhat unexpected given the older age TOI-7019b, because if we assume it has a very low metallicity (like its host star), it should have an even smaller radius than predicted by those models. If we instead take the models at face value, the observed radius would imply an age of approximately 1 Gyr for a metal-poor brown dwarf, which is significantly younger than our independent stellar age estimate of $12\pm2$Gyr (Section \ref{sec:toistar}).
Another possibility is that the Sonora models assume solar elemental abundance patterns; the enhancement in alpha-element abundances in the TOI-7019 system may have altered the atmospheric opacity, delaying cooling and contraction. 

% The discovery of TOI-7019b provides an immediate and critical empirical test for substellar evolutionary models in a new regime. As its host star provides a reliable metallicity constraint, we can directly compare the physical properties of TOI-7019b to theoretical predictions of substellar evolution models \citep{Morley24_Diamondback}. The Sonora Diamondback models at their lowest metallicity grid point ([Fe/H] = -0.5) predict a radius for a 60 \Mjup, 10 Gyr old brown dwarf that is marginally consistent with our measurement of $0.81 \pm 0.04$ \Rjup though TOI-7019b's radius falls on the upper end of the predicted range \citep{Morley24_Diamondback}. This is significant because TOI-7019b, at [Fe/H] = -0.8, is substantially more metal-poor than the model grid extends.

% The relationship between radius and metallicity is not expected to be linear: the opacity effects that drive metallicity-dependent changes may scale differently in the extremely metal-poor regime.  The fact that TOI-7019b's radius is on the high end of (or possibly slightly larger than) predictions for [Fe/H]~$= -0.5$ may suggest that the radius-metallicity relationship flattens at very low metallicities, or additional physics not captured in current models becomes important at such extreme metal deficiencies. Each possibility highlights the value of TOI-7019b as an empirical anchor for extending models into this unexplored parameter space.

Our discovery can also be contextualized within the growing sample of isolated metal-poor brown dwarfs identified through the Backyard Worlds citizen science project and other surveys. \cite{Schneider2020_ExtremeSD} and \cite{Meisner2020_BW} discovered extremely metal-poor field brown dwarfs reaching [Fe/H]~$ < -1.6$. \cite{Kirkpatrick2021_theaccidenht} and \cite{Lodieu2022} expanded this census, with more detailed characterization from \cite{Zhang2025}.  
However, these isolated objects suffer from fundamental limitations: without transits or dynamical companions, their masses must be inferred from evolutionary models assuming an age.
% , while their radii remain completely unconstrained. Additionally, metallicity estimates for isolated brown dwarfs rely on spectroscopic fitting of complex molecular features that can be degenerate with temperature and surface gravity. 
TOI-7019b overcomes these limitations by providing more direct measurements of mass, radius, and age. Measuring the metallicity of its solar-type host star is also more straightforward than for isolated brown dwarfs. While the brown dwarf's metallicity itself has not yet been measured, the observed properties we do have already make it uniquely valuable as an empirical anchor point for calibrating the very evolutionary and atmospheric models that must be relied upon to characterize the isolated population.  Of particular relevance is Wolf 1130 C \citep{Mace2018}, previously one of the most metal-poor brown dwarf companions at \feh $\approx -1.2$ (see \citealt{Zhang2025} for refined analysis), though as a wide companion it too lacks radius constraints. Additionally, revised metallicity measurements might place it at slightly smaller values around \mh $= 0.68 \pm 0.04$ \citep{Burgasser2025_wolf1130}. TOI-7019b thus represents the first brown dwarf with complete fundamental parameters (mass, radius, and age), all measured without relying on uncertain theoretical models of brown dwarfs, with a host star metallicity measurement.

% Furthermore, TOI-7019b's position near the predicted local radius minimum in the brown dwarf radius distribution (around 60-70 \Mjup \citep{Baraffe2003_MassRad, Marley2021_Sonora}) provides an additional constraint. At these masses, electron degeneracy pressure is thought to dominate thermal pressure in the interior. This causes the radius to decrease with increasing mass before hydrogen fusion can contribute to the energy support of the brown dwarf. It is therefore remarkable to observe this minimum in a metal-poor, old companion. 

\subsection{Formation Implications: Evidence for Gravitational Instability}

Various properties of TOI-7019b strongly favor a formation pathway more akin to disk fragmentation or stellar-like collapse than to core accretion. As shown by \citet{Schlaufman}, the correlation between host-star metallicity and companion mass provides a key diagnostic: companions below roughly 10 $M_\mathrm{Jup}$ preferentially orbit metal-rich stars, consistent with core accretion, whereas more massive companions show no such trend, implying metallicity-independent formation via gravitational instability (GI). Recent work by  \citet{Ma_Ge} may place the empirical transition closer to $\sim$40–45 $M_\mathrm{Jup}$, but even at this new cutoff, \citet{Vowell2025_BD} find similar results of lower mass companions preferentially orbiting more metal rich host stars, similar to their planet counterparts. At 61 $M_\mathrm{Jup}$, TOI-7019b still lies securely above this threshold, falling into the higher mass range of brown dwarfs demonstrated to favor gravitational instability .

Furthermore, assuming that the primordial disk metallicity scales linearly with stellar metallicity (e.g., \citealt{Yasui2010}), the disk from which TOI-7019b formed would have contained around $6 \times$ less solid material than a solar-metallicity disk if we assume the disk dust-to-gas fraction scales as $10^{-\mathrm{[Fe/H]}} = 10^{-0.8}$ as in \cite{Mordasini_2012}. This large depletion of solids would make core accretion extremely inefficient, as the timescale for building a massive core increases dramatically with decreasing solid surface density \citep{Mordasini_2012}. In contrast, gravitational instability, which depends primarily on disk mass and cooling rates rather than solid content, could still operate efficiently in such a metal-poor environment \citep{B0ss1997_formation, Boss2006_formation}.

Finally, the orbital eccentricity itself ($e = 0.403 \pm 0.002$) provides additional, independent clues. The eccentricity distribution of giant planets in the range of 0.1 to 1 AU peaks sharply at low $e$, particularly for metal-poor hosts \citep[e.g.,][]{Dawson2018_ecc_metallicity_planets}, whereas both massive ($>42.5$ \Msun) brown-dwarf companions and close stellar binaries exhibit much broader, often excited eccentricities \citep{Ma_Ge,Vowell2025_BD, Wu2025_eccentricity}. TOI-7019b’s eccentricity of 0.4 might be considered generally moderate, but it is on the highest end of planet-like giant planets at this separation \citep{Dawson2018_ecc_metallicity_planets}, while it lies near the peak of stellar binary formation \citep{Wu2025_eccentricity}. It is therefore more consistent with formation through fragmentation in a stellar-like process. Thus, both the system’s low metallicity and its eccentric orbit argue for a stellar-like formation rather than a planet-like core accretion origin.

\subsection{Context Within Galactic Planet Populations}

While planets and brown dwarfs may form through different mechanisms, TOI-7019b's discovery is a valuable contribution to our growing understanding of substellar companions across diverse galactic populations. 
Emerging evidence suggests that galactic formation environment may have an impact on planet formation: for example,
\cite{Zink2023_thickdiskoccurence} measured planet occurrence rates from K2 as a function of galactic velocity, finding tentative evidence for reduced small planet occurrence around stars with large galactic vertical oscillation amplitudes, and that this trend could not be explained purely by the effect of host star metallicity.
\citet{BoleyHaloPlanets_2021,Yoshida2022} used TESS to place upper limits on the occurrence of short-period giant planets around halo stars, although these limits were consistent with expectations of reduced giant planet occurrence with low stellar metallicity.
% Larger-scale surveys targeting even more ancient, metal-poor populations further highlight the extreme rarity of these giant worlds: for example, \citet{BoleyHaloPlanets_2021} performed a comprehensive search for hot Jupiters across more than 11,000 Galactic halo stars with TESS but found no new planets at that time in the range of $-2.0 < ~\feh < -0.6$.
At present, only a handful of substellar objects have been definitively shown to orbit stars outside the galactic thin disk, including the TOI-561 system, which hosts a super-Earth orbiting an [Fe/H]~$ = -0.41$ and thick disk kinematics \citep{Lacedelli2021_thickdiskplanet}.
Another example is WASP-21\,b, a hot Jupiter also found around a thick disk star \citep{Ciceri2013_wasp21b}, which shows that even close-in giant planets can form in these old, metal-poor environments (see also V. DiTomasso et al. 2025, submitted).
% Still, a few planets in these older populations have been confirmed, such as the TOI-561 system, which hosts a super-Earth despite having [Fe/H]~$ = -0.41$ and thick disk kinematics \citep{Lacedelli2021_thickdiskplanet}. Similarly, WASP-21b is a hot Jupiter also found around a thick disk star, which shows that even close-in giant planets can form in older environments \citep{Ciceri2013_wasp21b}.

TOI-7019b extends this work into the brown dwarf regime, where formation mechanisms may differ, but Galactic context remains important. Here, we are only just beginning to probe the formation of substellar objects in different galactic populations.

% If massive brown dwarfs form primarily via gravitational instability, they should be present even around the most metal-poor stars. TOI-7019b suggests that this prediction holds true.
% The existence of a massive brown dwarf companion in this ancient population has several implications. 

% \begin{itemize}
%     \item Disk Longevity: The formation of a 63.6 MJup companion requires a relatively massive protoplanetary disk. That such disks could form and survive long enough to produce brown dwarfs even in the turbulent environment of the early Galaxy suggests that disk formation is a robust process.
%     \item Chemical Evolution Context: With [ $\alpha / \mathrm{Fe}$ ] = +0.2, TOI-7019 formed during an epoch of rapid chemical enrichment dominated by Type II supernovae. The successful formation of a brown dwarf in this chemically distinct environment demonstrates that substellar formation is not sensitively dependent on detailed abundance patterns beyond overall metallicity.
%     \item Dynamical Survival: Despite around 9-10 Gyr of dynamical evolution and numerous passages through the Galactic disk, the TOI-7019 system remains intact. This places constraints on the orbital architecture, suggesting a relatively tight orbit consistent with our observed period of 48.3 days.
% \end{itemize}

% \begin{figure*}
%  \centering
%     \includegraphics[width=0.8\textwidth]{kinematics.png} 
%     \caption{Kinematics (TO VEDANT: ADD REAL CAPTION LATER)}
%     \label{fig:metals}
% \end{figure*}

\section{Summary \& Conclusions}\label{sec:conclusions}

We present the discovery and characterization of TOI-7019b, a transiting brown dwarf that represents a unique convergence of extremes in the substellar population. The combination of high mass (61 \Mjup) and extreme sub-solar metallicity (\feh~$= -0.8$) makes TOI-7019b a keystone brown dwarf for testing models of brown dwarf formation and evolution. Our key findings are as follows:

\begin{enumerate}

    \item TOI-7019 is a metal-poor (\feh $= -0.79$), alpha-enhanced (\afe $= 0.26$) star that kinematically and chemically belongs to the Milky Way's thick, high-$\alpha$ disk population (Figure~\ref{fig:discovery}). 

    \item Two independent methods --- a comparison of stellar parameters to isochrones, and the age-metallicity relation of the thick disk --- constrain the age of the TOI-7019 system to $\tau = 12 \pm 2$~Gyr (Figures~\ref{fig:kiel} and \ref{fig:amr}). This makes TOI-7019b the oldest known brown dwarf transiting a main sequence star for which such a precise age can be estimated.

    \item We combine TESS photometry, ground-based follow-up photometry, and high-resolution spectroscopy to measure the properties of TOI-7019's transiting companion, a remarkably dense brown dwarf (Figures~\ref{fig:TESS_lc} and \ref{fig:TRES_rv}).
    
    \item We measure \Mp = 61.3 $\pm$ 2.1 \Mjup and \Rp = 0.82 $\pm$ 0.04 \Rjup, placing TOI-7019b near the predicted radius minimum for brown dwarfs.
    A small radius is consistent with enhanced cooling expected in metal-poor atmospheres (Figure~\ref{fig:mr}).
    However, TOI-7019b remains somewhat larger than predicted by standard brown dwarf cooling tracks despite its old age, suggesting the need for incorporating different elemental abundance patterns in substellar models or improved calibration at the old ages.

    % \item TOI-7019 is by far the most metal-poor and ancient transiting brown dwarf host known, with a metallicity lower than the previous record holder by a factor of two (Figure~\ref{fig:metals}). 

    % \item  The existence of this massive brown dwarf in such a metal-poor environment supports formation via gravitational instability rather than core accretion, implying that protoplanetary disks may be long-lived. 
    
\end{enumerate}

The discovery of TOI-7019b opens several avenues for future investigation. Atmospheric characterization with JWST could reveal whether molecular features (H$_2$O, CH$_4$, CO) differ from solar-metallicity brown dwarfs of similar mass and effective temperature, though disentangling metallicity effects from other atmospheric parameters would be challenging. Continued radial velocity monitoring would refine the orbital parameters and search for long-term trends that might indicate additional companions.

More broadly, the combination of \textit{Gaia} kinematics with large spectroscopic surveys now enables systematic identification of substellar candidates in different Galactic populations. TOI-7019b demonstrates that brown dwarfs can form in the metal-poor thick disk, raising the question of whether they extend to even more extreme environments. The Galactic halo, with $\feh \lesssim -1.0$~dex, presents an intriguing frontier. Such objects, if they exist, would provide extreme test cases for substellar cooling models at metallicities well beyond current model grids. Even a single transiting brown dwarf around a halo star would extend the metallicity baseline by another factor of two beyond TOI-7019b. Continued exploration of substellar companions across the full diversity of Galactic stellar populations will reveal how planet formation operates in these extreme conditions.

% TOI-7019b demonstrates that the substellar population extends to environments radically different from the solar neighborhood, motivating continued exploration across the full diversity of Galactic stellar populations.

% Each additional discovery will tighten these constraints, strengthening our understanding of the relationship between radius, mass, metallicity, and age.

% Future atmospheric characterization with JWST and expanded surveys of thick disk stars promise to further illuminate the formation and evolution of substellar objects in our galaxy. 
% TOI-7019b stands as a reminder that in the era of Gaia and large spectroscopic surveys, combining exoplanet science with Galactic archaeology opens entirely new parameter spaces for discovery.

%% Please use the acknowledgment and contribution environments. This will 
%% be anonomyized when the "anonymous" style option is used. 
\begin{acknowledgments}
The authors thank Keivan Stassun for conducting an independent stellar analysis, which provided a useful verification of our stellar parameters. 
S.W.Y. gratefully acknowledges support from the Heising-Simons Foundation.

This paper includes data collected by the TESS mission. Funding for the TESS mission is provided by the NASA's Science Mission Directorate. This research has made use of the Exoplanet Follow-up Observation Program (ExoFOP; DOI: 10.26134/ExoFOP5) website, which is operated by the California Institute of Technology, under contract with the National Aeronautics and Space Administration under the Exoplanet Exploration Program.

\end{acknowledgments}

% \begin{contribution}
%%This section gives authors the space to recognize author contributions. The text inside this environment is NOT counted towards the total word quanta. At a minimum, manuscripts are expected to include this text:

% All authors contributed equally to the Terra Mater collaboration.

%% But authors are expected to provide more specific details, e.g. 
%%
%%SC was responsible for writing and submitting the manuscript.
%%WWM came up with the initial research concept and edited the manuscript.
%%OTS obtained the funding and edited the manuscript.
%%EBF provided the formal analysis and validation. He also edited the manuscript.
%%GEH Supervised the undergraduates, wrote the software and administers the project github and Zenodo repositories.
%%
%% Authors can use the Contributor Role Taxonomy (CRediT) at
%% for ideas on how write a good statement tailored to their needs.

% \end{contribution}

\facilities{TESS, FLWO:1.2m (KeplerCam), FLWO:1.5m (TRES)}

\software{astropy \citep{Astropy13,Astropy18,Astropy2022},  
          lightkurve \citep{Lightkurve18},
          juliet \citep{Espinoza2019_juliet}}

\bibliography{catalogs,instruments,software,refs}{}

@ARTICLE{Morley24_Diamondback,
       author = {{Morley}, Caroline V. and {Mukherjee}, Sagnick and {Marley}, Mark S. and {Fortney}, Jonathan J. and {Visscher}, Channon and {Lupu}, Roxana and {Gharib-Nezhad}, Ehsan and {Thorngren}, Daniel and {Freedman}, Richard and {Batalha}, Natasha},
        title = "{The Sonora Substellar Atmosphere Models. III. Diamondback: Atmospheric Properties, Spectra, and Evolution for Warm Cloudy Substellar Objects}",
      journal = {\apj},
         year = 2024,
        month = nov,
       volume = {975},
       number = {1},
          eid = {59},
        pages = {59},
          doi = {10.3847/1538-4357/ad71d5},
archivePrefix = {arXiv},
       eprint = {2402.00758},
 primaryClass = {astro-ph.SR},
       adsurl = {https://ui.adsabs.harvard.edu/abs/2024ApJ...975...59M}
}

@article{GaiaDR3_Vallenari2022,
  title = {Gaia {{Data Release}} 3. {{Summary}} of the Content and Survey Properties},
  author = {{Gaia Collaboration} and Vallenari, A. and Brown, A.G.A. and Prusti, T. and {et al.}},
  year = {2022},
  month = jun,
  journal = {Astronomy \& Astrophysics},
  issn = {0004-6361, 1432-0746},
  doi = {10.1051/0004-6361/202243940},
  langid = {english}
}

@article{GaiaEDR3_Brown2021,
  title = {Gaia {{Early Data Release}} 3 - {{Summary}} of the Contents and Survey Properties},
  author = {Brown, A. G. A. and Vallenari, A. and Prusti, T. and de Bruijne, J. H. J. and Babusiaux, C. and Biermann, M. and Creevey, O. L. and Evans, D. W. and Eyer, L. and Hutton, A. and Jansen, F. and Jordi, C. and Klioner, S. A. and Lammers, U. and Lindegren, L. and Luri, X. and Mignard, F. and Panem, C. and Pourbaix, D. and Randich, S. and Sartoretti, P. and Soubiran, C. and Walton, N. A. and Arenou, F. and {Bailer-Jones}, C. a. L. and Bastian, U. and Cropper, M. and Drimmel, R. and Katz, D. and Lattanzi, M. G. and van Leeuwen, F. and Bakker, J. and Cacciari, C. and Casta{\~n}eda, J. and Angeli, F. De and Ducourant, C. and Fabricius, C. and Fouesneau, M. and Fr{\'e}mat, Y. and Guerra, R. and Guerrier, A. and Guiraud, J. and Piccolo, A. Jean-Antoine and Masana, E. and Messineo, R. and Mowlavi, N. and Nicolas, C. and Nienartowicz, K. and Pailler, F. and Panuzzo, P. and Riclet, F. and Roux, W. and Seabroke, G. M. and Sordo, R. and Tanga, P. and Th{\'e}venin, F. and {Gracia-Abril}, G. and Portell, J. and Teyssier, D. and Altmann, M. and Andrae, R. and {Bellas-Velidis}, I. and Benson, K. and Berthier, J. and Blomme, R. and Brugaletta, E. and Burgess, P. W. and Busso, G. and Carry, B. and Cellino, A. and Cheek, N. and Clementini, G. and Damerdji, Y. and Davidson, M. and Delchambre, L. and Dell'Oro, A. and {Fern{\'a}ndez-Hern{\'a}ndez}, J. and Galluccio, L. and {Garc{\'i}a-Lario}, P. and {Garcia-Reinaldos}, M. and {Gonz{\'a}lez-N{\'u}{\~n}ez}, J. and Gosset, E. and Haigron, R. and Halbwachs, J.-L. and Hambly, N. C. and Harrison, D. L. and Hatzidimitriou, D. and Heiter, U. and Hern{\'a}ndez, J. and Hestroffer, D. and Hodgkin, S. T. and Holl, B. and Jan{\ss}en, K. and de Fombelle, G. Jevardat and Jordan, S. and {Krone-Martins}, A. and Lanzafame, A. C. and L{\"o}ffler, W. and Lorca, A. and Manteiga, M. and Marchal, O. and Marrese, P. M. and Moitinho, A. and Mora, A. and Muinonen, K. and Osborne, P. and Pancino, E. and Pauwels, T. and Petit, J.-M. and {Recio-Blanco}, A. and Richards, P. J. and Riello, M. and Rimoldini, L. and Robin, A. C. and Roegiers, T. and Rybizki, J. and Sarro, L. M. and Siopis, C. and Smith, M. and Sozzetti, A. and Ulla, A. and Utrilla, E. and van Leeuwen, M. and van Reeven, W. and Abbas, U. and Aramburu, A. Abreu and Accart, S. and Aerts, C. and Aguado, J. J. and Ajaj, M. and Altavilla, G. and {\'A}lvarez, M. A. and {Cid-Fuentes}, J. {\'A}lvarez and Alves, J. and Anderson, R. I. and Varela, E. Anglada and Antoja, T. and Audard, M. and Baines, D. and Baker, S. G. and {Balaguer-N{\'u}{\~n}ez}, L. and Balbinot, E. and Balog, Z. and Barache, C. and Barbato, D. and Barros, M. and Barstow, M. A. and Bartolom{\'e}, S. and Bassilana, J.-L. and Bauchet, N. and {Baudesson-Stella}, A. and Becciani, U. and Bellazzini, M. and Bernet, M. and Bertone, S. and Bianchi, L. and {Blanco-Cuaresma}, S. and Boch, T. and Bombrun, A. and Bossini, D. and Bouquillon, S. and Bragaglia, A. and Bramante, L. and Breedt, E. and Bressan, A. and Brouillet, N. and Bucciarelli, B. and Burlacu, A. and Busonero, D. and Butkevich, A. G. and Buzzi, R. and Caffau, E. and Cancelliere, R. and C{\'a}novas, H. and {Cantat-Gaudin}, T. and Carballo, R. and Carlucci, T. and Carnerero, M. I. and Carrasco, J. M. and Casamiquela, L. and Castellani, M. and {Castro-Ginard}, A. and Sampol, P. Castro and Chaoul, L. and Charlot, P. and Chemin, L. and Chiavassa, A. and Cioni, M.-R. L. and Comoretto, G. and Cooper, W. J. and Cornez, T. and Cowell, S. and Crifo, F. and Crosta, M. and Crowley, C. and Dafonte, C. and Dapergolas, A. and David, M. and David, P. and de Laverny, P. and Luise, F. De and March, R. De and Ridder, J. De and de Souza, R. and de Teodoro, P. and de Torres, A. and del Peloso, E. F. and del Pozo, E. and Delbo, M. and Delgado, A. and Delgado, H. E. and Delisle, J.-B. and Matteo, P. Di and Diakite, S. and Diener, C. and Distefano, E. and Dolding, C. and Eappachen, D. and Edvardsson, B. and Enke, H. and Esquej, P. and Fabre, C. and Fabrizio, M. and Faigler, S. and Fedorets, G. and Fernique, P. and Fienga, A. and Figueras, F. and Fouron, C. and Fragkoudi, F. and Fraile, E. and Franke, F. and Gai, M. and Garabato, D. and {Garcia-Gutierrez}, A. and {Garc{\'i}a-Torres}, M. and Garofalo, A. and Gavras, P. and Gerlach, E. and Geyer, R. and Giacobbe, P. and Gilmore, G. and Girona, S. and Giuffrida, G. and Gomel, R. and Gomez, A. and {Gonzalez-Santamaria}, I. and {Gonz{\'a}lez-Vidal}, J. J. and Granvik, M. and {Guti{\'e}rrez-S{\'a}nchez}, R. and Guy, L. P. and Hauser, M. and Haywood, M. and Helmi, A. and Hidalgo, S. L. and Hilger, T. and H{\l}adczuk, N. and Hobbs, D. and Holland, G. and Huckle, H. E. and Jasniewicz, G. and Jonker, P. G. and Campillo, J. Juaristi and Julbe, F. and Karbevska, L. and Kervella, P. and Khanna, S. and Kochoska, A. and Kontizas, M. and Kordopatis, G. and Korn, A. J. and {Kostrzewa-Rutkowska}, Z. and Kruszy{\'n}ska, K. and Lambert, S. and Lanza, A. F. and Lasne, Y. and Campion, J.-F. Le and Fustec, Y. Le and Lebreton, Y. and Lebzelter, T. and Leccia, S. and Leclerc, N. and {Lecoeur-Taibi}, I. and Liao, S. and Licata, E. and Lindstr{\o}m, E. P. and Lister, T. A. and Livanou, E. and Lobel, A. and Pardo, P. Madrero and Managau, S. and Mann, R. G. and Marchant, J. M. and Marconi, M. and Santos, M. M. S. Marcos and Marinoni, S. and Marocco, F. and Marshall, D. J. and Polo, L. Martin and {Mart{\'i}n-Fleitas}, J. M. and Masip, A. and Massari, D. and {Mastrobuono-Battisti}, A. and Mazeh, T. and McMillan, P. J. and Messina, S. and Michalik, D. and Millar, N. R. and Mints, A. and Molina, D. and Molinaro, R. and Moln{\'a}r, L. and Montegriffo, P. and Mor, R. and Morbidelli, R. and Morel, T. and Morris, D. and Mulone, A. F. and Munoz, D. and Muraveva, T. and Murphy, C. P. and Musella, I. and Noval, L. and Ord{\'e}novic, C. and Orr{\`u}, G. and Osinde, J. and Pagani, C. and Pagano, I. and Palaversa, L. and Palicio, P. A. and Panahi, A. and Pawlak, M. and Esteller, X. Pe{\~n}alosa and Penttil{\"a}, A. and Piersimoni, A. M. and Pineau, F.-X. and Plachy, E. and Plum, G. and Poggio, E. and Poretti, E. and Poujoulet, E. and Pr{\v s}a, A. and Pulone, L. and Racero, E. and Ragaini, S. and Rainer, M. and Raiteri, C. M. and Rambaux, N. and Ramos, P. and {Ramos-Lerate}, M. and Fiorentin, P. Re and Regibo, S. and Reyl{\'e}, C. and Ripepi, V. and Riva, A. and Rixon, G. and Robichon, N. and Robin, C. and Roelens, M. and Rohrbasser, L. and {Romero-G{\'o}mez}, M. and Rowell, N. and Royer, F. and Rybicki, K. A. and Sadowski, G. and Sell{\'e}s, A. Sagrist{\`a} and Sahlmann, J. and Salgado, J. and Salguero, E. and Samaras, N. and Gimenez, V. Sanchez and Sanna, N. and Santove{\~n}a, R. and Sarasso, M. and Schultheis, M. and Sciacca, E. and Segol, M. and Segovia, J. C. and S{\'e}gransan, D. and Semeux, D. and Shahaf, S. and Siddiqui, H. I. and Siebert, A. and Siltala, L. and Slezak, E. and Smart, R. L. and Solano, E. and Solitro, F. and Souami, D. and Souchay, J. and Spagna, A. and Spoto, F. and Steele, I. A. and Steidelm{\"u}ller, H. and Stephenson, C. A. and S{\"u}veges, M. and Szabados, L. and {Szegedi-Elek}, E. and Taris, F. and Tauran, G. and Taylor, M. B. and Teixeira, R. and Thuillot, W. and Tonello, N. and Torra, F. and Torra, J. and Turon, C. and Unger, N. and Vaillant, M. and van Dillen, E. and Vanel, O. and Vecchiato, A. and Viala, Y. and Vicente, D. and Voutsinas, S. and Weiler, M. and Wevers, T. and Wyrzykowski, {\L} and Yoldas, A. and Yvard, P. and Zhao, H. and Zorec, J. and Zucker, S. and Zurbach, C. and Zwitter, T.},
  year = {2021},
  month = may,
  journal = {Astronomy \& Astrophysics},
  volume = {649},
  pages = {A1},
  publisher = {{EDP Sciences}},
  issn = {0004-6361, 1432-0746},
  doi = {10.1051/0004-6361/202039657},
  langid = {english}
}

@article{GaiaEDR3_Riello2021,
  title = {Gaia {{Early Data Release}} 3 - {{Photometric}} Content and Validation},
  author = {Riello, M. and Angeli, F. De and Evans, D. W. and Montegriffo, P. and Carrasco, J. M. and Busso, G. and Palaversa, L. and Burgess, P. W. and Diener, C. and Davidson, M. and Rowell, N. and Fabricius, C. and Jordi, C. and Bellazzini, M. and Pancino, E. and Harrison, D. L. and Cacciari, C. and van Leeuwen, F. and Hambly, N. C. and Hodgkin, S. T. and Osborne, P. J. and Altavilla, G. and Barstow, M. A. and Brown, A. G. A. and Castellani, M. and Cowell, S. and Luise, F. De and Gilmore, G. and Giuffrida, G. and Hidalgo, S. and Holland, G. and Marinoni, S. and Pagani, C. and Piersimoni, A. M. and Pulone, L. and Ragaini, S. and Rainer, M. and Richards, P. J. and Sanna, N. and Walton, N. A. and Weiler, M. and Yoldas, A.},
  year = {2021},
  month = may,
  journal = {Astronomy \& Astrophysics},
  volume = {649},
  pages = {A3},
  publisher = {{EDP Sciences}},
  issn = {0004-6361, 1432-0746},
  doi = {10.1051/0004-6361/202039587},
  langid = {english}
}

@article{TMASS_Cutri2003,
  title = {{{VizieR Online Data Catalog}}: {{2MASS All}}-{{Sky Catalog}} of {{Point Sources}} ({{Cutri}}+ 2003)},
  shorttitle = {{{VizieR Online Data Catalog}}},
  author = {Cutri, R. M. and Skrutskie, M. F. and {van Dyk}, S. and Beichman, C. A. and Carpenter, J. M. and Chester, T. and Cambresy, L. and Evans, T. and Fowler, J. and Gizis, J. and Howard, E. and Huchra, J. and Jarrett, T. and Kopan, E. L. and Kirkpatrick, J. D. and Light, R. M. and Marsh, K. A. and McCallon, H. and Schneider, S. and Stiening, R. and Sykes, M. and Weinberg, M. and Wheaton, W. A. and Wheelock, S. and Zacarias, N.},
  year = {2003},
  month = jun,
  journal = {VizieR Online Data Catalog},
  pages = {II/246}
}

@article{WISE_Cutri2012,
  title = {{{VizieR Online Data Catalog}}: {{WISE All}}-{{Sky Data Release}} ({{Cutri}}+ 2012)},
  shorttitle = {{{VizieR Online Data Catalog}}},
  author = {Cutri, R. M. {et al.}},
  year = {2012},
  month = apr,
  journal = {VizieR Online Data Catalog},
  pages = {II/311}
}

@ARTICLE{TESS_Ricker15,
       author = {{Ricker}, George R. and {Winn}, Joshua N. and {Vanderspek}, Roland and {Latham}, David W. and {Bakos}, G{\'a}sp{\'a}r {\'A}. and {Bean}, Jacob L. and {Berta-Thompson}, Zachory K. and {Brown}, Timothy M. and {Buchhave}, Lars and {Butler}, Nathaniel R. and {Butler}, R. Paul and {Chaplin}, William J. and {Charbonneau}, David and {Christensen-Dalsgaard}, J{\o}rgen and {Clampin}, Mark and {Deming}, Drake and {Doty}, John and {De Lee}, Nathan and {Dressing}, Courtney and {Dunham}, Edward W. and {Endl}, Michael and {Fressin}, Francois and {Ge}, Jian and {Henning}, Thomas and {Holman}, Matthew J. and {Howard}, Andrew W. and {Ida}, Shigeru and {Jenkins}, Jon M. and {Jernigan}, Garrett and {Johnson}, John Asher and {Kaltenegger}, Lisa and {Kawai}, Nobuyuki and {Kjeldsen}, Hans and {Laughlin}, Gregory and {Levine}, Alan M. and {Lin}, Douglas and {Lissauer}, Jack J. and {MacQueen}, Phillip and {Marcy}, Geoffrey and {McCullough}, Peter R. and {Morton}, Timothy D. and {Narita}, Norio and {Paegert}, Martin and {Palle}, Enric and {Pepe}, Francesco and {Pepper}, Joshua and {Quirrenbach}, Andreas and {Rinehart}, Stephen A. and {Sasselov}, Dimitar and {Sato}, Bun'ei and {Seager}, Sara and {Sozzetti}, Alessandro and {Stassun}, Keivan G. and {Sullivan}, Peter and {Szentgyorgyi}, Andrew and {Torres}, Guillermo and {Udry}, Stephane and {Villasenor}, Joel},
        title = "{Transiting Exoplanet Survey Satellite (TESS)}",
      journal = {Journal of Astronomical Telescopes, Instruments, and Systems},
         year = 2015,
        month = jan,
       volume = {1},
          eid = {014003},
        pages = {014003},
          doi = {10.1117/1.JATIS.1.1.014003},
       adsurl = {https://ui.adsabs.harvard.edu/abs/2015JATIS...1a4003R}
}

@inproceedings{TESS_SPOC_Jenkins2016,
  title = {The {{TESS}} Science Processing Operations Center},
  booktitle = {Software and {{Cyberinfrastructure}} for {{Astronomy IV}}},
  author = {Jenkins, Jon M. and Twicken, Joseph D. and McCauliff, Sean and Campbell, Jennifer and Sanderfer, Dwight and Lung, David and {Mansouri-Samani}, Masoud and Girouard, Forrest and Tenenbaum, Peter and Klaus, Todd and Smith, Jeffrey C. and Caldwell, Douglas A. and Chacon, A. Dean and Henze, Christopher and Heiges, Cory and Latham, David W. and Morgan, Edward and Swade, Daryl and Rinehart, Stephen and Vanderspek, Roland},
  year = {2016},
  month = aug,
  volume = {9913},
  pages = {1232--1251},
  publisher = {{SPIE}},
  doi = {10.1117/12.2233418}
}

@article{TESS_QLP_Huang2020a,
  title = {Photometry of 10 {{Million Stars}} from the {{First Two Years}} of {{TESS Full Frame Images}}: Part {{I}}},
  shorttitle = {Photometry of 10 {{Million Stars}} from the {{First Two Years}} of {{TESS Full Frame Images}}},
  author = {Huang, Chelsea X. and Vanderburg, Andrew and P{\'a}l, Andras and Sha, Lizhou and Yu, Liang and Fong, Willie and Fausnaugh, Michael and Shporer, Avi and Guerrero, Natalia and Vanderspek, Roland and Ricker, George},
  year = {2020},
  month = nov,
  journal = {Research Notes of the AAS},
  volume = {4},
  number = {11},
  pages = {204},
  publisher = {{American Astronomical Society}},
  issn = {2515-5172},
  doi = {10.3847/2515-5172/abca2e},
  langid = {english}
}

@article{TESS_QLP_Huang2020b,
  title = {Photometry of 10 {{Million Stars}} from the {{First Two Years}} of {{TESS Full Frame Images}}: Part {{II}}},
  shorttitle = {Photometry of 10 {{Million Stars}} from the {{First Two Years}} of {{TESS Full Frame Images}}},
  author = {Huang, Chelsea X. and Vanderburg, Andrew and P{\'a}l, Andras and Sha, Lizhou and Yu, Liang and Fong, Willie and Fausnaugh, Michael and Shporer, Avi and Guerrero, Natalia and Vanderspek, Roland and Ricker, George},
  year = {2020},
  month = nov,
  journal = {Research Notes of the AAS},
  volume = {4},
  number = {11},
  pages = {206},
  publisher = {{American Astronomical Society}},
  issn = {2515-5172},
  doi = {10.3847/2515-5172/abca2d},
  langid = {english}
}

@article{TESS_QLP_Kunimoto2021,
  title = {Quick-Look {{Pipeline Lightcurves}} for 9.1 {{Million Stars Observed}} over the {{First Year}} of the {{TESS Extended Mission}}},
  author = {Kunimoto, Michelle and Huang, Chelsea and Tey, Evan and Fong, Willie and Hesse, Katharine and Shporer, Avi and Guerrero, Natalia and Fausnaugh, Michael and Vanderspek, Roland and Ricker, George},
  year = {2021},
  month = oct,
  volume = {5},
  number = {10},
  pages = {234},
  publisher = {{American Astronomical Society}},
  journal = {Research Notes of the AAS},
  issn = {2515-5172},
  doi = {10.3847/2515-5172/ac2ef0},
  langid = {english}
}

@article{TESS_QLP_Kunimoto2022b,
  title = {Quick-Look {{Pipeline Light Curves}} for 5.7 {{Million Stars Observed Over}} the {{Second Year}} of {{TESS}}' {{First Extended Mission}}},
  author = {Kunimoto, Michelle and Tey, Evan and Fong, Willie and Hesse, Katharine and Shporer, Avi and Fausnaugh, Michael and Vanderspek, Roland and Ricker, George},
  year = {2022},
  month = nov,
  journal = {Research Notes of the AAS},
  volume = {6},
  number = {11},
  pages = {236},
  publisher = {{The American Astronomical Society}},
  issn = {2515-5172},
  urldate = {2023-07-06},
  langid = {english}
}

@article{TESS_Faint_Kunimoto2022a,
  title = {The {{TESS Faint-star Search}}: 1617 {{TOIs}} from the {{TESS Primary Mission}}},
  shorttitle = {The {{TESS Faint-star Search}}},
  author = {Kunimoto, Michelle and Daylan, Tansu and Guerrero, Natalia and Fong, William and Bryson, Steve and Ricker, George R. and Fausnaugh, Michael and Huang, Chelsea X. and Sha, Lizhou and Shporer, Avi and Vanderburg, Andrew and Vanderspek, Roland K. and Yu, Liang},
  year = {2022},
  month = mar,
  journal = {The Astrophysical Journal Supplement Series},
  volume = {259},
  number = {2},
  pages = {33},
  publisher = {{American Astronomical Society}},
  issn = {0067-0049},
  doi = {10.3847/1538-4365/ac5688},
  langid = {english}
}

@article{TESS_TOIs_Guerrero2021,
  title = {The {{TESS Objects}} of {{Interest Catalog}} from the {{TESS Prime Mission}}},
  author = {Guerrero, Natalia M. and Seager, S. and Huang, Chelsea X. and Vanderburg, Andrew and Soto, Aylin Garcia and Mireles, Ismael and Hesse, Katharine and Fong, William and Glidden, Ana and Shporer, Avi and Latham, David W. and Collins, Karen A. and Quinn, Samuel N. and Burt, Jennifer and Dragomir, Diana and Crossfield, Ian and Vanderspek, Roland and Fausnaugh, Michael and Burke, Christopher J. and Ricker, George and Daylan, Tansu and Essack, Zahra and G{\"u}nther, Maximilian N. and Osborn, Hugh P. and Pepper, Joshua and Rowden, Pamela and Sha, Lizhou and Jr, Steven Villanueva and Yahalomi, Daniel A. and Yu, Liang and Ballard, Sarah and Batalha, Natalie M. and Berardo, David and Chontos, Ashley and Dittmann, Jason A. and Esquerdo, Gilbert A. and {Mikal-Evans}, Thomas and Jayaraman, Rahul and Krishnamurthy, Akshata and Louie, Dana R. and Mehrle, Nicholas and Niraula, Prajwal and Rackham, Benjamin V. and Rodriguez, Joseph E. and Rowden, Stephen J. L. and {Sousa-Silva}, Clara and Watanabe, David and Wong, Ian and Zhan, Zhuchang and Zivanovic, Goran and Christiansen, Jessie L. and Ciardi, David R. and Swain, Melanie A. and Lund, Michael B. and Mullally, Susan E. and Fleming, Scott W. and Rodriguez, David R. and Boyd, Patricia T. and Quintana, Elisa V. and Barclay, Thomas and Col{\'o}n, Knicole D. and Rinehart, S. A. and Schlieder, Joshua E. and Clampin, Mark and Jenkins, Jon M. and Twicken, Joseph D. and Caldwell, Douglas A. and Coughlin, Jeffrey L. and Henze, Chris and Lissauer, Jack J. and Morris, Robert L. and Rose, Mark E. and Smith, Jeffrey C. and Tenenbaum, Peter and Ting, Eric B. and Wohler, Bill and Bakos, G. {\'A} and Bean, Jacob L. and {Berta-Thompson}, Zachory K. and Bieryla, Allyson and Bouma, Luke G. and Buchhave, Lars A. and Butler, Nathaniel and Charbonneau, David and Doty, John P. and Ge, Jian and Holman, Matthew J. and Howard, Andrew W. and Kaltenegger, Lisa and Kane, Stephen R. and Kjeldsen, Hans and Kreidberg, Laura and Lin, Douglas N. C. and Minsky, Charlotte and Narita, Norio and Paegert, Martin and P{\'a}l, Andr{\'a}s and Palle, Enric and Sasselov, Dimitar D. and Spencer, Alton and Sozzetti, Alessandro and Stassun, Keivan G. and Torres, Guillermo and Udry, Stephane and Winn, Joshua N.},
  year = {2021},
  month = jun,
  journal = {The Astrophysical Journal Supplement Series},
  volume = {254},
  number = {2},
  pages = {39},
  publisher = {{IOP Publishing}},
  issn = {0067-0049},
  doi = {10.3847/1538-4365/abefe1},
  langid = {english}
}

@article{TESS_PDC_Stumpe2012,
  title = {Kepler {{Presearch Data Conditioning I}}\textemdash{{Architecture}} and {{Algorithms}} for {{Error Correction}} in {{Kepler Light Curves}}},
  author = {Stumpe, Martin C. and Smith, Jeffrey C. and Cleve, Jeffrey E. Van and Twicken, Joseph D. and Barclay, Thomas S. and Fanelli, Michael N. and Girouard, Forrest R. and Jenkins, Jon M. and Kolodziejczak, Jeffery J. and McCauliff, Sean D. and Morris, Robert L.},
  year = {2012},
  month = aug,
  journal = {Publications of the Astronomical Society of the Pacific},
  volume = {124},
  number = {919},
  pages = {985},
  publisher = {{IOP Publishing}},
  issn = {1538-3873},
  doi = {10.1086/667698},
  langid = {english}
}

@article{TESS_PDC_Smith2012,
  title = {Kepler {{Presearch Data Conditioning II}} - {{A Bayesian Approach}} to {{Systematic Error Correction}}},
  author = {Smith, Jeffrey C. and Stumpe, Martin C. and Cleve, Jeffrey E. Van and Jenkins, Jon M. and Barclay, Thomas S. and Fanelli, Michael N. and Girouard, Forrest R. and Kolodziejczak, Jeffery J. and McCauliff, Sean D. and Morris, Robert L. and Twicken, Joseph D.},
  year = {2012},
  month = sep,
  journal = {Publications of the Astronomical Society of the Pacific},
  volume = {124},
  number = {919},
  pages = {1000},
  publisher = {{IOP Publishing}},
  issn = {1538-3873},
  doi = {10.1086/667697},
  langid = {english}
}

@article{TESS_PDC_Stumpe2014,
  title = {Multiscale {{Systematic Error Correction}} via {{Wavelet}}-{{Based Bandsplitting}} in {{Kepler Data}}},
  author = {Stumpe, Martin C. and Smith, Jeffrey C. and Catanzarite, Joseph H. and Cleve, Jeffrey E. Van and Jenkins, Jon M. and Twicken, Joseph D. and Girouard, Forrest R.},
  year = {2014},
  month = jan,
  journal = {Publications of the Astronomical Society of the Pacific},
  volume = {126},
  number = {935},
  pages = {100},
  publisher = {{IOP Publishing}},
  issn = {1538-3873},
  doi = {10.1086/674989},
  langid = {english}
}

@PHDTHESIS{TRES_Furesz2008,
       author = {{F\H{u}r{\'e}sz}, Gabor},
        title = "{Design and Application of High Resolution and Multiobject Spectrographs: Dynamical Studies of Open Clusters}",
       school = {University of Szeged, Hungary},
         year = 2008,
}

@article{PHARO_Hayward2001,
  title = {{{PHARO}}: A {{Near}}-{{Infrared Camera}} for the {{Palomar Adaptive Optics System}}},
  shorttitle = {{{PHARO}}},
  author = {Hayward, T. L. and Brandl, B. and Pirger, B. and Blacken, C. and Gull, G. E. and Schoenwald, J. and Houck, J. R.},
  year = {2001},
  month = jan,
  journal = {Publications of the Astronomical Society of the Pacific},
  volume = {113},
  number = {779},
  pages = {105},
  publisher = {{IOP Publishing}},
  issn = {1538-3873},
  doi = {10.1086/317969},
  langid = {english}
}

@ARTICLE{PHARO_Furlan2017,
       author = {{Furlan}, E. and {Ciardi}, D.~R. and {Everett}, M.~E. and {Saylors}, M. and {Teske}, J.~K. and {Horch}, E.~P. and {Howell}, S.~B. and {van Belle}, G.~T. and {Hirsch}, L.~A. and {Gautier}, III, T.~N. and {Adams}, E.~R. and {Barrado}, D. and {Cartier}, K.~M.~S. and {Dressing}, C.~D. and {Dupree}, A.~K. and {Gilliland}, R.~L. and {Lillo-Box}, J. and {Lucas}, P.~W. and {Wang}, J.},
        title = "{The Kepler Follow-up Observation Program. I. A Catalog of Companions to Kepler Stars from High-Resolution Imaging}",
      journal = {\aj},
         year = 2017,
        month = feb,
       volume = {153},
       number = {2},
          eid = {71},
        pages = {71},
          doi = {10.3847/1538-3881/153/2/71},
archivePrefix = {arXiv},
       eprint = {1612.02392},
 primaryClass = {astro-ph.SR},
       adsurl = {https://ui.adsabs.harvard.edu/abs/2017AJ....153...71F}
}

@article{SAI_Safonov2017,
  title = {The Speckle Polarimeter of the 2.5-m Telescope: Design and Calibration},
  shorttitle = {The Speckle Polarimeter of the 2.5-m Telescope},
  author = {Safonov, B. S. and Lysenko, P. A. and Dodin, A. V.},
  year = {2017},
  month = may,
  journal = {Astronomy Letters},
  volume = {43},
  pages = {344--364},
  issn = {1063-7737},
  doi = {10.1134/S1063773717050036}
}

@ARTICLE{SAI_Strakhov2023,
       author = {{Strakhov}, I.~A. and {Safonov}, B.~S. and {Cheryasov}, D.~V.},
        title = "{Speckle Interferometry with CMOS Detector}",
      journal = {Astrophysical Bulletin},
         year = 2023,
        month = jun,
       volume = {78},
       number = {2},
        pages = {234-258},
          doi = {10.1134/S1990341323020104},
archivePrefix = {arXiv},
       eprint = {2305.00451},
 primaryClass = {astro-ph.IM},
       adsurl = {https://ui.adsabs.harvard.edu/abs/2023AstBu..78..234S}
}

@article{TRES_Quinn2012,
  title = {{{TWO}} ``b''s {{IN THE BEEHIVE}}: {{THE DISCOVERY OF THE FIRST HOT JUPITERS IN AN OPEN CLUSTER}}},
  shorttitle = {{{TWO}} ``b''s {{IN THE BEEHIVE}}},
  author = {Quinn, Samuel N. and White, Russel J. and Latham, David W. and Buchhave, Lars A. and Cantrell, Justin R. and Dahm, Scott E. and F{\textbackslash}Hur{\'e}sz, Gabor and Szentgyorgyi, Andrew H. and Geary, John C. and Torres, Guillermo and Bieryla, Allyson and Berlind, Perry and Calkins, Michael C. and Esquerdo, Gilbert A. and Stefanik, Robert P.},
  year = {2012},
  month = aug,
  journal = {The Astrophysical Journal},
  volume = {756},
  number = {2},
  pages = {L33},
  publisher = {{American Astronomical Society}},
  issn = {2041-8205},
  doi = {10.1088/2041-8205/756/2/L33},
  langid = {english}
}

@article{TRES_Buchhave2010,
  title = {{{HAT-P-16b}}: {{A 4MJPLANET TRANSITING A BRIGHT STAR ON AN ECCENTRIC ORBIT}},},
  shorttitle = {{{HAT-P-16b}}},
  author = {Buchhave, L. A. and Bakos, G. {\'A} and Hartman, J. D. and Torres, G. and Kov{\'a}cs, G. and Latham, D. W. and Noyes, R. W. and Esquerdo, G. A. and Everett, M. and Howard, A. W. and Marcy, G. W. and Fischer, D. A. and Johnson, J. A. and Andersen, J. and F{\textbackslash}Hur{\'e}sz, G. and Perumpilly, G. and Sasselov, D. D. and Stefanik, R. P. and B{\'e}ky, B. and L{\'a}z{\'a}r, J. and Papp, I. and S{\'a}ri, P.},
  year = {2010},
  month = aug,
  journal = {The Astrophysical Journal},
  volume = {720},
  number = {2},
  pages = {1118--1125},
  publisher = {{American Astronomical Society}},
  issn = {0004-637X},
  doi = {10.1088/0004-637X/720/2/1118},
  langid = {english}
}

@ARTICLE{Stassun2019,
       author = {{Stassun}, Keivan G. and {Oelkers}, Ryan J. and {Paegert}, Martin and {Torres}, Guillermo and {Pepper}, Joshua and {De Lee}, Nathan and {Collins}, Kevin and {Latham}, David W. and {Muirhead}, Philip S. and {Chittidi}, Jay and {Rojas-Ayala}, B{\'a}rbara and {Fleming}, Scott W. and {Rose}, Mark E. and {Tenenbaum}, Peter and {Ting}, Eric B. and {Kane}, Stephen R. and {Barclay}, Thomas and {Bean}, Jacob L. and {Brassuer}, C.~E. and {Charbonneau}, David and {Ge}, Jian and {Lissauer}, Jack J. and {Mann}, Andrew W. and {McLean}, Brian and {Mullally}, Susan and {Narita}, Norio and {Plavchan}, Peter and {Ricker}, George R. and {Sasselov}, Dimitar and {Seager}, S. and {Sharma}, Sanjib and {Shiao}, Bernie and {Sozzetti}, Alessandro and {Stello}, Dennis and {Vanderspek}, Roland and {Wallace}, Geoff and {Winn}, Joshua N.},
        title = "{The Revised TESS Input Catalog and Candidate Target List}",
      journal = {\aj},
         year = 2019,
        month = oct,
       volume = {158},
       number = {4},
          eid = {138},
        pages = {138},
          doi = {10.3847/1538-3881/ab3467},
archivePrefix = {arXiv},
       eprint = {1905.10694},
 primaryClass = {astro-ph.SR},
       adsurl = {https://ui.adsabs.harvard.edu/abs/2019AJ....158..138S}
}

@ARTICLE{Yasui2010,
       author = {{Yasui}, Chikako and {Kobayashi}, Naoto and {Tokunaga}, Alan T. and {Saito}, Masao and {Tokoku}, Chihiro},
        title = "{Short Lifetime of Protoplanetary Disks in Low-metallicity Environments}",
      journal = {\apjl},
         year = 2010,
        month = nov,
       volume = {723},
       number = {1},
        pages = {L113-L116},
          doi = {10.1088/2041-8205/723/1/L113},
archivePrefix = {arXiv},
       eprint = {1010.1668},
 primaryClass = {astro-ph.SR},
       adsurl = {https://ui.adsabs.harvard.edu/abs/2010ApJ...723L.113Y}
}

@ARTICLE{Burgasser2025_wolf1130,
       author = {{Burgasser}, Adam J. and {Gonzales}, Eileen C. and {Beiler}, Samuel A. and {Visscher}, Channon and {Burningham}, Ben and {Mace}, Gregory N. and {Faherty}, Jacqueline K. and {Zhang}, Zenghua and {Sousa-Silva}, Clara and {Lodieu}, Nicolas and {Metchev}, Stanimir A. and {Meisner}, Aaron and {Cushing}, Michael and {Schneider}, Adam C. and {Suarez}, Genaro and {Hsu}, Chih-Chun and {Gerasimov}, Roman and {Aganze}, Christian and {Theissen}, Christopher A.},
        title = "{Observation of undepleted phosphine in the atmosphere of a low-temperature brown dwarf}",
      journal = {arXiv e-prints},
         year = 2025,
        month = oct,
          eid = {arXiv:2510.03916},
        pages = {arXiv:2510.03916},
          doi = {10.48550/arXiv.2510.03916},
archivePrefix = {arXiv},
       eprint = {2510.03916},
 primaryClass = {astro-ph.SR},
       adsurl = {https://ui.adsabs.harvard.edu/abs/2025arXiv251003916B}
}

@ARTICLE{Hawkins_widebinary,
       author = {{Hawkins}, Keith and {Lucey}, Madeline and {Ting}, Yuan-Sen and {Ji}, Alexander and {Katzberg}, Dustin and {Thompson}, Megan and {El-Badry}, Kareem and {Teske}, Johanna and {Nelson}, Tyler and {Carrillo}, Andreia},
        title = "{Identical or fraternal twins? The chemical homogeneity of wide binaries from Gaia DR2}",
      journal = {\mnras},
         year = 2020,
        month = feb,
       volume = {492},
       number = {1},
        pages = {1164-1179},
          doi = {10.1093/mnras/stz3132},
archivePrefix = {arXiv},
       eprint = {1912.08895},
 primaryClass = {astro-ph.SR},
       adsurl = {https://ui.adsabs.harvard.edu/abs/2020MNRAS.492.1164H}
}

@ARTICLE{Kratter2010_gi,
       author = {{Kratter}, Kaitlin M. and {Murray-Clay}, Ruth A. and {Youdin}, Andrew N.},
        title = "{The Runts of the Litter: Why Planets Formed Through Gravitational Instability Can Only Be Failed Binary Stars}",
      journal = {\apj},
         year = 2010,
        month = feb,
       volume = {710},
       number = {2},
        pages = {1375-1386},
          doi = {10.1088/0004-637X/710/2/1375},
archivePrefix = {arXiv},
       eprint = {0909.2644},
 primaryClass = {astro-ph.EP},
       adsurl = {https://ui.adsabs.harvard.edu/abs/2010ApJ...710.1375K}
}

@ARTICLE{Thorngren2016_enrichment,
       author = {{Thorngren}, Daniel P. and {Fortney}, Jonathan J. and {Murray-Clay}, Ruth A. and {Lopez}, Eric D.},
        title = "{The Mass-Metallicity Relation for Giant Planets}",
      journal = {\apj},
         year = 2016,
        month = nov,
       volume = {831},
       number = {1},
          eid = {64},
        pages = {64},
          doi = {10.3847/0004-637X/831/1/64},
archivePrefix = {arXiv},
       eprint = {1511.07854},
 primaryClass = {astro-ph.EP},
       adsurl = {https://ui.adsabs.harvard.edu/abs/2016ApJ...831...64T}
}

@ARTICLE{Atreya2016_saturn,
       author = {{Atreya}, Sushil K. and {Crida}, Aurelien and {Guillot}, Tristan and {Lunine}, Jonathan I. and {Madhusudhan}, Nikku and {Mousis}, Olivier},
        title = "{The Origin and Evolution of Saturn, with Exoplanet Perspective}",
      journal = {arXiv e-prints},
         year = 2016,
        month = jun,
          eid = {arXiv:1606.04510},
        pages = {arXiv:1606.04510},
          doi = {10.48550/arXiv.1606.04510},
archivePrefix = {arXiv},
       eprint = {1606.04510},
 primaryClass = {astro-ph.EP},
       adsurl = {https://ui.adsabs.harvard.edu/abs/2016arXiv160604510A}
}

@ARTICLE{Wu2025_eccentricity,
       author = {{Wu}, Yanqin and {Hadden}, Sam and {Dewberry}, Janosz and {El-Badry}, Kareem and {Matzner}, Christopher D.},
        title = "{Eccentricities of Close Stellar Binaries}",
      journal = {\apjl},
         year = 2025,
        month = mar,
       volume = {982},
       number = {1},
          eid = {L34},
        pages = {L34},
          doi = {10.3847/2041-8213/adb751},
archivePrefix = {arXiv},
       eprint = {2411.09905},
 primaryClass = {astro-ph.SR},
       adsurl = {https://ui.adsabs.harvard.edu/abs/2025ApJ...982L..34W}
}

@ARTICLE{Dawson2018_ecc_metallicity_planets,
       author = {{Dawson}, Rebekah I. and {Murray-Clay}, Ruth A.},
        title = "{Giant Planets Orbiting Metal-rich Stars Show Signatures of Planet-Planet Interactions}",
      journal = {\apjl},
         year = 2013,
        month = apr,
       volume = {767},
       number = {2},
          eid = {L24},
        pages = {L24},
          doi = {10.1088/2041-8205/767/2/L24},
archivePrefix = {arXiv},
       eprint = {1302.6244},
 primaryClass = {astro-ph.EP},
       adsurl = {https://ui.adsabs.harvard.edu/abs/2013ApJ...767L..24D}
}

@ARTICLE{Haywood2013,
       author = {{Haywood}, Misha and {Di Matteo}, Paola and {Lehnert}, Matthew D. and {Katz}, David and {G{\'o}mez}, Ana},
        title = "{The age structure of stellar populations in the solar vicinity. Clues of a two-phase formation history of the Milky Way disk}",
      journal = {\aap},
         year = 2013,
        month = dec,
       volume = {560},
          eid = {A109},
        pages = {A109},
          doi = {10.1051/0004-6361/201321397},
archivePrefix = {arXiv},
       eprint = {1305.4663},
 primaryClass = {astro-ph.GA},
       adsurl = {https://ui.adsabs.harvard.edu/abs/2013A&A...560A.109H}
}

@ARTICLE{Bonaca2020,
       author = {{Bonaca}, Ana and {Conroy}, Charlie and {Cargile}, Phillip A. and {Naidu}, Rohan P. and {Johnson}, Benjamin D. and {Zaritsky}, Dennis and {Ting}, Yuan-Sen and {Caldwell}, Nelson and {Han}, Jiwon Jesse and {van Dokkum}, Pieter},
        title = "{Timing the Early Assembly of the Milky Way with the H3 Survey}",
      journal = {\apjl},
         year = 2020,
        month = jul,
       volume = {897},
       number = {1},
          eid = {L18},
        pages = {L18},
          doi = {10.3847/2041-8213/ab9caa},
archivePrefix = {arXiv},
       eprint = {2004.11384},
 primaryClass = {astro-ph.GA},
       adsurl = {https://ui.adsabs.harvard.edu/abs/2020ApJ...897L..18B}
}

@ARTICLE{Ciceri2013_wasp21b,
       author = {{Ciceri}, S. and {Mancini}, L. and {Southworth}, J. and {Nikolov}, N. and {Bozza}, V. and {Bruni}, I. and {Calchi Novati}, S. and {D'Ago}, G. and {Henning}, Th.},
        title = "{Simultaneous follow-up of planetary transits: revised physical properties for the planetary systems HAT-P-16 and WASP-21}",
      journal = {\aap},
         year = 2013,
        month = sep,
       volume = {557},
          eid = {A30},
        pages = {A30},
          doi = {10.1051/0004-6361/201321669},
archivePrefix = {arXiv},
       eprint = {1307.5874},
 primaryClass = {astro-ph.EP},
       adsurl = {https://ui.adsabs.harvard.edu/abs/2013A&A...557A..30C}
}

@ARTICLE{BoleyHaloPlanets_2021,
       author = {{Boley}, Kiersten M. and {Wang}, Ji and {Zinn}, Joel C. and {Collins}, Karen A. and {Collins}, Kevin I. and {Gan}, Tianjun and {Li}, Ting S.},
        title = "{Searching For Transiting Planets Around Halo Stars. II. Constraining the Occurrence Rate of Hot Jupiters}",
      journal = {\aj},
         year = 2021,
        month = sep,
       volume = {162},
       number = {3},
          eid = {85},
        pages = {85},
          doi = {10.3847/1538-3881/ac0e2d},
archivePrefix = {arXiv},
       eprint = {2106.13242},
 primaryClass = {astro-ph.EP},
       adsurl = {https://ui.adsabs.harvard.edu/abs/2021AJ....162...85B}
}

@ARTICLE{Zink2023_thickdiskoccurence,
       author = {{Zink}, Jon K. and {Hardegree-Ullman}, Kevin K. and {Christiansen}, Jessie L. and {Petigura}, Erik A. and {Boley}, Kiersten M. and {Bhure}, Sakhee and {Rice}, Malena and {Yee}, Samuel W. and {Isaacson}, Howard and {Fernandes}, Rachel B. and {Howard}, Andrew W. and {Blunt}, Sarah and {Lubin}, Jack and {Chontos}, Ashley and {Pidhorodetska}, Daria and {MacDougall}, Mason G.},
        title = "{Scaling K2. VI. Reduced Small-planet Occurrence in High-galactic-amplitude Stars}",
      journal = {\aj},
         year = 2023,
        month = jun,
       volume = {165},
       number = {6},
          eid = {262},
        pages = {262},
          doi = {10.3847/1538-3881/acd24c},
archivePrefix = {arXiv},
       eprint = {2305.13389},
 primaryClass = {astro-ph.EP},
       adsurl = {https://ui.adsabs.harvard.edu/abs/2023AJ....165..262Z}
}

@ARTICLE{Lacedelli2021_thickdiskplanet,
       author = {{Lacedelli}, G. and {Malavolta}, L. and {Borsato}, L. and {Piotto}, G. and {Nardiello}, D. and {Mortier}, A. and {Stalport}, M. and {Collier Cameron}, A. and {Poretti}, E. and {Buchhave}, L.~A. and {L{\'o}pez-Morales}, M. and {Nascimbeni}, V. and {Wilson}, T.~G. and {Udry}, S. and {Latham}, D.~W. and {Bonomo}, A.~S. and {Damasso}, M. and {Dumusque}, X. and {Jenkins}, J.~M. and {Lovis}, C. and {Rice}, K. and {Sasselov}, D. and {Winn}, J.~N. and {Andreuzzi}, G. and {Cosentino}, R. and {Charbonneau}, D. and {Di Fabrizio}, L. and {Martnez Fiorenzano}, A.~F. and {Ghedina}, A. and {Harutyunyan}, A. and {Lienhard}, F. and {Micela}, G. and {Molinari}, E. and {Pagano}, I. and {Pepe}, F. and {Phillips}, D.~F. and {Pinamonti}, M. and {Ricker}, G. and {Scandariato}, G. and {Sozzetti}, A. and {Watson}, C.~A.},
        title = "{An unusually low density ultra-short period super-Earth and three mini-Neptunes around the old star TOI-561}",
      journal = {\mnras},
         year = 2021,
        month = mar,
       volume = {501},
       number = {3},
        pages = {4148-4166},
          doi = {10.1093/mnras/staa3728},
archivePrefix = {arXiv},
       eprint = {2009.02332},
 primaryClass = {astro-ph.EP},
       adsurl = {https://ui.adsabs.harvard.edu/abs/2021MNRAS.501.4148L}
}

@ARTICLE{Mordasini_2012,
       author = {{Mordasini}, C. and {Alibert}, Y. and {Benz}, W. and {Klahr}, H. and {Henning}, T.},
        title = "{Extrasolar planet population synthesis . IV. Correlations with disk metallicity, mass, and lifetime}",
      journal = {\aap},
         year = 2012,
        month = may,
       volume = {541},
          eid = {A97},
        pages = {A97},
          doi = {10.1051/0004-6361/201117350},
archivePrefix = {arXiv},
       eprint = {1201.1036},
 primaryClass = {astro-ph.EP},
       adsurl = {https://ui.adsabs.harvard.edu/abs/2012A&A...541A..97M}
}

@ARTICLE{Ma_Ge,
       author = {{Ma}, Bo and {Ge}, Jian},
        title = "{Statistical properties of brown dwarf companions: implications for different formation mechanisms}",
      journal = {\mnras},
         year = 2014,
        month = apr,
       volume = {439},
       number = {3},
        pages = {2781-2789},
          doi = {10.1093/mnras/stu134},
archivePrefix = {arXiv},
       eprint = {1303.6442},
 primaryClass = {astro-ph.EP},
       adsurl = {https://ui.adsabs.harvard.edu/abs/2014MNRAS.439.2781M}
}

@ARTICLE{B0ss1997_formation,
       author = {{Boss}, A.~P.},
        title = "{Giant planet formation by gravitational instability.}",
      journal = {Science},
         year = 1997,
        month = jan,
       volume = {276},
        pages = {1836-1839},
          doi = {10.1126/science.276.5320.1836},
       adsurl = {https://ui.adsabs.harvard.edu/abs/1997Sci...276.1836B}
}

@ARTICLE{Boss2006_formation,
       author = {{Boss}, Alan P.},
        title = "{Rapid Formation of Gas Giant Planets around M Dwarf Stars}",
      journal = {\apj},
         year = 2006,
        month = may,
       volume = {643},
       number = {1},
        pages = {501-508},
          doi = {10.1086/501522},
archivePrefix = {arXiv},
       eprint = {astro-ph/0601486},
 primaryClass = {astro-ph},
       adsurl = {https://ui.adsabs.harvard.edu/abs/2006ApJ...643..501B}
}

@ARTICLE{Mordasini_Formation,
       author = {{Mordasini}, C. and {Alibert}, Y. and {Benz}, W. and {Klahr}, H. and {Henning}, T.},
        title = "{Extrasolar planet population synthesis . IV. Correlations with disk metallicity, mass, and lifetime}",
      journal = {\aap},
         year = 2012,
        month = may,
       volume = {541},
          eid = {A97},
        pages = {A97},
          doi = {10.1051/0004-6361/201117350},
archivePrefix = {arXiv},
       eprint = {1201.1036},
 primaryClass = {astro-ph.EP},
       adsurl = {https://ui.adsabs.harvard.edu/abs/2012A&A...541A..97M}
}

@ARTICLE{Alibert2004_formation,
       author = {{Alibert}, Y. and {Mordasini}, C. and {Benz}, W.},
        title = "{Migration and giant planet formation}",
      journal = {\aap},
         year = 2004,
        month = apr,
       volume = {417},
        pages = {L25-L28},
          doi = {10.1051/0004-6361:20040053},
archivePrefix = {arXiv},
       eprint = {astro-ph/0403574},
 primaryClass = {astro-ph},
       adsurl = {https://ui.adsabs.harvard.edu/abs/2004A&A...417L..25A}
}

@ARTICLE{RiceArmitage_Formation,
       author = {{Rice}, W.~K.~M. and {Armitage}, Philip J.},
        title = "{On the Formation Timescale and Core Masses of Gas Giant Planets}",
      journal = {\apjl},
         year = 2003,
        month = nov,
       volume = {598},
       number = {1},
        pages = {L55-L58},
          doi = {10.1086/380390},
archivePrefix = {arXiv},
       eprint = {astro-ph/0310191},
 primaryClass = {astro-ph},
       adsurl = {https://ui.adsabs.harvard.edu/abs/2003ApJ...598L..55R}
}

@ARTICLE{Sahlmann2011,
       author = {{Sahlmann}, J. and {S{\'e}gransan}, D. and {Queloz}, D. and {Udry}, S. and {Santos}, N.~C. and {Marmier}, M. and {Mayor}, M. and {Naef}, D. and {Pepe}, F. and {Zucker}, S.},
        title = "{Search for brown-dwarf companions of stars}",
      journal = {\aap},
         year = 2011,
        month = jan,
       volume = {525},
          eid = {A95},
        pages = {A95},
          doi = {10.1051/0004-6361/201015427},
archivePrefix = {arXiv},
       eprint = {1009.5991},
 primaryClass = {astro-ph.EP},
       adsurl = {https://ui.adsabs.harvard.edu/abs/2011A&A...525A..95S}
}

@ARTICLE{Grether_BDDesert,
       author = {{Grether}, Daniel and {Lineweaver}, Charles H.},
        title = "{How Dry is the Brown Dwarf Desert? Quantifying the Relative Number of Planets, Brown Dwarfs, and Stellar Companions around Nearby Sun-like Stars}",
      journal = {\apj},
         year = 2006,
        month = apr,
       volume = {640},
       number = {2},
        pages = {1051-1062},
          doi = {10.1086/500161},
archivePrefix = {arXiv},
       eprint = {astro-ph/0412356},
 primaryClass = {astro-ph},
       adsurl = {https://ui.adsabs.harvard.edu/abs/2006ApJ...640.1051G}
}

@ARTICLE{Mace2018,
       author = {{Mace}, Gregory N. and {Mann}, Andrew W. and {Skiff}, Brian A. and {Sneden}, Christopher and {Kirkpatrick}, J. Davy and {Schneider}, Adam C. and {Kidder}, Benjamin and {Gosnell}, Natalie M. and {Kim}, Hwihyun and {Mulligan}, Brian W. and {Prato}, L. and {Jaffe}, Daniel},
        title = "{Wolf 1130: A Nearby Triple System Containing a Cool, Ultramassive White Dwarf}",
      journal = {\apj},
         year = 2018,
        month = feb,
       volume = {854},
       number = {2},
          eid = {145},
        pages = {145},
          doi = {10.3847/1538-4357/aaa8dd},
archivePrefix = {arXiv},
       eprint = {1802.04803},
 primaryClass = {astro-ph.SR},
       adsurl = {https://ui.adsabs.harvard.edu/abs/2018ApJ...854..145M}
}

@ARTICLE{Zhang2025,
       author = {{Zhang}, J. -Y. and {Lodieu}, N. and {Mart{\'\i}n}, E.~L. and {Zapatero Osorio}, M.~R. and {B{\'e}jar}, V.~J.~S. and {Ivanov}, V.~D. and {Boffin}, H.~M.~J. and {Shahbaz}, T. and {Pavlenko}, Ya. V. and {Rebolo}, R. and {Gauza}, B. and {Sedighi}, N. and {Quezada}, C.},
        title = "{Optical constraints on the coldest metal-poor population}",
      journal = {\aap},
         year = 2025,
        month = jun,
       volume = {698},
          eid = {A141},
        pages = {A141},
          doi = {10.1051/0004-6361/202453246},
archivePrefix = {arXiv},
       eprint = {2412.04393},
 primaryClass = {astro-ph.SR},
       adsurl = {https://ui.adsabs.harvard.edu/abs/2025A&A...698A.141Z}
}

@ARTICLE{Lodieu2022,
       author = {{Lodieu}, N. and {Zapatero Osorio}, M.~R. and {Mart{\'\i}n}, E.~L. and {Rebolo L{\'o}pez}, R. and {Gauza}, B.},
        title = "{Physical properties and trigonometric distance of the peculiar dwarf WISE J181005.5{\ensuremath{-}}101002.3}",
      journal = {\aap},
         year = 2022,
        month = jul,
       volume = {663},
          eid = {A84},
        pages = {A84},
          doi = {10.1051/0004-6361/202243516},
archivePrefix = {arXiv},
       eprint = {2206.13097},
 primaryClass = {astro-ph.SR},
       adsurl = {https://ui.adsabs.harvard.edu/abs/2022A&A...663A..84L}
}

@ARTICLE{Kirkpatrick2021_theaccidenht,
       author = {{Kirkpatrick}, J. Davy and {Marocco}, Federico and {Caselden}, Dan and {Meisner}, Aaron M. and {Faherty}, Jacqueline K. and {Schneider}, Adam C. and {Kuchner}, Marc J. and {Casewell}, S.~L. and {Gelino}, Christopher R. and {Cushing}, Michael C. and {Eisenhardt}, Peter R. and {Wright}, Edward L. and {Schurr}, Steven D.},
        title = "{The Enigmatic Brown Dwarf WISEA J153429.75-104303.3 (a.k.a. ``The Accident'')}",
      journal = {\apjl},
         year = 2021,
        month = jul,
       volume = {915},
       number = {1},
          eid = {L6},
        pages = {L6},
          doi = {10.3847/2041-8213/ac0437},
archivePrefix = {arXiv},
       eprint = {2106.13408},
 primaryClass = {astro-ph.SR},
       adsurl = {https://ui.adsabs.harvard.edu/abs/2021ApJ...915L...6K}
}

@ARTICLE{Meisner2020_BW,
       author = {{Meisner}, Aaron M. and {Faherty}, Jacqueline K. and {Kirkpatrick}, J. Davy and {Schneider}, Adam C. and {Caselden}, Dan and {Gagn{\'e}}, Jonathan and {Kuchner}, Marc J. and {Burgasser}, Adam J. and {Casewell}, Sarah L. and {Debes}, John H. and {Artigau}, {\'E}tienne and {Bardalez Gagliuffi}, Daniella C. and {Logsdon}, Sarah E. and {Kiman}, Rocio and {Allers}, Katelyn and {Hsu}, Chih-chun and {Wisniewski}, John P. and {Allen}, Michaela B. and {Beaulieu}, Paul and {Colin}, Guillaume and {Durantini Luca}, Hugo A. and {Goodman}, Sam and {Gramaize}, L{\'e}opold and {Hamlet}, Leslie K. and {Hinckley}, Ken and {Kiwy}, Frank and {Martin}, David W. and {Pendrill}, William and {Rothermich}, Austin and {Sainio}, Arttu and {Sch{\"u}mann}, J{\"o}rg and {Andersen}, Nikolaj Stevnbak and {Tanner}, Christopher and {Thakur}, Vinod and {Th{\'e}venot}, Melina and {Walla}, Jim and {W{\k{e}}dracki}, Zbigniew and {Aganze}, Christian and {Gerasimov}, Roman and {Theissen}, Christopher and {Backyard Worlds: Planet 9 Collaboration}},
        title = "{Spitzer Follow-up of Extremely Cold Brown Dwarfs Discovered by the Backyard Worlds: Planet 9 Citizen Science Project}",
      journal = {\apj},
         year = 2020,
        month = aug,
       volume = {899},
       number = {2},
          eid = {123},
        pages = {123},
          doi = {10.3847/1538-4357/aba633},
archivePrefix = {arXiv},
       eprint = {2008.06396},
 primaryClass = {astro-ph.SR},
       adsurl = {https://ui.adsabs.harvard.edu/abs/2020ApJ...899..123M}
}

@ARTICLE{Schneider2020_ExtremeSD,
       author = {{Schneider}, Adam C. and {Burgasser}, Adam J. and {Gerasimov}, Roman and {Marocco}, Federico and {Gagn{\'e}}, Jonathan and {Goodman}, Sam and {Beaulieu}, Paul and {Pendrill}, William and {Rothermich}, Austin and {Sainio}, Arttu and {Kuchner}, Marc J. and {Caselden}, Dan and {Meisner}, Aaron M. and {Faherty}, Jacqueline K. and {Mamajek}, Eric E. and {Hsu}, Chih-Chun and {Greco}, Jennifer J. and {Cushing}, Michael C. and {Kirkpatrick}, J. Davy and {Bardalez-Gagliuffi}, Daniella and {Logsdon}, Sarah E. and {Allers}, Katelyn and {Debes}, John H. and {Backyard Worlds: Planet 9 Collaboration}},
        title = "{WISEA J041451.67-585456.7 and WISEA J181006.18-101000.5: The First Extreme T-type Subdwarfs?}",
      journal = {\apj},
         year = 2020,
        month = jul,
       volume = {898},
       number = {1},
          eid = {77},
        pages = {77},
          doi = {10.3847/1538-4357/ab9a40},
archivePrefix = {arXiv},
       eprint = {2007.03836},
 primaryClass = {astro-ph.SR},
       adsurl = {https://ui.adsabs.harvard.edu/abs/2020ApJ...898...77S}
}

@ARTICLE{Hayden2015_agemetals,
       author = {{Hayden}, Michael R. and {Bovy}, Jo and {Holtzman}, Jon A. and {Nidever}, David L. and {Bird}, Jonathan C. and {Weinberg}, David H. and {Andrews}, Brett H. and {Majewski}, Steven R. and {Allende Prieto}, Carlos and {Anders}, Friedrich and {Beers}, Timothy C. and {Bizyaev}, Dmitry and {Chiappini}, Cristina and {Cunha}, Katia and {Frinchaboy}, Peter and {Garc{\'\i}a-Her{\'n}andez}, D.~A. and {Garc{\'\i}a P{\'e}rez}, Ana E. and {Girardi}, L{\'e}o and {Harding}, Paul and {Hearty}, Fred R. and {Johnson}, Jennifer A. and {M{\'e}sz{\'a}ros}, Szabolcs and {Minchev}, Ivan and {O'Connell}, Robert and {Pan}, Kaike and {Robin}, Annie C. and {Schiavon}, Ricardo P. and {Schneider}, Donald P. and {Schultheis}, Mathias and {Shetrone}, Matthew and {Skrutskie}, Michael and {Steinmetz}, Matthias and {Smith}, Verne and {Wilson}, John C. and {Zamora}, Olga and {Zasowski}, Gail},
        title = "{Chemical Cartography with APOGEE: Metallicity Distribution Functions and the Chemical Structure of the Milky Way Disk}",
      journal = {\apj},
         year = 2015,
        month = aug,
       volume = {808},
       number = {2},
          eid = {132},
        pages = {132},
          doi = {10.1088/0004-637X/808/2/132},
archivePrefix = {arXiv},
       eprint = {1503.02110},
 primaryClass = {astro-ph.GA},
       adsurl = {https://ui.adsabs.harvard.edu/abs/2015ApJ...808..132H}
}

@ARTICLE{Imig2023_agemetals,
       author = {{Imig}, Julie and {Price}, Cathryn and {Holtzman}, Jon A. and {Stone-Martinez}, Alexander and {Majewski}, Steven R. and {Weinberg}, David H. and {Johnson}, Jennifer A. and {Allende Prieto}, Carlos and {Beaton}, Rachael L. and {Beers}, Timothy C. and {Bizyaev}, Dmitry and {Blanton}, Michael R. and {Brownstein}, Joel R. and {Cunha}, Katia and {Fern{\'a}ndez-Trincado}, Jos{\'e} G. and {Feuillet}, Diane K. and {Hasselquist}, Sten and {Hayes}, Christian R. and {J{\"o}nsson}, Henrik and {Lane}, Richard R. and {Lian}, Jianhui and {M{\'e}sz{\'a}ros}, Szabolcs and {Nidever}, David L. and {Robin}, Annie C. and {Shetrone}, Matthew and {Smith}, Verne and {Wilson}, John C.},
        title = "{A Tale of Two Disks: Mapping the Milky Way with the Final Data Release of APOGEE}",
      journal = {\apj},
         year = 2023,
        month = sep,
       volume = {954},
       number = {2},
          eid = {124},
        pages = {124},
          doi = {10.3847/1538-4357/ace9b8},
archivePrefix = {arXiv},
       eprint = {2307.13887},
 primaryClass = {astro-ph.GA},
       adsurl = {https://ui.adsabs.harvard.edu/abs/2023ApJ...954..124I}
}

@ARTICLE{Imig2025,
       author = {{Imig}, Julie and {Holtzman}, Jon A. and {Zasowski}, Gail and {Lian}, Jianhui and {Boardman}, Nicholas F. and {Stone-Martinez}, Alexander and {Mackereth}, J. Ted and {Prescott}, Moire K.~M. and {Beaton}, Rachael L. and {Beers}, Timothy C. and {Bizyaev}, Dmitry and {Blanton}, Michael R. and {Cunha}, Katia and {Fern{\'a}ndez-Trincado}, Jos{\'e} G. and {Fielder}, Catherine E. and {Hasselquist}, Sten and {Hayes}, Christian R. and {Haywood}, Misha and {J{\"o}nsson}, Henrik and {Lane}, Richard R. and {Majewski}, Steven R. and {M{\'e}sz{\'a}ros}, Szabolcs and {Minchev}, Ivan and {Nidever}, David L. and {Nitschelm}, Christian and {Sobeck}, Jennifer},
        title = "{A Galactic Self-portrait: Density Structure and Integrated Properties of the Milky Way Disk}",
      journal = {\apj},
         year = 2025,
        month = sep,
       volume = {990},
       number = {2},
          eid = {203},
        pages = {203},
          doi = {10.3847/1538-4357/adf723},
archivePrefix = {arXiv},
       eprint = {2507.17629},
 primaryClass = {astro-ph.GA},
       adsurl = {https://ui.adsabs.harvard.edu/abs/2025ApJ...990..203I}
}

@ARTICLE{Bensby2014,
       author = {{Bensby}, T. and {Feltzing}, S. and {Oey}, M.~S.},
        title = "{Exploring the Milky Way stellar disk. A detailed elemental abundance study of 714 F and G dwarf stars in the solar neighbourhood}",
      journal = {\aap},
         year = 2014,
        month = feb,
       volume = {562},
          eid = {A71},
        pages = {A71},
          doi = {10.1051/0004-6361/201322631},
archivePrefix = {arXiv},
       eprint = {1309.2631},
 primaryClass = {astro-ph.GA},
       adsurl = {https://ui.adsabs.harvard.edu/abs/2014A&A...562A..71B}
}

@ARTICLE{Bensby2003,
       author = {{Bensby}, T. and {Feltzing}, S. and {Lundstr{\"o}m}, I.},
        title = "{Elemental abundance trends in the Galactic thin and thick disks as traced by nearby F and G dwarf stars}",
      journal = {\aap},
         year = 2003,
        month = nov,
       volume = {410},
        pages = {527-551},
          doi = {10.1051/0004-6361:20031213},
       adsurl = {https://ui.adsabs.harvard.edu/abs/2003A&A...410..527B}
}

@INPROCEEDINGS{Chabrier2001,
       author = {{Chabrier}, G.},
        title = "{Baryonic Mass Budget in the Galaxy}",
    booktitle = {Dynamics of Star Clusters and the Milky Way},
         year = 2001,
       editor = {{Deiters}, S. and {Fuchs}, B. and {Just}, A. and {Spurzem}, R. and {Wielen}, R.},
       series = {Astronomical Society of the Pacific Conference Series},
       volume = {228},
        month = jan,
        pages = {249},
       adsurl = {https://ui.adsabs.harvard.edu/abs/2001ASPC..228..249C}
}

@ARTICLE{Burrows2001_BDTheory,
       author = {{Burrows}, Adam and {Hubbard}, W.~B. and {Lunine}, J.~I. and {Liebert}, James},
        title = "{The theory of brown dwarfs and extrasolar giant planets}",
      journal = {Reviews of Modern Physics},
         year = 2001,
        month = jul,
       volume = {73},
       number = {3},
        pages = {719-765},
          doi = {10.1103/RevModPhys.73.719},
archivePrefix = {arXiv},
       eprint = {astro-ph/0103383},
 primaryClass = {astro-ph},
       adsurl = {https://ui.adsabs.harvard.edu/abs/2001RvMP...73..719B}
}

@ARTICLE{Burrows2011,
       author = {{Burrows}, Adam and {Heng}, Kevin and {Nampaisarn}, Thane},
        title = "{The Dependence of Brown Dwarf Radii on Atmospheric Metallicity and Clouds: Theory and Comparison with Observations}",
      journal = {\apj},
         year = 2011,
        month = jul,
       volume = {736},
       number = {1},
          eid = {47},
        pages = {47},
          doi = {10.1088/0004-637X/736/1/47},
archivePrefix = {arXiv},
       eprint = {1102.3922},
 primaryClass = {astro-ph.SR},
       adsurl = {https://ui.adsabs.harvard.edu/abs/2011ApJ...736...47B}
}

@ARTICLE{Phillips2020_ATMO,
       author = {{Phillips}, M.~W. and {Tremblin}, P. and {Baraffe}, I. and {Chabrier}, G. and {Allard}, N.~F. and {Spiegelman}, F. and {Goyal}, J.~M. and {Drummond}, B. and {H{\'e}brard}, E.},
        title = "{A new set of atmosphere and evolution models for cool T-Y brown dwarfs and giant exoplanets}",
      journal = {\aap},
         year = 2020,
        month = may,
       volume = {637},
          eid = {A38},
        pages = {A38},
          doi = {10.1051/0004-6361/201937381},
archivePrefix = {arXiv},
       eprint = {2003.13717},
 primaryClass = {astro-ph.SR},
       adsurl = {https://ui.adsabs.harvard.edu/abs/2020A&A...637A..38P}
}

@ARTICLE{Baraffe2003_MassRad,
       author = {{Baraffe}, I. and {Chabrier}, G. and {Barman}, T.~S. and {Allard}, F. and {Hauschildt}, P.~H.},
        title = "{Evolutionary models for cool brown dwarfs and extrasolar giant planets. The case of HD 209458}",
      journal = {\aap},
         year = 2003,
        month = may,
       volume = {402},
        pages = {701-712},
          doi = {10.1051/0004-6361:20030252},
archivePrefix = {arXiv},
       eprint = {astro-ph/0302293},
 primaryClass = {astro-ph},
       adsurl = {https://ui.adsabs.harvard.edu/abs/2003A&A...402..701B}
}

@ARTICLE{Larsen_prevrec,
       author = {{Larsen}, Alexander and {Swaby}, Tera N. and {Kobulnicky}, Henry A. and {Ca{\~n}as}, Caleb I. and {Kanodia}, Shubham and {Libby-Roberts}, Jessica and {Monson}, Andrew and {Gupta}, Arvind F. and {Cochran}, William and {Mahadevan}, Suvrath and {Bender}, Chad and {Diddams}, Scott A. and {Halverson}, Samuel and {Lin}, Andrea S.~J. and {Moe}, Maxwell and {Ninan}, Joe and {Robertson}, Paul and {Roy}, Arpita and {Schwab}, Christian and {Stefansson}, Gudmundur},
        title = "{Searching for GEMS: Discovery and Characterization of Two Brown Dwarfs Around M Dwarfs}",
      journal = {\aj},
         year = 2025,
        month = may,
       volume = {169},
       number = {5},
          eid = {246},
        pages = {246},
          doi = {10.3847/1538-3881/adbb54},
archivePrefix = {arXiv},
       eprint = {2501.16554},
 primaryClass = {astro-ph.EP},
       adsurl = {https://ui.adsabs.harvard.edu/abs/2025AJ....169..246L}
}

@ARTICLE{Schlaufman,
       author = {{Schlaufman}, Kevin C.},
        title = "{Evidence of an Upper Bound on the Masses of Planets and Its Implications for Giant Planet Formation}",
      journal = {\apj},
         year = 2018,
        month = jan,
       volume = {853},
       number = {1},
          eid = {37},
        pages = {37},
          doi = {10.3847/1538-4357/aa961c},
archivePrefix = {arXiv},
       eprint = {1801.06185},
 primaryClass = {astro-ph.EP},
       adsurl = {https://ui.adsabs.harvard.edu/abs/2018ApJ...853...37S}
}

@ARTICLE{Carmichael2023,
       author = {{Carmichael}, Theron W.},
        title = "{Improved radius determinations for the transiting brown dwarf population in the era of Gaia and TESS}",
      journal = {\mnras},
         year = 2023,
        month = mar,
       volume = {519},
       number = {4},
        pages = {5177-5190},
          doi = {10.1093/mnras/stac3720},
archivePrefix = {arXiv},
       eprint = {2212.02502},
 primaryClass = {astro-ph.SR},
       adsurl = {https://ui.adsabs.harvard.edu/abs/2023MNRAS.519.5177C}
}

@ARTICLE{Vowell2025_BD,
       author = {{Vowell}, Noah and {Rodriguez}, Joseph E. and {Latham}, David W. and {Quinn}, Samuel N. and {Schulte}, Jack and {Eastman}, Jason D. and {Bieryla}, Allyson and {Barkaoui}, Khalid and {Ciardi}, David R. and {Collins}, Karen A. and {Girardin}, Eric and {H{\'e}brard}, Guillaume and {Heldridge}, Elisabeth and {Jafariyazani}, Marziye and {Kotten}, Brooke and {Mancini}, Luigi and {Murgas}, Felipe and {Narita}, Norio and {Radford}, D.~J. and {Relles}, Howard M. and {Shporer}, Avi and {Soares-Furtado}, Melinda and {Strakhov}, Ivan A. and {Ziegler}, Carl and {Boisse}, Isabelle and {Brice{\~n}o}, C{\'e}sar and {Calkins}, Michael L. and {Clark}, Catherine A. and {Collins}, Kevin I. and {de Leon}, Jerome and {Esquerdo}, Gilbert A. and {Fajardo-Acosta}, Sergio B. and {Forveille}, Thierry and {Fukui}, Akihiko and {Watkins}, Cristilyn N. and {He}, Ruixuan and {Heidari}, Neda and {Horne}, Keith and {Jenkins}, Jon M. and {Mann}, Andrew W. and {Naponiello}, Luca and {Palle}, Enric and {Schwarz}, Richard P. and {Seager}, S. and {Southworth}, John and {Srdoc}, Gregor and {Swift}, Jonathan J. and {Winn}, Joshua N.},
        title = "{Eleven New Transiting Brown Dwarfs and Very-low-mass Stars from TESS}",
      journal = {\aj},
         year = 2025,
        month = aug,
       volume = {170},
       number = {2},
          eid = {68},
        pages = {68},
          doi = {10.3847/1538-3881/addd17},
archivePrefix = {arXiv},
       eprint = {2501.09795},
 primaryClass = {astro-ph.EP},
       adsurl = {https://ui.adsabs.harvard.edu/abs/2025AJ....170...68V}
}

@ARTICLE{Marley2021_Sonora,
       author = {{Marley}, Mark S. and {Saumon}, Didier and {Visscher}, Channon and {Lupu}, Roxana and {Freedman}, Richard and {Morley}, Caroline and {Fortney}, Jonathan J. and {Seay}, Christopher and {Smith}, Adam J.~R.~W. and {Teal}, D.~J. and {Wang}, Ruoyan},
        title = "{The Sonora Brown Dwarf Atmosphere and Evolution Models. I. Model Description and Application to Cloudless Atmospheres in Rainout Chemical Equilibrium}",
      journal = {\apj},
         year = 2021,
        month = oct,
       volume = {920},
       number = {2},
          eid = {85},
        pages = {85},
          doi = {10.3847/1538-4357/ac141d},
archivePrefix = {arXiv},
       eprint = {2107.07434},
 primaryClass = {astro-ph.SR},
       adsurl = {https://ui.adsabs.harvard.edu/abs/2021ApJ...920...85M}
}

@ARTICLE{Pass2025_UberMS,
       author = {{Pass}, Emily K. and {Cargile}, Phillip A. and {DiTomasso}, Victoria and {Rodr{\'\i}guez Mart{\'\i}nez}, Romy and {Charbonneau}, David and {Latham}, David W. and {Vanderburg}, Andrew and {Bieryla}, Allyson and {Quinn}, Samuel N. and {Buchhave}, Lars A.},
        title = "{Metallicities from High-Resolution TRES Spectra with The Payne and uberMS: Performance Benchmarks and Literature Comparison}",
      journal = {arXiv e-prints},
         year = 2025,
        month = jun,
          eid = {arXiv:2506.18961},
        pages = {arXiv:2506.18961},
          doi = {10.48550/arXiv.2506.18961},
archivePrefix = {arXiv},
       eprint = {2506.18961},
 primaryClass = {astro-ph.SR},
       adsurl = {https://ui.adsabs.harvard.edu/abs/2025arXiv250618961P}
}

@ARTICLE{Andrae2023,
       author = {{Andrae}, Ren{\'e} and {Rix}, Hans-Walter and {Chandra}, Vedant},
        title = "{Robust Data-driven Metallicities for 175 Million Stars from Gaia XP Spectra}",
      journal = {ApJS},
         year = 2023,
        month = jul,
       volume = {267},
       number = {1},
          eid = {8},
        pages = {8},
          doi = {10.3847/1538-4365/acd53e},
archivePrefix = {arXiv},
       eprint = {2302.02611},
 primaryClass = {astro-ph.SR},
       adsurl = {https://ui.adsabs.harvard.edu/abs/2023ApJS..267....8A}
}

@ARTICLE{Majewski2017,
       author = {{Majewski}, Steven R. and {Schiavon}, Ricardo P. and {Frinchaboy}, Peter M. and {Allende Prieto}, Carlos and {Barkhouser}, Robert and {Bizyaev}, Dmitry and {Blank}, Basil and {Brunner}, Sophia and {Burton}, Adam and {Carrera}, Ricardo and {Chojnowski}, S. Drew and {Cunha}, K{\'a}tia and {Epstein}, Courtney and {Fitzgerald}, Greg and {Garc{\'\i}a P{\'e}rez}, Ana E. and {Hearty}, Fred R. and {Henderson}, Chuck and {Holtzman}, Jon A. and {Johnson}, Jennifer A. and {Lam}, Charles R. and {Lawler}, James E. and {Maseman}, Paul and {M{\'e}sz{\'a}ros}, Szabolcs and {Nelson}, Matthew and {Nguyen}, Duy Coung and {Nidever}, David L. and {Pinsonneault}, Marc and {Shetrone}, Matthew and {Smee}, Stephen and {Smith}, Verne V. and {Stolberg}, Todd and {Skrutskie}, Michael F. and {Walker}, Eric and {Wilson}, John C. and {Zasowski}, Gail and {Anders}, Friedrich and {Basu}, Sarbani and {Beland}, Stephane and {Blanton}, Michael R. and {Bovy}, Jo and {Brownstein}, Joel R. and {Carlberg}, Joleen and {Chaplin}, William and {Chiappini}, Cristina and {Eisenstein}, Daniel J. and {Elsworth}, Yvonne and {Feuillet}, Diane and {Fleming}, Scott W. and {Galbraith-Frew}, Jessica and {Garc{\'\i}a}, Rafael A. and {Garc{\'\i}a-Hern{\'a}ndez}, D. An{\'\i}bal and {Gillespie}, Bruce A. and {Girardi}, L{\'e}o and {Gunn}, James E. and {Hasselquist}, Sten and {Hayden}, Michael R. and {Hekker}, Saskia and {Ivans}, Inese and {Kinemuchi}, Karen and {Klaene}, Mark and {Mahadevan}, Suvrath and {Mathur}, Savita and {Mosser}, Beno{\^\i}t and {Muna}, Demitri and {Munn}, Jeffrey A. and {Nichol}, Robert C. and {O'Connell}, Robert W. and {Parejko}, John K. and {Robin}, A.~C. and {Rocha-Pinto}, Helio and {Schultheis}, Matthias and {Serenelli}, Aldo M. and {Shane}, Neville and {Silva Aguirre}, Victor and {Sobeck}, Jennifer S. and {Thompson}, Benjamin and {Troup}, Nicholas W. and {Weinberg}, David H. and {Zamora}, Olga},
        title = "{The Apache Point Observatory Galactic Evolution Experiment (APOGEE)}",
      journal = {AJ},
         year = 2017,
        month = sep,
       volume = {154},
       number = {3},
          eid = {94},
        pages = {94},
          doi = {10.3847/1538-3881/aa784d},
archivePrefix = {arXiv},
       eprint = {1509.05420},
 primaryClass = {astro-ph.IM},
       adsurl = {https://ui.adsabs.harvard.edu/abs/2017AJ....154...94M}
}

@ARTICLE{Tayar2022,
       author = {{Tayar}, Jamie and {Claytor}, Zachary R. and {Huber}, Daniel and {van Saders}, Jennifer},
        title = "{A Guide to Realistic Uncertainties on the Fundamental Properties of Solar-type Exoplanet Host Stars}",
      journal = {ApJ},
         year = 2022,
        month = mar,
       volume = {927},
       number = {1},
          eid = {31},
        pages = {31},
          doi = {10.3847/1538-4357/ac4bbc},
archivePrefix = {arXiv},
       eprint = {2012.07957},
 primaryClass = {astro-ph.EP},
       adsurl = {https://ui.adsabs.harvard.edu/abs/2022ApJ...927...31T}
}

@ARTICLE{Park2024,
       author = {{Park}, Minjung and {Conroy}, Charlie and {Johnson}, Benjamin D. and {Leja}, Joel and {Dotter}, Aaron and {Cargile}, Phillip A.},
        title = "{$\alpha$-MC: Self-consistent $\alpha$-enhanced stellar population models covering a wide range of age, metallicity, and wavelength}",
      journal = {arXiv},
         year = 2024,
        month = oct,
          eid = {arXiv:2410.21375},
        pages = {arXiv:2410.21375},
          doi = {10.48550/arXiv.2410.21375},
archivePrefix = {arXiv},
       eprint = {2410.21375},
 primaryClass = {astro-ph.GA},
       adsurl = {https://ui.adsabs.harvard.edu/abs/2024arXiv241021375P}
}

@article{Bailer-Jones2020,
    title = {{Estimating Distances from Parallaxes. V. Geometric and Photogeometric Distances to 1.47 Billion Stars in Gaia Early Data Release 3}},
    year = {2021},
    journal = {The Astronomical Journal},
    author = {Bailer-Jones, C. A. L. and Rybizki, J. and Fouesneau, M. and Demleitner, M. and Andrae, R.},
    number = {3},
    pages = {147},
    volume = {161},
    url = {http://arxiv.org/abs/2012.05220},
    doi = {10.3847/1538-3881/abd806},
    issn = {0004-6256},
    arxivId = {2012.05220}
}

@article{Astraatmadja2016,
    title = {{ Estimating Distances From Parallaxes. Ii. Performance of Bayesian Distance Estimators on a Gaia -Like Catalogue }},
    year = {2016},
    journal = {The Astrophysical Journal},
    author = {Astraatmadja, Tri L. and Bailer-Jones, Coryn A. L.},
    number = {2},
    pages = {137},
    volume = {832},
    publisher = {IOP Publishing},
    url = {http://dx.doi.org/10.3847/0004-637X/832/2/137},
    doi = {10.3847/0004-637x/832/2/137},
    issn = {1538-4357},
    arxivId = {1609.03424}
}

@ARTICLE{Hayden2015,
       author = {{Hayden}, Michael R. and {Bovy}, Jo and {Holtzman}, Jon A. and {Nidever}, David L. and {Bird}, Jonathan C. and {Weinberg}, David H. and {Andrews}, Brett H. and {Majewski}, Steven R. and {Allende Prieto}, Carlos and {Anders}, Friedrich and {Beers}, Timothy C. and {Bizyaev}, Dmitry and {Chiappini}, Cristina and {Cunha}, Katia and {Frinchaboy}, Peter and {Garc{\'\i}a-Her{\'n}andez}, D.~A. and {Garc{\'\i}a P{\'e}rez}, Ana E. and {Girardi}, L{\'e}o and {Harding}, Paul and {Hearty}, Fred R. and {Johnson}, Jennifer A. and {M{\'e}sz{\'a}ros}, Szabolcs and {Minchev}, Ivan and {O'Connell}, Robert and {Pan}, Kaike and {Robin}, Annie C. and {Schiavon}, Ricardo P. and {Schneider}, Donald P. and {Schultheis}, Mathias and {Shetrone}, Matthew and {Skrutskie}, Michael and {Steinmetz}, Matthias and {Smith}, Verne and {Wilson}, John C. and {Zamora}, Olga and {Zasowski}, Gail},
        title = "{Chemical Cartography with APOGEE: Metallicity Distribution Functions and the Chemical Structure of the Milky Way Disk}",
      journal = {ApJ},
         year = 2015,
        month = aug,
       volume = {808},
       number = {2},
          eid = {132},
        pages = {132},
          doi = {10.1088/0004-637X/808/2/132},
archivePrefix = {arXiv},
       eprint = {1503.02110},
 primaryClass = {astro-ph.GA},
       adsurl = {https://ui.adsabs.harvard.edu/abs/2015ApJ...808..132H}
}

@ARTICLE{AstropyCollaboration2013,
       author = {{Astropy Collaboration} and {Robitaille}, Thomas P. and {Tollerud}, Erik J. and {Greenfield}, Perry and {Droettboom}, Michael and {Bray}, Erik and {Aldcroft}, Tom and {Davis}, Matt and {Ginsburg}, Adam and {Price-Whelan}, Adrian M. and {Kerzendorf}, Wolfgang E. and {Conley}, Alexander and {Crighton}, Neil and {Barbary}, Kyle and {Muna}, Demitri and {Ferguson}, Henry and {Grollier}, Fr{\'e}d{\'e}ric and {Parikh}, Madhura M. and {Nair}, Prasanth H. and {Unther}, Hans M. and {Deil}, Christoph and {Woillez}, Julien and {Conseil}, Simon and {Kramer}, Roban and {Turner}, James E.~H. and {Singer}, Leo and {Fox}, Ryan and {Weaver}, Benjamin A. and {Zabalza}, Victor and {Edwards}, Zachary I. and {Azalee Bostroem}, K. and {Burke}, D.~J. and {Casey}, Andrew R. and {Crawford}, Steven M. and {Dencheva}, Nadia and {Ely}, Justin and {Jenness}, Tim and {Labrie}, Kathleen and {Lim}, Pey Lian and {Pierfederici}, Francesco and {Pontzen}, Andrew and {Ptak}, Andy and {Refsdal}, Brian and {Servillat}, Mathieu and {Streicher}, Ole},
        title = "{Astropy: A community Python package for astronomy}",
      journal = {A\&A},
         year = 2013,
        month = oct,
       volume = {558},
          eid = {A33},
        pages = {A33},
          doi = {10.1051/0004-6361/201322068},
archivePrefix = {arXiv},
       eprint = {1307.6212},
 primaryClass = {astro-ph.IM},
       adsurl = {https://ui.adsabs.harvard.edu/abs/2013A&A...558A..33A}
}

@ARTICLE{AstropyCollaboration2018,
       author = {{Astropy Collaboration} and {Price-Whelan}, A.~M. and {Sip{\H{o}}cz}, B.~M. and {G{\"u}nther}, H.~M. and {Lim}, P.~L. and {Crawford}, S.~M. and {Conseil}, S. and {Shupe}, D.~L. and {Craig}, M.~W. and {Dencheva}, N. and {Ginsburg}, A. and {VanderPlas}, J.~T. and {Bradley}, L.~D. and {P{\'e}rez-Su{\'a}rez}, D. and {de Val-Borro}, M. and {Aldcroft}, T.~L. and {Cruz}, K.~L. and {Robitaille}, T.~P. and {Tollerud}, E.~J. and {Ardelean}, C. and {Babej}, T. and {Bach}, Y.~P. and {Bachetti}, M. and {Bakanov}, A.~V. and {Bamford}, S.~P. and {Barentsen}, G. and {Barmby}, P. and {Baumbach}, A. and {Berry}, K.~L. and {Biscani}, F. and {Boquien}, M. and {Bostroem}, K.~A. and {Bouma}, L.~G. and {Brammer}, G.~B. and {Bray}, E.~M. and {Breytenbach}, H. and {Buddelmeijer}, H. and {Burke}, D.~J. and {Calderone}, G. and {Cano Rodr{\'\i}guez}, J.~L. and {Cara}, M. and {Cardoso}, J.~V.~M. and {Cheedella}, S. and {Copin}, Y. and {Corrales}, L. and {Crichton}, D. and {D'Avella}, D. and {Deil}, C. and {Depagne}, {\'E}. and {Dietrich}, J.~P. and {Donath}, A. and {Droettboom}, M. and {Earl}, N. and {Erben}, T. and {Fabbro}, S. and {Ferreira}, L.~A. and {Finethy}, T. and {Fox}, R.~T. and {Garrison}, L.~H. and {Gibbons}, S.~L.~J. and {Goldstein}, D.~A. and {Gommers}, R. and {Greco}, J.~P. and {Greenfield}, P. and {Groener}, A.~M. and {Grollier}, F. and {Hagen}, A. and {Hirst}, P. and {Homeier}, D. and {Horton}, A.~J. and {Hosseinzadeh}, G. and {Hu}, L. and {Hunkeler}, J.~S. and {Ivezi{\'c}}, {\v{Z}}. and {Jain}, A. and {Jenness}, T. and {Kanarek}, G. and {Kendrew}, S. and {Kern}, N.~S. and {Kerzendorf}, W.~E. and {Khvalko}, A. and {King}, J. and {Kirkby}, D. and {Kulkarni}, A.~M. and {Kumar}, A. and {Lee}, A. and {Lenz}, D. and {Littlefair}, S.~P. and {Ma}, Z. and {Macleod}, D.~M. and {Mastropietro}, M. and {McCully}, C. and {Montagnac}, S. and {Morris}, B.~M. and {Mueller}, M. and {Mumford}, S.~J. and {Muna}, D. and {Murphy}, N.~A. and {Nelson}, S. and {Nguyen}, G.~H. and {Ninan}, J.~P. and {N{\"o}the}, M. and {Ogaz}, S. and {Oh}, S. and {Parejko}, J.~K. and {Parley}, N. and {Pascual}, S. and {Patil}, R. and {Patil}, A.~A. and {Plunkett}, A.~L. and {Prochaska}, J.~X. and {Rastogi}, T. and {Reddy Janga}, V. and {Sabater}, J. and {Sakurikar}, P. and {Seifert}, M. and {Sherbert}, L.~E. and {Sherwood-Taylor}, H. and {Shih}, A.~Y. and {Sick}, J. and {Silbiger}, M.~T. and {Singanamalla}, S. and {Singer}, L.~P. and {Sladen}, P.~H. and {Sooley}, K.~A. and {Sornarajah}, S. and {Streicher}, O. and {Teuben}, P. and {Thomas}, S.~W. and {Tremblay}, G.~R. and {Turner}, J.~E.~H. and {Terr{\'o}n}, V. and {van Kerkwijk}, M.~H. and {de la Vega}, A. and {Watkins}, L.~L. and {Weaver}, B.~A. and {Whitmore}, J.~B. and {Woillez}, J. and {Zabalza}, V. and {Astropy Contributors}},
        title = "{The Astropy Project: Building an Open-science Project and Status of the v2.0 Core Package}",
      journal = {AJ},
         year = 2018,
        month = sep,
       volume = {156},
       number = {3},
          eid = {123},
        pages = {123},
          doi = {10.3847/1538-3881/aabc4f},
archivePrefix = {arXiv},
       eprint = {1801.02634},
 primaryClass = {astro-ph.IM},
       adsurl = {https://ui.adsabs.harvard.edu/abs/2018AJ....156..123A}
}

@INPROCEEDINGS{Kurucz1992,
       author = {{Kurucz}, R.~L.},
        title = "{Model Atmospheres for Population Synthesis}",
    booktitle = {The Stellar Populations of Galaxies},
         year = 1992,
       editor = {{Barbuy}, Beatriz and {Renzini}, Alvio},
       series = {IAU Symposium},
       volume = {149},
        month = jan,
        pages = {225},
       adsurl = {https://ui.adsabs.harvard.edu/abs/1992IAUS..149..225K}
}

@ARTICLE{Woody2025,
       author = {{Woody}, Rebecca and {Conroy}, Charlie and {Cargile}, Phillip and {Bonaca}, Ana and {Chandra}, Vedant and {Han}, Jiwon Jesse and {Johnson}, Benjamin D. and {Naidu}, Rohan P. and {Ting}, Yuan-Sen},
        title = "{The Rapid Formation of the Metal-poor Milky Way}",
      journal = {ApJ},
         year = 2025,
        month = jan,
       volume = {978},
       number = {2},
          eid = {152},
        pages = {152},
          doi = {10.3847/1538-4357/ad968e},
archivePrefix = {arXiv},
       eprint = {2409.04529},
 primaryClass = {astro-ph.GA},
       adsurl = {https://ui.adsabs.harvard.edu/abs/2025ApJ...978..152W}
}

@article{gala,
  doi = {10.21105/joss.00388},
  url = {https://doi.org/10.21105%2Fjoss.00388},
  year = 2017,
  month = {oct},
  publisher = {The Open Journal},
  volume = {2},
  number = {18},
  author = {Adrian M. Price-Whelan},
  title = {Gala: A Python package for galactic dynamics},
  journal = {The Journal of Open Source Software}}

@misc{adrian_price_whelan_2020_4159870,
  author       = {Adrian Price-Whelan and
                  Brigitta Sipőcz and
                  Daniel Lenz and
                  Johnny Greco and
                  Nathaniel Starkman and
                  Dan Foreman-Mackey and
                  P. L. Lim and
                  Semyeong Oh and
                  Sergey Koposov and
                  Syrtis Major},
  title        = {adrn/gala: v1.3},
  month        = oct,
  year         = 2020,
  publisher    = {Zenodo},
  version      = {v1.3},
  doi          = {10.5281/zenodo.4159870},
  url          = {https://doi.org/10.5281/zenodo.4159870}
}

@ARTICLE{Reid2004,
       author = {{Reid}, M.~J. and {Brunthaler}, A.},
        title = "{The Proper Motion of Sagittarius A*. II. The Mass of Sagittarius A*}",
      journal = {ApJ},
         year = 2004,
        month = dec,
       volume = {616},
       number = {2},
        pages = {872-884},
          doi = {10.1086/424960},
archivePrefix = {arXiv},
       eprint = {astro-ph/0408107},
 primaryClass = {astro-ph},
       adsurl = {https://ui.adsabs.harvard.edu/abs/2004ApJ...616..872R}
}

@ARTICLE{Drimmel2018,
       author = {{Drimmel}, Ronald and {Poggio}, Eloisa},
        title = "{On the Solar Velocity}",
      journal = {RNAAS},
         year = 2018,
        month = nov,
       volume = {2},
       number = {4},
          eid = {210},
        pages = {210},
          doi = {10.3847/2515-5172/aaef8b},
       adsurl = {https://ui.adsabs.harvard.edu/abs/2018RNAAS...2..210D}
}

@ARTICLE{Eilers2019,
       author = {{Eilers}, Anna-Christina and {Hogg}, David W. and {Rix}, Hans-Walter and {Ness}, Melissa K.},
        title = "{The Circular Velocity Curve of the Milky Way from 5 to 25 kpc}",
      journal = {ApJ},
         year = 2019,
        month = jan,
       volume = {871},
       number = {1},
          eid = {120},
        pages = {120},
          doi = {10.3847/1538-4357/aaf648},
archivePrefix = {arXiv},
       eprint = {1810.09466},
 primaryClass = {astro-ph.GA},
       adsurl = {https://ui.adsabs.harvard.edu/abs/2019ApJ...871..120E}
}

@BOOK{Kurucz1993,
       author = {{Kurucz}, Robert L.},
        title = "{SYNTHE spectrum synthesis programs and line data}",
         year = 1993,
       adsurl = {https://ui.adsabs.harvard.edu/abs/1993sssp.book.....K}
}
\bibliographystyle{aasjournalv7}



\appendix

\section{High-Contrast Imaging}

To search for stellar companions that could dilute the transit signal or contaminate the radial velocity measurements, we obtained high angular resolution imaging of TOI-7019. These observations, described in Section \ref{imaging}, rule out the presence of bright stellar companions within several arcseconds of the target star, confirming that the observed transit and RV signals originate from a single brown dwarf companion to TOI-7019.

\begin{figure}[!h]
\plotone{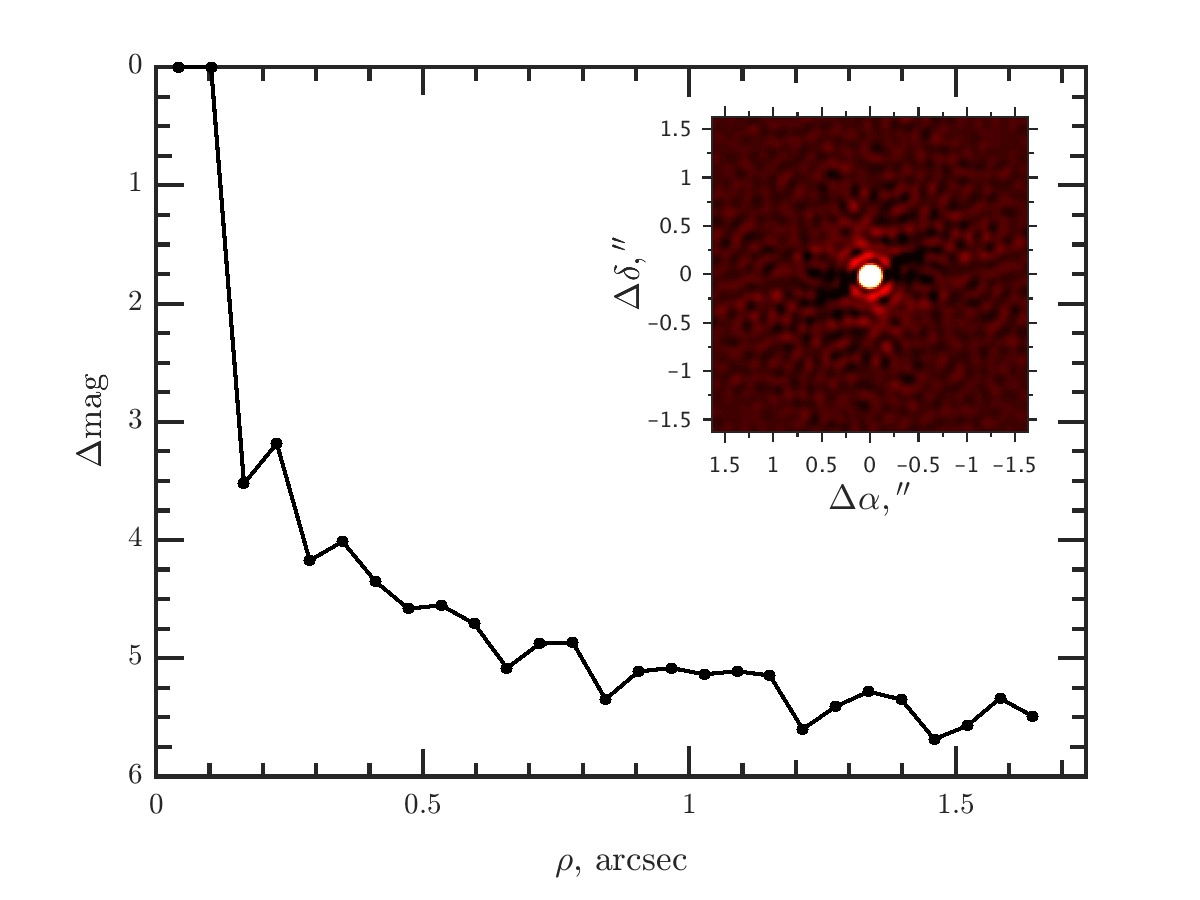}
\plotone{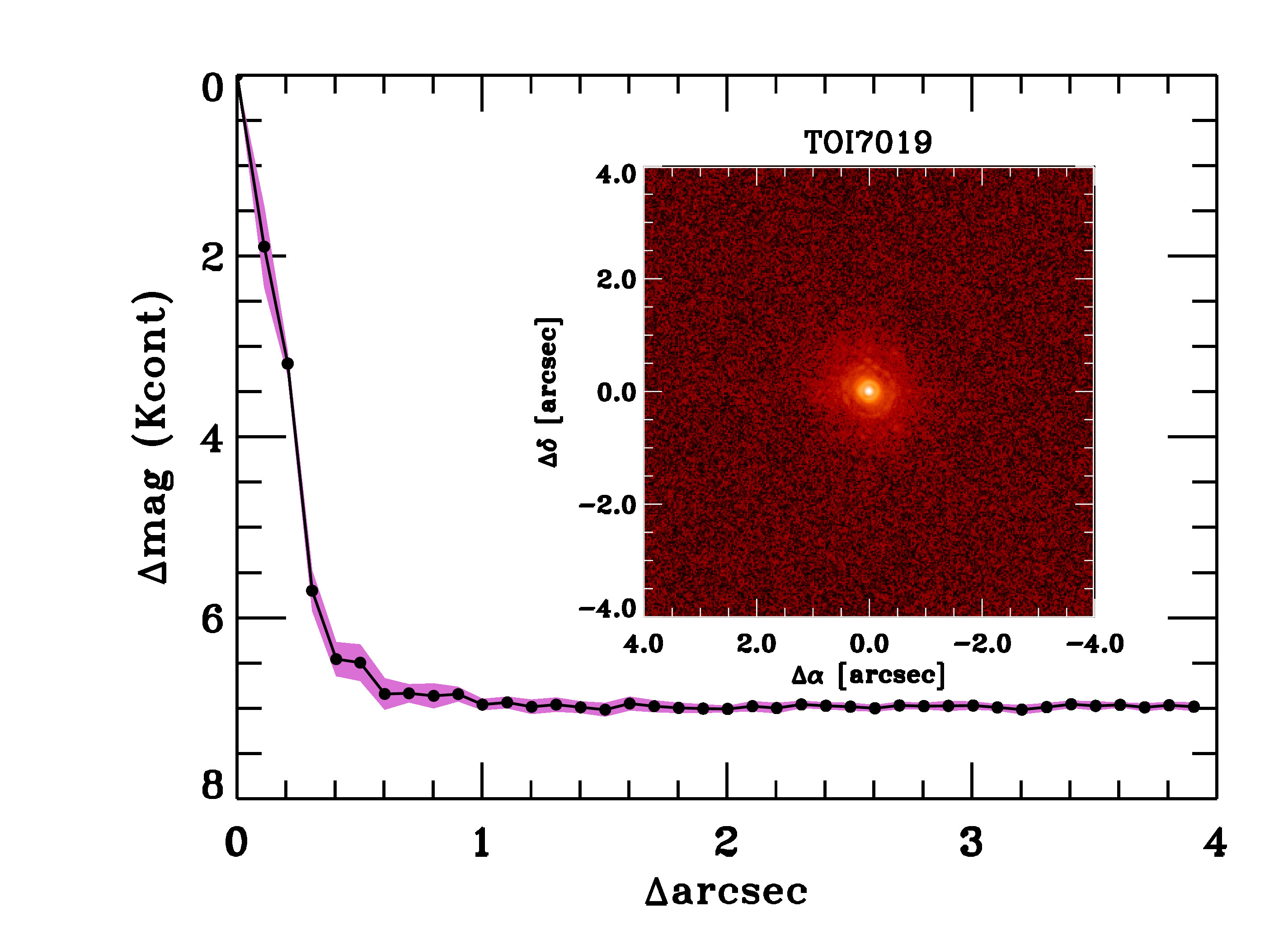}
\caption{\textbf{Top:} Speckle imaging observations of TOI-7019 from SAI. The solid line shows the contrast limits from the observations, while the inset figure shows the reconstructed image.
\textbf{Bottom:} Palomar/PHARO adaptive optics imaging observations of TOI-7019. The solid line and shaded region shows the contrast limit and 5$\sigma$ confidence intervals for excluding faint companions. \label{fig:ao_imaging}}
\end{figure}

\section{Observation Log}

This section provides the complete observational data used in our analysis. Table \ref{tab:tres_rvs} lists all TRES radial velocity measurements obtained for TOI-7019, while Table \ref{tab:tess} summarizes the TESS photometric observations across multiple sectors, indicating which sectors contained detected transits.

\begin{deluxetable}{ccc}[!h]
\tablehead{
\colhead{\bjdtdb} & \colhead{RV (\ms)} & \colhead{$\sigma(\mathrm{RV})$ (\ms)}
}
\tablecaption{TRES Radial Velocity Measurements \label{tab:tres_rvs}}
\startdata
2460780.954943 & 7055.5 & 46.7 \\
2460792.882175 & 8340.7 & 37.9 \\
2460798.880485 & 7839.3 & 38.8 \\
2460803.816594 & 6888.6 & 38.7 \\
2460810.805074 & 4077.1 & 72.8 \\
2460814.930774 & 773.0 & 52.5 \\
2460818.853537 & -81.4 & 43.2 \\
2460822.789941 & 3555.4 & 37.2 \\
2460832.786568 & 7858.1 & 34.4 \\
2460839.751297 & 8240.2 & 49.6 \\
2460846.832255 & 7950.3 & 37.2 \\
2460854.809310 & 6085.8 & 54.3 \\
2460862.886387 & 985.9 & 36.1 \\
2460871.740670 & 4161.1 & 66.2 \\
2460879.693059 & 7644.6 & 51.0%
\enddata
\tablecomments{The systemic velocity offset in this reference frame, fitted from the joint analysis, is 
$\mu_{\mathrm{TRES}} = 5656 \pm 17$~\ms. The absolute barycentric systemic velocity, independently calibrated using minor planet observations, is 
$\gamma_{\mathrm{bary}} = -144.366 \pm 0.023$~\kms (statistical).}
\end{deluxetable}

\begin{deluxetable}{cccc}
\tablehead{
\colhead{Sector} & \colhead{Cadence} & \colhead{Pipeline} & \colhead{Transit Detected}
}
\tablecaption{TESS Observations of TOI-7019 \label{tab:tess}}
\startdata
14, 18, 25, 26      & 1800s &  QLP & 14, 26 \\
40, 41, 48, 52--55  &  600s &  QLP & 40, 53, 55 \\
58, 74, 75, 79--82  &  200s &  QLP & 58, 74, 79, 81 \\
85                  &  120s & SPOC & 85 \\
\enddata
\end{deluxetable}

\section{\texttt{juliet} Priors}
Table \ref{tab:priors} lists the prior distributions used in our joint transit and RV analysis with juliet (Section 3.4

\begin{deluxetable}{lcc}
\tablehead{
\colhead{Parameter} & \colhead{Prior} & \colhead{Values}
}
\tablecaption{Priors used in the \texttt{juliet} fit. \label{tab:priors}}
\startdata
\multicolumn{3}{c}{\textit{Transit Parameters (TESS)}} \\
\hline
Orbital Period, $P$ (days) & Uniform & [47, 49] \\
Time of Transit Center, $t_0$ (\bjdtdb)\tablenotemark{a} & Uniform & [2941, 2944] \\
% Parameterization constants\tablenotemark{b}, $r_1, r_2$ & Uniform & [0, 1] \\
Planet-to-star radius ratio\tablenotemark{b}, $p = R_p/R_*$ & Uniform & [0.01, 0.2] \\
Impact parameter\tablenotemark{b},  & Uniform & [0, 1] \\
Limb-darkening coefficients\tablenotemark{c}, $q_{1,\mathrm{TESS}}, q_{2,\mathrm{TESS}}$ & Uniform & [0, 1] \\
% Eccentricity, $e$ & Uniform & [0, 0.6] \\
$\sqrt{e}\cos\omega$\tablenotemark{d} & Uniform & [-0.8, 0.8]\\
$\sqrt{e}\sin\omega$\tablenotemark{d} & Uniform & [-0.8, 0.8]\\
Argument of Periastron, $\omega$ (deg) & Uniform & [0, 360] \\
Scaled Semi-major Axis, $a/R_*$ & Uniform & [1, 100] \\
% Dilution factor, $D_\mathrm{TESS}$ & Normal & 0.0411\\
Dilution factor\tablenotemark{e}, $D_\mathrm{TESS}$ & Normal & $\mathcal{N}(1.0, 0.00429)$\\
Out-of-transit flux, $F_{0, \mathrm{TESS}}$ & Normal & $\mathcal{N}(0.0, 0.1)$ \\
Photometric jitter, $\sigma_{w, \mathrm{TESS}}$ (ppm) & Log-uniform & [0.1, 10000] \\
\hline
\multicolumn{3}{c}{\textit{Ground-Based Photometry Parameters}} \\
\hline
Limb-darkening coefficients (R, B, i bands), $q_1, q_2$ & Uniform & [0, 1] \\
Dilution factors (R, B, i bands) & Fixed & 1.0 \\
Out-of-transit flux (R, B, i bands), $F_0$ & Normal & $\mathcal{N}(0.0, 0.1)$ \\
Photometric jitter (R, B, i bands), $\sigma_w$ (ppm) & Log-uniform & [0.1, 10000] \\
Detrending coefficients (R, B, i bands), $\theta_0, \theta_1$ & Uniform & [-0.01, 0.01] \\
\hline
\multicolumn{3}{c}{\textit{Radial Velocity Parameters (TRES)}} \\
\hline
RV semi-amplitude, $K$ (\ms) & Uniform & [4000, 5000] \\
Center-of-mass velocity, $\mu_{\mathrm{TRES}}$ (\ms)\tablenotemark{f} & Uniform & [0, 10000] \\
RV jitter, $\sigma_{w, \mathrm{TRES}}$ (\ms) & Log-uniform & [1, 150] \\
\enddata
\tablenotetext{a}{BJD Offset of 2457000}
\tablenotetext{b}{We parameterize the transit using $p = R_p/R_*$ and impact parameter $b$ directly}
\tablenotetext{c}{Quadratic limb-darkening coefficients, parameterized according to \cite{Kipping2013_limbdarkening}.}
\tablenotetext{d}{Eccentricity is parameterized using $\sqrt{e}\cos\omega$ and $\sqrt{e}\sin\omega$ to avoid the eccentric anomaly solver failing at high eccentricities. An upper limit of $e < 0.8$ was enforced during sampling.}
\tablenotetext{e}{TESS dilution prior accounts for 4.5\% contamination in the 
TESS aperture, with 10\% uncertainty on this correction.}
\tablenotetext{f}{Center-of-mass velocity with respect to the zero-point of the relative TRES RVs reported in Table \ref{tab:tres_rvs}.}
\end{deluxetable}

\end{document}